\documentclass[aps,superscriptaddress,prc,reprint,nofootinbib,showkeys]{revtex4-1}
\pdfoutput=1
\usepackage{graphicx}
\usepackage{dcolumn}
\usepackage{bm}

\usepackage{amsmath,amsfonts,amsthm,amssymb}
\usepackage{graphics}
\usepackage{cancel}
\usepackage{multirow}
\usepackage{longtable}
\usepackage{color}
\usepackage[normalem]{ulem} 

\def\be{\begin{equation}}
\def\ee{\end{equation}}
\newcommand{\bel}[1]{\begin{eqnarray}\label{#1}}
\newcommand{\eel}{\end{eqnarray}}
\def\barr{\begin{array}}
	\def\earr{\end{array}}
\def\beq{\begin{eqnarray}}
\def\eeq{\end{eqnarray}}
\def\bfig{\begin{figure}}
	\def\efig{\end{figure}}

\newcommand{\nn}{\nonumber}
\newcommand{\f}[2]{\frac{#1}{#2}}
\newcommand{\onehalf}{{\nicefrac{1}{2}}}

\newcommand{\p}{\partial}

\def\LR{\left(} 
\def\RR{\right)}

\newcommand{\ch}[1]{\cosh#1}

\def\GLW{{\rm GLW}}

\usepackage{amssymb}

\usepackage{graphicx}
\usepackage{amsmath}
\usepackage{subfigure}
\usepackage{bbold}
\usepackage{yfonts}
\usepackage[colorlinks=true,linktocpage=true,linkcolor=blue,citecolor=blue]{hyperref}
\usepackage{placeins}
\usepackage{bm} 
\usepackage{nicefrac}
\usepackage{slashed}
\usepackage{marginnote}

\newcommand{\bea}{\begin{eqnarray}}
\newcommand{\eea}{\end{eqnarray}}

\def\LB{\left(}
\def\RB{\right)}
\def\LSB{\left[}
\def\RSB{\right]}

\newcommand{\EQ}[1]{Eq.~(\ref{#1})}
\newcommand{\EQn}[1]{(\ref{#1})}

\newcommand{\CIT}[1]{Ref.~\cite{#1}} 
 
\newcommand{\CITn}[1]{\cite{#1}}

\newcommand{\Lv}{{\boldsymbol L}}
\newcommand{\Sv}{{\boldsymbol S}}
\newcommand{\Jv}{{\boldsymbol J}}

\newcommand{\pv}{{\boldsymbol p}}

\newcommand{\sv}{{\boldsymbol s}}

\def\omnU{\omega^{\mu\nu}}

\def\epsUabgd{\epsilon^{\alpha \beta \gamma \delta}}

\def\S0iU{{\Sigma}^{0i}}

\def\n0{n_{0}}
\def\e0{\varepsilon_{0}}
\def\P0{P_{0}}

\renewcommand\sout{\bgroup\markoverwith{\textcolor{blue}{\rule[0.5ex]{2pt}{0.4pt}}}\ULon}

\DeclareMathOperator{\sech}{sech}

\begin{document}
	
	\preprint{APS/123-QED}

	\title{Dissipative Spin Dynamics in Relativistic Matter}
	\author{Samapan Bhadury}
	\email{samapan.bhadury@niser.ac.in}
	\affiliation{School of Physical Sciences, National Institute of Science Education and Research, HBNI, Jatni-752050, India}
	\author{Wojciech Florkowski}
	\email{wojciech.florkowski@uj.edu.pl}
	\affiliation{Institute of Theoretical Physics, Jagiellonian University, ul. St. \L ojasiewicza 11, 30-348 Krakow, Poland}
	\author{Amaresh Jaiswal}
	\email{a.jaiswal@niser.ac.in}
	\affiliation{School of Physical Sciences, National Institute of Science Education and Research, HBNI, Jatni-752050, India}
	\author{Avdhesh Kumar} 
	\email{avdhesh.5000@gmail.com} 
	\affiliation{School of Physical Sciences, National Institute of Science Education and Research, HBNI, Jatni-752050, India}
	\author{Radoslaw Ryblewski} 
	\email{radoslaw.ryblewski@ifj.edu.pl}
	\affiliation{Institute of Nuclear Physics Polish Academy of Sciences, PL-31-342 Krakow, Poland}
	\date{\today}
	
\begin{abstract}
Using classical description of spin degrees of freedom, we extend recent formulation of the perfect-fluid  hydrodynamics for spin-polarized fluids to the case including dissipation. Our work is based on the analysis of classical kinetic equations for massive particles with spin $\onehalf$, with the collision terms treated in the relaxation time approximation. The kinetic-theory framework determines the structure of viscous and diffusive terms and allows to explicitly calculate a complete set of new kinetic coefficients that characterize dissipative spin dynamics.
\end{abstract}
     
\date{\today}


\maketitle

\section{Introduction}
\label{intro}
In non-central ultra-relativistic heavy-ion collisions, the two colliding nuclei carry large amount of orbital angular momentum $\Lv$. Soon after the initial impact, a~substantial part of $\Lv$ is deposited in the interaction zone and can be further transformed to the spin part $\Sv$ (with the total angular momentum $\Jv=\Lv+\Sv$ being conserved). The latter can be reflected in the spin polarization of the particles emitted at freeze-out. To verify this phenomenon, the spin polarization of various particles ($\Lambda$, $K^*$, $\phi$) produced in relativistic heavy-ion collisions has been recently measured by the STAR~\cite{STAR:2017ckg,Adam:2018ivw}, ALICE~\cite{Acharya:2019vpe} and HADES~\cite{Kornas:2019} experiments. 

On the theoretical side, first predictions of a non-zero global spin polarization of the $\Lambda$ hyperons, based on perturbative-QCD calculations and the spin-orbit interaction, were made in Refs.~\cite{Liang:2004ph,Liang:2004xn} and \cite{Voloshin:2004ha}, respectively (see also \CIT{Betz:2007kg}). In these works, a substantial polarization effect of the order of 10\% was found. Subsequently, using relativistic hydrodynamics with local thermodynamic equilibrium of the spin degrees of freedom~\cite{Becattini:2007sr,Becattini:2013vja,Becattini:2013fla,Becattini:2007nd,Becattini:2016gvu,Becattini:2015ska,Karpenko:2016jyx,Xie:2017upb,Pang:2016igs,Becattini:2017gcx}, a~smaller polarization of about 1\% was predicted, an effect which was eventually confirmed by STAR~\cite{STAR:2017ckg,Adam:2018ivw}.

Interestingly, the same hydrodynamic models~\cite{Becattini:2017gcx,Becattini:2020ngo} are not able to describe the experimentally measured longitudinal polarization of $\Lambda$'s~\cite{Niida:2018hfw,Adam:2019srw}. For example, the oscillation of the longitudinal polarization of the $\Lambda$ hyperons measured as a function of the azimuthal angle in the transverse plane~\cite{Niida:2018hfw} has an opposite sign compared to the results obtained with relativistic hydrodynamics with thermalized spin degrees of freedom. This issue is at the moment the subject of very intensive investigations \cite{Li:2017dan,Li:2017slc,Pang:2016igs,Xie:2017upb,Sun:2017xhx,Sun:2018bjl,Fang:2016vpj,Florkowski:2019qdp,Florkowski:2019voj,Gao:2020vbh,Li:2020vwh,Liu:2020bbd,Liu:2019krs,Ayala:2020ndx,Ivanov:2020qqe,Liu:2020ymh,Huang:2020xyr,Deng:2020ygd,Montenegro:2020paq,Ivanov:2019ern}. 

The relativistic hydrodynamic models (perfect or viscous) that have been used so far to describe the global spin polarization of the $\Lambda$ and $\bar\Lambda$ hyperons~\cite{Becattini:2016gvu,Karpenko:2016jyx,Becattini:2017gcx} make use of the fact that spin polarization effects are governed by the thermal vorticity tensor
\bea
\varpi_{\mu \nu}= -\frac{1}{2} (\partial_\mu \beta_\nu-\partial_\nu \beta_\mu).
\label{eq:thermvor}
\eea
Here the four-vector $\beta_\mu$ is defined in the standard way as the ratio of the fluid flow vector $u_\mu$ and the local temperature $T$, i.e., $\beta_\mu = u_\mu/T$. One can notice that the use of \EQn{eq:thermvor} does not require any modifications of the existing hydrodynamic codes as spin effects are determined solely by the form of $u^\mu$ and $T$.

However, on the general thermodynamic grounds~\cite{Becattini:2018duy}, it is expected that the spin polarization effects may be  governed by the tensor $\omega_{\mu \nu}$ (called below the spin polarization tensor) that can be independent of the thermal vorticity \EQn{eq:thermvor}.  This suggests that a completely new hydrodynamic approach including spin dynamics can be constructed, with the spin polarization tensor $\omega_{\mu \nu}$ treated as an independent hydrodynamical variable. In this context, the concept of local spin equilibrium also changes as one no longer requires that $\omega_{\mu \nu} = \varpi_{\mu \nu}$ to have zero entropy production. 

First steps to formulate the perfect-fluid version of hydrodynamics of spin polarized fluids that incorporates the spin polarization tensor $\omega_{\mu\nu}$ have already been made in a series of publications~\cite{Florkowski:2017ruc,Florkowski:2017dyn,Becattini:2018duy}, for a recent summary see~\CIT{Florkowski:2018fap}. However, only in a very recent work~\CITn{Bhadury:2020puc}, the dissipation effects in such systems have been explicitly considered, see also Refs.~\cite{Weickgenannt:2020aaf,Speranza:2020ilk,Hattori:2019ahi,Yang:2020hri,Hattori:2019lfp, Shi:2020htn,Gallegos:2020otk}. 

In this work we continue and significantly extend the results obtained in \CIT{Bhadury:2020puc}. In order to identify the structure of dissipative terms, we use classical kinetic theory for particles with spin $\onehalf$.  The collision terms are treated in the relaxation time approximation (RTA) according to the prescription defined in \CIT{Bhadury:2020puc} and, for the sake of simplicity, we restrict our considerations to the Boltzmann statistics. The kinetic-theory framework determines the structure of viscous and diffusive terms and allows to explicitly calculate a set of new kinetic coefficients that characterize dissipative spin dynamics. These coefficients describe coupling between a non-equilibrium part of the spin tensor and thermodynamic forces such as the expansion scalar, shear flow tensor, the gradient of chemical potential divided by temperature, and, finally, the gradient of the spin polarization tensor.  

\smallskip
The structure of the paper is as follows: In Sec. II we recall the formulation of the perfect-fluid hydrodynamics with spin. Our presentation is based on the classical concept of spin and classical distribution functions in an extended phase space. In Sec. III we introduce kinetic equations with the collision terms treated in the relaxation time approximation and derive the form of the dissipative corrections. This section contains also the explicit form of the new, spin-related kinetic coefficients. We conclude and summarize in Sec. IV. The paper is closed with several appendices where details of our straightforward but quite lengthy calculations are given. We use natural units and the metric tensor with the signature $(+---)$.

\section{Formulation of perfect fluid hydrodynamics for spin polarized fluids}
\label{sec:1}
\subsection{Spin-dependent equilibrium distribution function}
\label{subsec:1}

We start with the classical treatment of massive particles with spin-$\onehalf$ and introduce their internal angular momentum $s^{\alpha\beta}$~\cite{Mathisson:1937zz}. It is connected with the particle four-momentum $p_\gamma$ and spin four-vector~$s_\delta$~\cite{Itzykson:1980rh}  by the following relation~\footnote{We follow here the sign conventions used in our previous publications, e.g., in \cite{Florkowski:2018fap}. We note that they are different from those used in \cite{Weickgenannt:2019dks}. }
\bea
s^{\alpha\beta} = \f{1}{m} \epsUabgd p_\gamma s_\delta,
\label{eq:salbe}
\eea
where $m$ is the particle mass. Equation \EQn{eq:salbe} implies that $s^{\alpha\beta} = -s^{\beta\alpha}$ and $p_\alpha s^{\alpha\beta}  = 0$. Moreover,  assuming that the four-vectors $p$ and  $s$ are orthogonal to each other we find
\bea
s^{\alpha} = \f{1}{2m} \epsUabgd p_\beta s_{\gamma \delta}.
\label{eq:sabinv}
\eea
In the particle rest frame (PRF), where the four-momentum of a particle is $p^\mu = (m,0,0,0)$, the spin four-vector $s^\alpha$ has only spatial components, i.e., $s^\alpha = (0,\sv_*)$, with the length of the spin vector defined by $-s^2 = |\sv_*|^2={\mathfrak{s}}^2 =  \f{1}{2} \left( 1+ \f{1}{2}  \right)$. 

Identification of the so-called collisional invariants of the Boltzmann equation allows us to construct the equilibrium distribution functions $f^\pm_{s, \rm eq}(x,p,s)$ for particles and antiparticles~\cite{Florkowski:2018fap,Bhadury:2020puc},
\begin{equation}
f^\pm_{s, \rm eq}(x,p,s) =  f_{\rm eq}^{\pm}(x,p)\exp\left[\frac{1}{2} \omega_{\mu\nu} (x) s^{\mu\nu} \right].
\label{eq:feqxps}
\end{equation}
Here $f_{\rm eq}^{\pm}(x,p)=\exp\left[-p^\mu\beta_\mu(x)\pm\xi(x)\right]$ is the J\"uttner distribution, with $\xi$ and $\beta_{\mu}$ traditionally defined as ratios of chemical potential $\mu$ to temperature $T$ and four-velocity $u_\mu$ to temperature $T$, i.e., $\xi=\mu/T$ and $\beta_\mu=u_\mu/T$.~\footnote{We note that since we always consider particles being on the mass shell ($p^0 = E_p = \sqrt{\pv^2+m^2}$) the distribution $f(x,p,s)$ is in fact a function of $\pv$ only.} The spin polarization tensor $\omega_{\mu\nu}$ has been introduced in Sec.~\ref{intro}. It plays a crucial role in our formalism and can be interpreted as the (tensor) potential conjugated to the spin angular momentum.

Before we proceed further we note that in our approach $s^{\mu\nu}$ is dimensionless (measured in units of $\hbar$) and so is $\omega_{\mu\nu}$. Consequently, we can make expansions in $\omega_{\mu\nu}$ and, in fact, most of our results will be valid in the leading order of $\omega_{\mu\nu}$.

Ordinary phase-space equilibrium distribution functions can be obtained by integrating out the spin degrees of freedom present in $f^\pm_{s, \rm eq}(x,p,s)$, 
\begin{equation}
\int \mathrm{dS} \, f^\pm_{s, \rm eq}(x,p,s) = f^\pm_{\rm eq}(x,p), 
\label{a}
\end{equation}
where~\cite{Florkowski:2018fap}
\bea
\mathrm{dS} &=& \frac{m}{\pi {\mathfrak{s}}} \, \mathrm{d}^4s~\delta(s\cdot s + {{\mathfrak{s}}}^2)~\delta(p\cdot s)\label{eq:dS}.
\eea
Different properties of spin integrals done with the integration measure \EQn{eq:dS} are collected in Appendix \ref{sec:appspin}.

\subsection{Perfect fluid hydrodynamics for spin polarized fluids}
\label{subsec:2}

For a system of particles and anti-particles with spin degrees of freedom included only through degeneracy factors, the relevant conserved quantities are the energy-momentum tensor ($T^{\mu\nu}$) and charge current ($N^\mu$). If spin is explicitly included, one has to consider an additional conserved quantity, namely, the angular-momentum tensor ($J^{\lambda, \mu\nu}$)~\cite{Florkowski:2018fap,Florkowski:2018ahw}. This is connected with the fact that the total angular momentum conservation law for particles with spin has a non-trivial form. 

The total angular-momentum tensor ($J^{\lambda, \mu\nu}$) can be written as a sum of the orbital ($L^{\lambda, \mu\nu}$) and spin ($S^{\lambda, \mu\nu}$) parts. The latter is known as the spin tensor. It is well known that there are various equivalent forms of the energy-momentum and spin tensors that can be used to define system's dynamics \cite{Hehl:1976vr,Speranza:2020ilk,Tinti:2020gyh}. The forms used in this work agree with the definitions introduced by de Groot, van Leeuwen, and van Weert in \cite{DeGroot:1980dk}. To emphasize this fact we sometimes use the acronym GLW.

The structures of $T^{\mu\nu}$, $N^\mu$ and $S^{\lambda, \mu\nu}$ can be connected to the behaviour of microscopic constituents of the system through the moments of the  phase-space distribution functions  $f_{\rm eq}(x,p,s)$. Using the equilibrium distributions $f_{\rm eq}(x,p,s)$ defined above, the hydrodynamic quantities such as charge current, energy-momentum tensor, and the spin tensor can be obtained in the similar way as in standard hydrodynamics. 

\subsection{Charge current}
\label{subsubsec:1}
The equilibrium charge current is defined by the formula
\beq
N^\mu_{\rm eq} 
&=& \!\int \! \mathrm{dP}~\mathrm{dS} \, \, p^\mu \, \left[f^+_{s, \rm eq}(x,p,s)\!-\!f^-_{s, \rm eq}(x,p,s) \right],
\label{eq:Neq-sp0}
\eeq
where the invariant momentum integration measure $\mathrm{dP}$ is 
\bea
\mathrm{dP} &=& \frac{d^3p}{(2 \pi )^3 E_p},
\label{eq:dP}
\eea
while the measure $\mathrm{dS}$ is defined by \EQ{eq:dS}. Using the equilibrium functions (\ref{eq:feqxps}) we obtain
\bea
N^\mu_{\rm eq} = 2 \sinh(\xi) \int \mathrm{dP} \, p^\mu 
e^{- p \cdot \beta}
\int \mathrm{dS} \exp\LB \f{1}{2}  \omega_{\alpha \beta} s^{\alpha\beta} \RB.\nn\\
\label{eq:Neq-sp1}
\eea
Since for large values of the spin polarization tensor the system becomes anisotropic in the momentum space and requires special treatment~\cite{Florkowski:2010cf,Martinez:2010sc}, in most of our calculations we consider only the case of small values of $\omega$. In this case the last exponential function in \EQn{eq:Neq-sp1} can be expanded up to linear order and we find
\bea
N^\mu_{\rm eq} &=& 2 \sinh(\xi) \int \mathrm{dP} \, p^\mu \, e^{- p \cdot \beta} \int \mathrm{dS} \, \left(1 +  \f{1}{2}  \omega_{\alpha \beta} s^{\alpha\beta}\right). \nn\\ 
\label{eq:Neq-sp21}
\eea
After carrying out integration first over spin and then over momentum we get
\bel{Nmu}
N^\alpha_{\rm eq} = n u^\alpha,
\eel
where 
\bel{nden}
n = 4 \, \sinh(\xi)\, n_0(T)
\label{eq:n_eql}
\eel
is the charge density~\cite{Florkowski:2017ruc}.
In Eq.~(\ref{nden}) the quantity $n_0(T)$ is the number density of spinless, neutral massive Boltzmann particles which is defined by the thermal average
\bel{avdef}
n_0(T)= \langle u\cdot p\rangle_0 ,
\eel
where
\bel{avdef1.1}
\langle \cdots \rangle_0 \equiv \int \mathrm{dP}  (\cdots) \,  e^{- \beta \cdot p}.
\eel
The explicit calculation gives
\bea
n_{0}(T) &=& \int \mathrm{dP} \, (u\cdot p)\,  e^{- \beta \cdot p} =I_{{10}}^{(0)} \nn \\
&=& \frac{1}{2 \pi ^2} T^3 z^2  K_2 (z) \,,
\eea
with $z\equiv m/T$. Thermodynamic integrals $I_{nq}^{(r)}$ are defined in Appendix \ref{sec:thermint}.

\subsection{Energy-momentum tensor}
\label{subsubsec:2}

The energy-momentum tensor is defined as the second moment in momentum space,
\bea
T^{\mu \nu}_{\rm eq}
&=& \int \mathrm{dP}~\mathrm{dS} \, \, p^\mu p^\nu \, \left[f^+_{s, \rm eq}(x,p,s) + f^-_{s, \rm eq}(x,p,s) \right]. \nn\\
\label{eq:Teq-sp02}
\eea
Using Eq.~(\ref{eq:feqxps}) we can rewrite this formula as
\bea
T^{\mu \nu}_{\rm eq}
&=& 2 \cosh(\xi) \int \mathrm{dP} \, p^\mu p^\nu \, e^{- p \cdot \beta}
\int \mathrm{dS} \exp\LB  \f{1}{2}  \omega_{\alpha \beta} s^{\alpha\beta} \RB. \nn\\
\label{eq:Teq-sp03}
\eea
Considering the case of small $\omega$ and carrying out integration over spin and momentum space we get
\bel{Tmn}
T^{\alpha\beta}_{\rm eq}(x) &=& \varepsilon u^\alpha u^\beta - P \Delta^{\alpha\beta},
\eel
where
\bel{enden}
\varepsilon = 4 \,\cosh(\xi) \, \e0(T)
\eel
and
\bel{prs}
P = 4 \, \cosh(\xi) \, \P0(T),
\label{eq:P_eql}
\eel
respectively~\cite{Florkowski:2017ruc}. The auxiliary quantities $\e0(T)$ and $\P0(T)$ are defined as follows
\bel{enden0}
\e0(T) &=& \langle(u\cdot p)^2\rangle_0
\eel
and
\bel{prs0}
\P0(T) = -(1/3) \langle \,  p\cdot p - (u\cdot p)^2\,   \rangle_0. 
\eel
Similarly to $n_0(T)$, they describe the energy density and pressure of spinless, neutral massive Boltzmann particles. In Eq.~(\ref{Tmn}), the tensor $\Delta^{\alpha\beta} = g^{\alpha\beta} - u^{\alpha} u^{\beta}$ is an operator projecting on the space orthogonal to the fluid four-velocity $u^{\mu}$. For the reader's convenience, the properties of this and other projectors are listed in Appendix~\ref{sec:projectors}.

With the help of thermodynamic integrals $I_{nq}^{(r)}$  defined in Appendix \ref{sec:thermint} one obtains
\bea
\varepsilon_0(T) &=& \int \mathrm{dP}\,(u\cdot p)^2 e^{- \beta \cdot p}=I^{(0)}_{20} \nn \\
&=& \frac{1}{2 \pi ^2} T^4 z^2 \left[3 K_2 (z)+z K_1 (z)\right]
\eea
and
\bea
P_0(T) &=& -\frac{1}{3}\Delta_{\mu\nu} \int \mathrm{dP}\, p^{\mu}p^{\nu} e^{- \beta \cdot p} \nn \\ 
&=& -\frac{1}{3}\int \mathrm{dP}\,\left[ p\cdot p-(u\cdot p)^2\right] e^{- \beta \cdot p}=-I^{(0)}_{21} \nn \\
&=&  \frac{1}{2 \pi ^2}{T^4 z^2  K_2 (z)}= n_0 (T) T.
\label{eq:p0n0T}
\eea

\subsection{Spin tensor}
\label{subsubsec:3}

Now we come to the fundamental object in our formalism, namely, the spin tensor. We adopt
the following definition \cite{Florkowski:2018fap}
\beq
S^{\lambda, \mu\nu}_{\rm eq} &=& \int  \mathrm{dP}~\mathrm{dS} \, \, p^\lambda \, s^{\mu \nu} 
\left[f^+_{s, \rm eq}(x,p,s) + f^-_{s, \rm eq}(x,p,s) \right] \nn \\
&=& 2 \cosh(\xi) \int  \mathrm{dP} \, p^\lambda \exp\LB - p \cdot \beta \RB \label{eq:Seq-sp01} \\
&& \hspace{1cm} \times
\int  \mathrm{dS} \, s^{\mu \nu} \, \exp\LB \f{1}{2}  \omega_{\alpha \beta} s^{\alpha\beta} \RB. \nonumber
\eeq
Expanding the exponential function in the last line, in the leading order in $\omega$ we obtain
\bea
&&\hspace{-0.cm}  \int  \mathrm{dS} \, s^{\mu \nu}  \, \exp\LB  \f{1}{2}  \omega_{\alpha \beta} s^{\alpha\beta}\RB =
\int  \mathrm{dS} \, s^{\mu \nu}  \, \left(1 +   \f{1}{2}  \omega_{\alpha \beta} s^{\alpha\beta}\right)
\nn \\
&&\hspace{1.5cm}=\f{2}{3 m^2} {{\mathfrak{s}}}^2 \LB m^2 \omnU + 2 p^\alpha p^{[\mu} \omega^{\nu ]}_{\,\,\alpha} \RB.
\label{eq:spinint3}
\eea
Using \EQ{eq:spinint3} in \EQ{eq:Seq-sp01} we find
\bea
S^{\lambda, \mu\nu}_{\rm eq} 
&=& \f{4 {{\mathfrak{s}}}^2}{3 m^2} \cosh(\xi) \!\!\int \!\! \mathrm{dP} \, p^\lambda \, e^{ - p \cdot \beta }
\LB m^2 \omnU \!+\! 2 p^\alpha p^{[\mu} \omega^{\nu ]}_{\,\,\alpha} \RB. \nn\\
\label{eq:Seq-sp1}
\eea
It is interesting to observe that the last result agrees with the formula $S^{\lambda , \mu \nu }_{\rm GLW}$ obtained in the semiclassical expansion of the Wigner functions~\CITn{Florkowski:2018ahw}. This fact supports our use of the definition \EQn{eq:Seq-sp01}.

After carrying out the momentum integration we get
\bea
S^{\lambda, \mu\nu}_{\rm eq}=S^{\lambda , \mu \nu }_{\rm GLW}
&=&  {\cal C} \left( n_{0}(T) u^\lambda \omega^{\mu\nu}  +  S^{\lambda , \mu \nu }_{\Delta\GLW} \right).
\label{eq:Smunulambda_de_Groot2}
\eea 
Here ${\cal C}= (4/3)\mathfrak{s}^2 \ch({\xi)}$ and the auxiliary tensor $S^{\lambda , \mu \nu }_{\Delta\GLW}$ is given by the expression
\beq
S^{\alpha, \beta\gamma}_{\Delta\GLW} 
&=&  {\cal A}_{0} \, u^\alpha u^\delta u^{[\beta} \omega^{\gamma]}_{~\delta} \label{SDeltaGLW} \\
&+&{\cal B}_{0} \, \Big( 
u^{[\beta} \Delta^{\alpha\delta} \omega^{\gamma]}_{~\delta}
+ u^\alpha \Delta^{\delta[\beta} \omega^{\gamma]}_{~\delta}
+ u^\delta \Delta^{\alpha[\beta} \omega^{\gamma]}_{~\delta}\Big),\nn
\eeq
where
\beq 
{\cal B}_{0} &=&-\frac{2}{z^2}  \frac{\varepsilon_{0}(T)+P_{0}(T)}{T}=-\frac{2}{z^2} s_{0}(T)\label{coefB}
\eeq
and
\beq
{\cal A}_{0} &=&\frac{6}{z^2} s_{0}(T) +2 n_{0} (T) = -3{\cal B}_{0} +2 n_{0}(T),\nn\\
\label{coefA}
\eeq
with $s_0$ being the entropy density of spinless, neutral, massive Boltzmann particles satisfying thermodynamic relation $s_0 =\LR \varepsilon_0+ P_0\RR / T$. 

We note that since our energy-momentum tensor is symmetric, the spin tensor is separately conserved. The conservation of the spin tensor gives six additional equations which are required to determine the space-time evolution of $\omega$. We note that this situation may change if non-local effects are included, for a very recent discussion of this point see Refs.~\cite{Weickgenannt:2020aaf,Speranza:2020ilk}.

\subsection{Entropy Current}
\label{subsubsec:4}

To construct the entropy current we adopt the Boltzmann definition
\bea
H^\mu &=& -\!\!\int \! \mathrm{dP}~\mathrm{dS} \, p^\mu
\LSB 
f^+_{s, \rm eq} \LB \ln f^+_{s, \rm eq} -1 \RB 
\right. \nn \\
&& \left. \hspace{1.5cm}
+ 
f^-_{s, \rm eq} \LB \ln f^-_{s, \rm eq} -1 \RB \RSB.
\label{eq:H1}
\eea 
Using Eqs.~(\ref{eq:feqxps}), (\ref{eq:Neq-sp0}), (\ref{eq:Teq-sp02}), and  (\ref{eq:Seq-sp01}), we find
\bea
H^\mu = \beta_\alpha T^{\mu \alpha}_{\rm eq} - \f{1}{2} \omega_{\alpha\beta} S^{\mu, \alpha \beta}_{\rm eq} - \xi N^\mu_{\rm eq} + P \beta^\mu.
\label{eq:H2}
\eea 
Using Eqs.~\eqref{eq:n_eql}, \eqref{eq:P_eql} and \eqref{eq:p0n0T}, one can obtain the relation, $P \beta^\mu = (\cosh(\xi)/\sinh(\xi))  N^\mu_{\rm eq}$.
Using Eq.~(\ref{eq:H2}) as well as the conservation laws for charge, energy-momentum and spin we obtain the following expression,
\bea
\p_\mu H^\mu &=& \left( \p_\mu \beta_\alpha \right) T^{\mu \alpha}_{\rm eq} 
-\f{1}{2} \left( \p_\mu \omega_{\alpha\beta} \right) S^{\mu, \alpha \beta}_{\rm eq}
\nn \\
&& 
- \left(\p_\mu \xi \right)  N^\mu_{\rm eq} + \p_\mu (P \beta^\mu) .\nn\\
\label{eq:H3}
\eea 
Now starting from the definition (\ref{eq:Neq-sp0}), applying the conservation laws for charge current, and using Eqs. (\ref{eq:Neq-sp0}), (\ref{eq:Teq-sp02}), (\ref{eq:Seq-sp01}), and the definition of $P\beta^{\mu}$ given above, one can easily show that the right-hand side of the last equation is zero (see Appendix \ref{sec:entropy} for details of the proof), i.e., the entropy current is conserved,
\bea 
\p_\mu H^\mu = 0.
\label{eq:entcon}
\eea
It should be emphasized that the last result is exact in the sense that it does not depend on the expansion in $\omega$. Moreover, we see that the contributions to the entropy production coming from the spin polarization tensor are quadratic. This means that there is no effect on the entropy production from the polarization in the linear order. This suggests that we can neglect the effects of polarization on the global evolution of matter, provided we restrict our considerations to the linear terms. For both the conserved charge and the energy-momentum tensor the corrections start with the second order, hence, as long as we restrict ourselves to the linear terms in $\omega$, we can first solve the system of standard hydrodynamic equations (which are not affected by polarization in the linear order) and subsequently determine the spin evolution (linear in $\omega)$ on top of such a hydrodynamic background.
%
\section{Formulation of dissipative hydrodynamics for spin polarized fluids}
\label{sec:3}

The formalism presented in the previous section is already well established and may be treated as the definition of the perfect-fluid hydrodynamics with spin. In the next section, we include dissipation effects. This will be done with the help of the relaxation time approximation used for the collision terms in the classical kinetic equations, as originally introduced in Ref.~\cite{Bhadury:2020puc}.

\subsection{Classical RTA kinetic equation}
\label{sec:2.1}
In the absence of mean fields, the distribution function satisfies the equation
\bea
p^\mu \partial_\mu f^\pm_s(x,p,s) =C[f^\pm_s(x,p,s)],
\label{RTA_spin}
\eea
where $C[f^\pm_s(x,p,s)]$ is the collision term. In the relaxation time approximation, the collision term has the form~\CITn{Bhadury:2020puc}
\begin{equation}
C[f^\pm_s(x,p,s)] =  p \cdot u
\, \frac{f^\pm_{s, \rm eq}(x,p,s)-f_s^\pm(x,p,s)}{\tau_{\rm eq}}.
\label{eq:col}
\end{equation}
We consider now a simple Chapman-Enskog expansion of the single particle distribution function about its equilibrium value in powers of space-time gradients 
\begin{equation}
f^\pm_s(x,p,s) =f^\pm_{s, \rm eq}(x,p,s)+\delta f^\pm_s(x,p,s).
\label{cha-ensk}
\end{equation}
In the above equation $\delta f^\pm_s(x,p,s)$ is a deviation from the equilibrium single-particle distribution function and, in principle, can be of any order in space-time gradients. 

 Using Eqs.~(\ref{eq:col}) and~(\ref{cha-ensk}) in Eq.~(\ref{RTA_spin}), and keeping only the first-order terms in space-time gradients, we get

\begin{equation}\label{RTA_spin2}
p^\mu \partial_\mu f^\pm_{s, \rm eq}(x,p,s) =-p \cdot u
\, \frac{\delta f^\pm_s(x,p,s)}{\tau_{\rm eq}}.
\end{equation}
%
After substituting equilibrium distribution function (\ref{eq:feqxps})  in Eq.~(\ref{RTA_spin2}) we obtain (in linear order in $\omega$)
\begin{widetext}
\beq
\delta f^\pm_s &=&- \frac{\tau_{\rm eq}}{(u \cdot p)}e^{\pm\xi - p\cdot \beta} \bigg[ \Big(\pm p^\mu\partial_\mu\xi - p^\lambda p^\mu\partial_\mu\beta_\lambda\Big) \bigg(1 + \frac{1}{2}s^{\alpha\beta}\omega_{\alpha\beta}\bigg)+\frac{1}{2} p^\mu s^{\alpha\beta}(\partial_\mu\omega_{\alpha\beta})\bigg].\label{del feq lim}
\eeq
\end{widetext}
The corrections $\delta f^\pm_s$ result in dissipative effects in the conserved quantities such as charge current, energy-momentum tensor, and spin tensor. We discuss them now starting from the simplest case of the charge current. The details of rather lengthy calculations are given in Appendix~\ref{sec:appD}.

\subsection{Conserved hydrodynamic quantities and dissipative corrections}

Taking the appropriate moments of  the transport equation (\ref{RTA_spin}), the following equations for the charge current ($N^{\mu}$), energy-momentum tensor ($T^{\mu\nu}$) and spin tensor ($S^{\lambda , \mu \nu}$) can be obtained
\beq
\p_\mu N^{\mu}(x)&=&-u_{\mu}\left(\frac{N^{\mu}(x)-N^{\mu}_{\rm eq}(x)}{\tau_{\rm eq}}\right), \label{eq:cc2}\\
\p_\mu T^{\mu\nu}(x)&=&-u_{\mu}\left(\frac{T^{\mu\nu}(x)-T^{\mu\nu}_{\rm eq}(x)}{\tau_{\rm eq}}\right), \label{eq:emt2}\\
\p_\lambda S^{\lambda , \mu \nu}(x)&=&-u_{\lambda}\left(\frac{S^{\lambda , \mu \nu}(x)-S^{\lambda , \mu \nu}_{\rm eq}(x)}{\tau_{\rm eq}}\right),\label{eq:st2}
\eeq
respectively.

Conservation of the charge current ($\p_\mu N^{\mu}=0$), energy-momentum tensor ($\p_\mu T^{\mu\nu}=0$), and spin tensor ($\p_\lambda S^{\lambda , \mu \nu}=0$) implies that the quantities on the right-hand sides of Eqs.~(\ref{eq:cc2})--(\ref{eq:st2}) should be zero, i.e., we must have
\beq
u_{\mu}\delta N^{\mu}&=&0,\label{eq:lm1}\\
	u_{\mu}\delta T^{\mu\nu}&=&0,\label{eq:lm2}\\
	u_{\lambda}\delta S^{\lambda , \mu \nu}&=& 0, \label{eq:lm3}
\eeq
where  $\delta N^\mu$, $\delta T^{\mu\nu}$, and $\delta S^{\lambda,\mu\nu}$ are defined in terms of the non-equilibrium parts of the distribution functions:
\begin{align}
&\delta N^\mu = \int \mathrm{dP}~\mathrm{dS}~p^\mu (\delta f^+_s-\delta f^-_s),\label{eq:dN}\\
&\delta T^{\mu\nu} = \int \mathrm{dP}~\mathrm{dS}~p^\mu p^\nu (\delta f^+_s+\delta f^-_s)\label{eq:dT},\\
&\delta S^{\lambda,\mu\nu}=\int \mathrm{dP~dS}~p^{\lambda} s^{\mu\nu} (\delta f^+_s+\delta f^-_s).\label{eq_dspin}
\end{align}
Note that Eqs.~(\ref{eq:lm1}) and  (\ref{eq:lm2}),  satisfied by the corrections $\delta N^\mu$ and $\delta T^{\mu\nu}$, are known in the literature as the Landau matching conditions. They are used (and needed) to determine the values of the chemical potential, temperature, and three independent components of the flow four-vector appearing in the equilibrium distributions defined by Eq.~(\ref{eq:feqxps}) --- altogether Eqs.~(\ref{eq:lm1}) and  (\ref{eq:lm2}) are five independent equations for five unknown functions. A novel feature of our approach is that we introduce an additional matching condition given by Eq.~(\ref{eq:lm3}). These are in fact six equations that allow us to determine six independent components of the spin polarization tensor $\omega_{\mu\nu}$. Below we refer to the complete set of  Eqs.~(\ref{eq:lm1})--(\ref{eq:lm3}) as to the Landau matching conditions.

The conserved quantities obtained from the moments of the transport equations (\ref{RTA_spin}) can be further tensor decomposed in terms of the hydrodynamic degrees of freedom. The charge current is decomposed into two parts
\beq
N^\mu &=&\int  \mathrm{dP}~ \mathrm{dS}~p^{\mu} \left[f^+_s(x, p, s) - f^-_s(x, p, s)\right]\nn\\&=&N^\mu_{\rm eq}+\delta N^{\mu}=nu^{\mu}+\nu^\mu\label{Nmu1}.
\eeq
In this decomposition, the quantity $\nu^\mu$ is known as the charge diffusion current. The presence of the dissipative corrections implies that the form of the energy-momentum tensor is
\bea
T^{\mu \nu}&=&\int  \mathrm{dP}~ \mathrm{dS}~p^{\mu} 
p^{\nu} \left[f^+_s(x, p, s) + f^-_s(x, p, s)\right]\nn\\
&=& T^{\mu \nu}_{\rm eq}+\delta T^{\mu \nu} \nn \\
&=& \varepsilon u^\mu u^\nu - P \Delta^{\mu\nu}+\pi^{\mu\nu}-\Pi \Delta^{\mu\nu}. \label{Tmunu1}
\eea
In this decomposition, $\varepsilon, P, \pi^{\mu\nu}$, and $\Pi$ are energy density, equilibrium pressure, shear stress tensor, and bulk pressure,  respectively. We use here the Landau frame, where $T^{\mu\nu}u_{\nu}=\varepsilon u^\mu$. Finally, we define the correction to the spin tensor by the decomposition
\beq
S^{\lambda,\mu\nu} &=&\int \mathrm{dP~dS}~p^{\lambda} s^{\mu\nu}\left[f^+_s(x, p, s) + f^-_s(x, p, s)\right]\nn\\&=&S^{\lambda,\mu\nu}_{\rm eq}+\delta S^{\lambda,\mu\nu}.\label{Slmunu}
\eeq

The non-equilibrium quantities $n$, $\varepsilon$, $P$ can be obtained by the Landau matching conditions, namely
\bea
n &=& n_{\rm {eq}} = u_{\mu } N_{\rm {eq}}^{\mu } \label{eq:no_den} \\
&=&u_{\mu } \int  \mathrm{dP}~\mathrm{dS} \, p^{\mu }\left[f^{+}_{s, \rm {eq}}(x,p,s)-f^{-}_{s, \rm {eq}}(x,p,s)\right],  \nn
\eea
\bea
\varepsilon &=& \varepsilon_{\rm {eq}} = u_{\mu }u_{\nu} T^{\mu\nu}_{\rm {eq}}  \label{eq:end} \\
&=&u_{\mu} u_{\nu} \int  \mathrm{dP}~\mathrm{dS} \, 
p^{\mu }p^{\nu}\left[f^{+}_{s, \rm {eq}}(x,p,s) + f^{-}_{s, \rm {eq}}(x,p,s)\right] \nn
\eea
and
\bea
P &=& P_{\rm {eq}} = -\frac{1}{3}\Delta_{\mu\nu} 
T^{\mu\nu}_{\rm {eq}} \label{eq:P} \\
&=&-\frac{\Delta_{\mu\nu}}{3}
\int  \mathrm{dP}~\mathrm{dS} \, p^{\mu}p^{\nu}\left[f^{+}_{s, \rm {eq}}(x,p,s) + f^{-}_{s, \rm {eq}}(x,p,s)\right].\nn
\eea
After carrying out integration over spin and momentum, Eqs.~(\ref{eq:no_den}), (\ref{eq:end}), and (\ref{eq:P}) yield the same results as Eqs.~(\ref{nden}), (\ref{enden}), and (\ref{prs}). Here we also note that the choice of Landau frame and matching conditions enforces the following constraints on the dissipative currents
\beq
u_\mu \nu^\mu&=&0, \nn\\
u_\mu \pi^{\mu\nu}&=&0. \nn\\
u_\lambda \delta S^{\lambda,\mu\nu}&=&0. \label{matching_conditions}
\eeq
\subsection{Convective derivatives of hydrodynamic variables}
\label{Sbc}

An intermediate step in the calculation of standard kinetic coefficients is the derivation of expressions for the convective derivatives of the hydrodynamic variables $\xi$, $\beta$, and $u^\mu$. The convective derivatives are space-time derivatives taken along the streamlines of the fluid. We denote them by a dot or the letter $D$, for example, 
\beq
{\dot \xi} = D \xi 
= u^\mu \p_\mu \xi.
\eeq
With spin degrees of freedom included, one has to calculate the convective derivative of the spin polarization tensor $\omega_{\mu\nu}$ as well. In this section we describe the necessary steps needed to determine all those derivatives. The details of the calculations, which are quite lengthy due to complicated tensor structures, are given in the Appendices \ref{Ac}-\ref{sec:LanCon}.

Using the conservation laws for energy and momentum ($\partial_\mu T^{\mu \nu}= 0$) as well as charge  ($\partial_\mu N^{\mu \nu}= 0$), we get the following equations that dictate the evolution of $T$, $u^\mu$, and $\mu$,~\footnote{Equations~ (\ref{eq:con_e})--(\ref{eq:con_n}) do not include the spin polarization tensor, if we consider only linear terms in $\omega$.}
\begin{align}
&\dot{\varepsilon} + (\varepsilon + P + \Pi)~\theta -\pi^{\mu\nu}\sigma_{\mu\nu} = 0, \label{eq:con_e}\\
&(\varepsilon + P) \dot{u}^\alpha - \nabla^\alpha P + \Delta^\alpha_\mu \partial_\nu \pi^{\mu\nu} = 0, \label{eq:con_P}\\
&\dot{n} + n\theta + \partial_\mu n^\mu = 0. \label{eq:con_n}
\end{align}
Here we use the following notation:  $\theta=\partial_{\mu}u^{\mu}$ is the expansion scalar, $\nabla^\mu = \Delta^{\mu\nu} \partial_\nu$ denotes the transverse gradient, and $\sigma ^{\mu\nu}=\frac{1}{2}\left(\nabla ^{\mu }u^{\nu }+\nabla ^{\nu }u^{\mu}\right)-\frac{1}{3}\Delta ^{\mu\nu}\left.(\nabla ^{\lambda }u_{\lambda }\right)$ is the shear flow tensor. In order to determine the space-time evolution of the spin polarization tensor, the above system of equations should be supplemented by the conservation of the spin tensor, 
\begin{align}
\partial_\lambda S^{\lambda,\mu\nu}  = 0. \label{eq:con_S}
\end{align}

Keeping only the terms up to the first order in velocity gradients,  the conservation equations (\ref{eq:con_e}), (\ref{eq:con_P}), (\ref{eq:con_n}), and (\ref{eq:con_S}) are reduced to
\begin{align}
&\dot{\varepsilon} + (\varepsilon + P)~\theta= 0,\label{eq:con_e1}\\
&(\varepsilon + P) \dot{u}^\alpha - \nabla^\alpha P = 0,\label{eq:con_P1}\\
&\dot{n} + n\theta = 0,\label{eq:con_n1}\\
&\partial_\lambda S^{\lambda,\mu\nu}_{\rm {eq}}  = 0, \label{eq:con_S1}
\end{align}
respectively.
Furthermore, from Eqs.~(\ref{nden}) and (\ref{enden}) we obtain
\beq
\dot{n}&=& 4 \cosh (\xi )\dot{\xi } I_{10}^{(0)} +4 \sinh (\xi )\dot{I}_{10}^{(0)}, \label{eq:dotn}\\
\dot{\varepsilon }&=&4 \sinh (\xi)\dot{\xi} I_{20}^{(0)}+4\cosh (\xi)\dot{I}_{20}^{(0)}.\label{eq:enddot}
\eeq
Using Eq.~(\ref{eq:ir3}) that connects derivatives of the thermodynamic integrals, the above equations can be written as
\beq
\dot{n}&=& 4 \cosh (\xi )\dot{\xi } I_{10}^{(0)} -4 \sinh (\xi) \dot{\beta } I_{20}^{(0)} , \label{eq:dotn1.1}\\
\dot{\varepsilon }&=&4 \sinh (\xi)\dot{\xi} I_{20}^{(0)}-4\cosh (\xi) \dot{\beta }I_{30}^{(0)}. \label{eq:enddot1.1}
\eeq
Substituting $n$, $\varepsilon$, $P$, $\dot{n}$ and $\dot{\varepsilon }$ from Eqs. (\ref{nden}), (\ref{enden}), (\ref{prs}), (\ref{eq:dotn1.1}) and (\ref{eq:enddot1.1})  in Eqs. (\ref{eq:con_e1}) and (\ref{eq:con_n1}) we get
\beq
  \sinh (\xi) \dot{\xi} I_{20}^{(0)}\!-\!  \cosh (\xi) \dot{\beta }  I_{30}^{(0)} &=&\!- \cosh (\xi)\left(I_{20}^{(0)}\!-\!I_{21}^{(0)}\right)\!\theta, \nn\\\label{eq:enddot1}\\
  \cosh (\xi ) \dot{\xi } I_{10}^{(0)}\!-\!  \sinh (\xi) \dot{\beta}I_{20}^{(0)}&=&-  \sinh (\xi)I_{10}^{(0)} \theta. \label{eq:dotn1}
\eeq
Using the relations: $I_{20}^{(0)}=\varepsilon_{0}$, $I_{21}^{(0)}=-P_{0}=-n_0 T$, $ I_{10}^{(0)}=n_{0}$, and $I_{30}^{(0)}=\frac{1}{\beta }{\left(3 \left(P_0+\varepsilon _0\right)+z^2 P_0\right)}$, and solving  Eqs.~(\ref{eq:enddot1}) and (\ref{eq:dotn1}) for $\dot{\xi}$ and $\dot{\beta}$ we can get
\beq
	\dot{\xi} &=&\xi_\theta \, \theta, 
	\label{eq:xidot1} \\
		\dot{\beta}&=& \beta_\theta \, \theta, \label{eq:betadot1}
	\eeq
where
\beq
\xi_\theta &=&\frac{\sinh (\xi ) \cosh (\xi ) \left[\varepsilon _0^2-n_0 T \left(\left(3+z^2\right)P_0+2 \varepsilon _0\right)\right]}{n_0 T \cosh ^2(\xi ) \left( \left(3+z^2\right)P_0+3 \varepsilon _0\right)-\varepsilon _0^2 \sinh ^2(\xi )}, \nn\\ \label{eq:xith} \\
\beta_\theta&=& \frac{n_0 \left(\cosh ^2(\xi )P_0 +\varepsilon _0\right)}{n_0 T \cosh ^2(\xi ) \left(\left(3+z^2\right)P_0+3 \varepsilon _0\right)-\varepsilon _0^2 \sinh ^2(\xi )}.\nn\\ \label{eq:Dmunu}
\eeq
Substituting into Eq.~(\ref{eq:con_P1}) the energy density $\varepsilon$ and  pressure $P$ defined by Eqs.~(\ref{eq:end}) and (\ref{eq:P})  we get
\beq
  \cosh(\xi)\left(I_{20}^{(0)}-I_{21}^{(0)}\right) \dot{u}^{\alpha }&=&- \sinh (\xi )\left(\nabla ^{\alpha }\xi \right)I_{21}^{(0)}\nn\\&&\hspace{-0.5cm}- \cosh (\xi)\left(\nabla ^{\alpha }I_{21}^{(0)}\right). \label{eq:dotu1.1}
\eeq
Now we can write
\beq
\nabla ^{\alpha }I_{21}^{(0)}&=&\nabla ^{\alpha }\left(\frac{1}{3}\Delta _{\mu \nu }\int {\rm dP} \,p^{\mu } p^{\nu } e^{-p^{\lambda }\beta _{\lambda }}\right)\nn\\
&=&\frac{1}{3}\Delta _{\mu \nu }\left(-\nabla ^{\alpha }\beta_{\lambda}\right)\int {\rm dP}\, p^{\mu } p^{\nu } p^{\lambda} e^{-p^{\lambda }\beta _{\lambda }}\nn\\
&=&-\frac{1}{3}\Delta _{\mu \nu }\left(\frac{1}{T}\nabla ^{\alpha }u_{\lambda }-\frac{u_{\lambda }}{T^2}\nabla ^{\alpha }T\right) \bigg[I_{30}^{(0)} u^{\lambda } u^{\mu} u^{\nu}\nn\\&+&I_{31}^{(0)} \left(\Delta ^{\lambda \mu } u^{\nu }+\Delta ^{\nu \lambda } u^{\mu }+\Delta ^{\mu \nu } u^{\lambda }\right)\bigg]\nn\\
&=&-\left(-\frac{u_{\lambda }}{T^2}\nabla ^{\alpha }T\right) u^{\lambda }I_{31}^{(0)}\nn\\
&=&\left(-\nabla^{\alpha}\beta\right)I_{31}^{(0)}\label{eq:I21r}
\eeq
and using Eq.~(\ref{eq:I21r}) in Eq.~(\ref{eq:dotu1.1}) we obtain
\beq
  \cosh(\xi)\left(I_{20}^{(0)}-I_{21}^{(0)}\right) \dot{u}^{\alpha }&=&-\, \sinh (\xi )\left(\nabla ^{\alpha }\xi \right)I_{21}^{(0)}\nn\\&&\hspace{-0.5cm}+\, \cosh (\xi)\left(\nabla^{\alpha}\beta\right)I_{31}^{(0)}.
 \label{eq:dotu1.2}
\eeq
Now from the recurrence relation (\ref{eq:ir2}) we obtain
\beq
I_{31}^{(0)}&=&-\frac{1}{\beta }\left(I_{20}^{(0)}-I_{21}^{(0)}\right)=-\frac{1}{\beta }{(\varepsilon _0 + P_0)}, \nn\\ I_{21}^{(0)}&=&-P_0=-\frac{n_0}{\beta}.  
\eeq
Using the above expressions for $I_{31}^{(0)}$ and $I_{21}^{(0)}$ in Eq.~(\ref{eq:dotu1.2}), the following equation for $\dot{u}^{\mu}$ can be derived
\beq
\beta  \dot{u}^{\alpha }&=&\frac{n_0 \tanh (\xi )}{ \varepsilon _0+P_0 }\left(\nabla ^{\alpha }\xi \right)-\left(\nabla ^{\alpha }\beta \right). \label{eq:udot1}
\eeq


Now we turn to the equilibrium spin tensor. With the help of  Eq.~(\ref{eq:Smunulambda_de_Groot2}) it can be written as
\bea
&& S^{\lambda, \mu\nu}_{\rm eq} =\frac{4 \mathfrak{s}^2}{3}\cosh (\xi )I_{10}^{(0)}u^{\lambda }\omega ^{\mu \nu }  \\
&& 
+\frac{4 \mathfrak{s}^2}{3 m^2}\cosh (\xi )\bigg[2I_{30}^{(0)}u^{\lambda }u^{\alpha }u^{[\mu }\omega ^{\nu ]}{}_{\alpha }\nn\\
&& 
+ 2I_{31}^{(0)}\left(\Delta ^{\lambda \alpha }u^{[\mu }\omega ^{\nu ]}{}_{\alpha }+u^{\lambda }\Delta ^{\alpha [\mu }\omega ^{\nu ]}{}_{\alpha }+u^{\alpha }\Delta ^{\lambda [\mu }\omega ^{\nu ]}{}_{\alpha }\right)\bigg]. \nn
\eea 
The above equation can further be simplified as
\bea
S^{\lambda, \mu\nu}_{\rm eq}&=&\frac{4 \mathfrak{s}^2}{3}\cosh (\xi )I_{10}^{(0)}u^{\lambda }\omega ^{\mu \nu }
\label{eq:Smunulambda_de_Groot2.1}\\
&+&\frac{8 \mathfrak{s}^2}{3 m^2}\cosh (\xi )\bigg[\left(I_{30}^{(0)}-3 I_{31}^{(0)}\right)u^{\lambda }u^{\alpha }u^{[\mu }\omega ^{\nu ]}{}_{\alpha }\nn\\&+&I_{31}^{(0)}\left(u^{[\mu }\omega ^{\nu ]\lambda }-\omega ^{\mu \nu } u^{\lambda }+u^{\alpha }g^{\lambda [\mu }\omega ^{\nu ]}{}_{\alpha }\right)\bigg].
\nn
\eea 
Substituting Eq.~(\ref{eq:Smunulambda_de_Groot2.1}) into Eq.~(\ref{eq:con_S1}), and using Eqs.~(\ref{eq:xidot1}), (\ref{eq:betadot1}), and (\ref{eq:udot1}), the following dynamical equation for the spin polarization tensor $\omega^{\mu\nu}$ can be obtained 
\beq
\dot{\omega }^{\mu \nu}&=& D_{\Pi }^{\mu \nu }\theta +\left(\nabla ^{\alpha }\xi \right)D_n^{[\mu \nu ]}{}_{\alpha }+D_{\pi }^{[\nu }{}_{\lambda }\sigma ^{\lambda\mu ]}\nn\\&&+D_{\text{$\Sigma $1}}^{\alpha }\nabla ^{[\mu }\omega ^{\nu ]}{}_{\alpha }+D_{\text{$\Sigma $2}}^{[\mu \nu ]\alpha }\nabla ^{\lambda }\omega _{\alpha \lambda }. \label{eq:dotomega}
\eeq
For details see Appendix \ref{Ac}, where the explicit expressions for various $D$-coefficients are given.

Note that while deriving the dynamical equation (\ref{eq:dotomega}), we initially encounter the term $u_{\nu}\dot{\omega }^{\mu \nu }$ in the expression for $\dot{\omega }^{\mu \nu }$. To eliminate this term we derive another dynamical equation for $u_{\nu}\dot{\omega }^{\mu \nu }$ by taking projection of Eq.~(\ref{eq:con_S1}) along $u_\nu$. The dynamical equation for $u_{\nu}\dot{\omega }^{\mu \nu }$ is given by the expression
\beq
u_{\nu }\dot{\omega }^{\mu \nu }&=&C_{\Pi }^{\mu }\theta +C_{n{}{\lambda }}^{\mu }{}(\nabla ^{\lambda }\xi )+C_{\pi{}{\alpha }}\sigma ^{\alpha \mu }+C_{\text{$\Sigma$} {}{\nu }}^{\mu } \nabla _{\lambda }\omega ^{\nu \lambda }. \nn\\
\eeq
The explicit expression for various $C$-coefficients appearing above are also given in Appendix \ref{Ac}. See also Appendix \ref{sec:LanCon}, where the Landau matching conditions are presented in more detail.
%
%
%

\subsection{Transport coefficients}
\label{sec:trcoeff}

The dissipative forces arise due to non-zero gradients in the system. In the present case, we will confine ourselves only to first order in gradients and hence the dissipative parts of $T^{\mu\nu}$, $N^\mu$, and $S^{\lambda,\mu\nu}$, i.e., $\delta T^{\mu\nu}$, $\delta N^\mu$, and $\delta S^{\lambda,\mu\nu}$, respectively, must be first order in gradients too. The shear stress $(\pi^{\mu\nu})$, bulk viscous pressure ($\Pi$) and particle diffusion current $(n^\mu)$ can be found from $\delta T^{\mu\nu}$ and $\delta N^\mu$ as:
\begin{align}
\pi^{\mu\nu} = \Delta^{\mu\nu}_{\alpha\beta}\,\delta T^{\alpha\beta}, \quad \Pi = -\frac{1}{3}\Delta_{\alpha\beta}\,\delta T^{\alpha\beta}, \quad \nu^\mu = \Delta^\mu_\alpha~\delta N^\alpha.
\end{align}
Hence, using Eqs.~(\ref{eq:dT}) and (\ref{eq:dN}), the above  dissipative quantities can be written as:
\begin{align}
&\pi^{\mu\nu} = \Delta^{\mu\nu}_{\alpha\beta} \int \mathrm{dP}~\mathrm{dS}~p^\alpha p^\beta (\delta f^+_s+\delta f^-_s),\label{eq:shear}\\
&\Pi = - \frac{1}{3}\Delta_{\alpha\beta} \int \mathrm{dP}~\mathrm{dS}~p^\alpha p^\beta (\delta f^+_s+\delta f^-_s),\label{eq:bulk}\\
&\nu^\mu = \Delta^\mu_\alpha \int \mathrm{dP}~\mathrm{dS}~p^\alpha (\delta f^+_s-\delta f^-_s).\label{eq:nu}
\end{align}
Evaluating the expressions defined by Eqs.~(\ref{eq:shear}), (\ref{eq:bulk}), and (\ref{eq:nu}),  the dissipative quantities are found to be (see Appendix \ref{Ag})
\begin{align}
\pi^{\mu\nu} &= 2 \tau_{\rm eq}\,\beta_\pi \sigma^{\mu\nu}, \nn \\ \Pi &= - \tau_{\rm eq}\,\beta_\Pi \theta, \nn \\
\nu^\mu &= \tau_{\rm eq}~\beta_n \nabla^{\mu} \xi.
\end{align}
Here, coefficients, $\beta_\pi$, $\beta_\Pi$ and $\beta_n$ are the first-order transport coefficients which for massive particles with finite chemical potential are found to be
\begin{widetext}
\begin{align}
&\beta_\pi = 4~I^{(1)}_{42} \cosh({\xi}),\\
&\beta _{\Pi }=4\Bigg\{\frac{n_0 \cosh (\xi )}{\beta} \left[ \frac{ {\sinh}^2(\xi ) \left(\varepsilon _0 \left(P_0+\varepsilon _0\right)-n_0 T \left(P_0 \left(z^2+3\right)+3 \varepsilon _0\right)\right)}{ \varepsilon _0^2 \sinh ^2(\xi )-n_0 T \cosh ^2(\xi ) \left(P_0 \left(z^2+3\right)+3 \varepsilon _0\right) }\right]\nn\\
&~~~-\frac{n_0 \cosh (\xi )}{\beta } \left[\frac{ \left(P_0+\varepsilon _0\right) \left(P_0 \cosh ^2(\xi )+\varepsilon _0\right)}{n_0 T \cosh ^2(\xi ) \left(P_0 \left(z^2+3\right)+3 \varepsilon _0\right)-\varepsilon _0^2 \sinh ^2(\xi )}\right]+\frac{5\beta }{3}I^{(1)}_{42}\Bigg\},\\
&\beta_n = 4 \bigg[\bigg(\frac{n_0 \tanh(\xi)}{\varepsilon_0 + P_0}\bigg)~I^{(0)}_{21}~\sinh({\xi}) - I^{(1)}_{21}~\cosh({\xi})\bigg].
\end{align}
Similarly, using Eq.~(\ref{del feq lim}) in (\ref{eq_dspin}) and then carrying out integration over spin and momentum variables we get, 	
\begin{eqnarray}
\delta S^{\lambda,\mu\nu} &=& \tau_{\rm eq} \Big[ B^{\lambda,\mu\nu}_{\Pi}\, \theta + B^{\kappa\lambda,\mu\nu}_{n}\, (\nabla_\kappa \xi) + B_{\pi }^{(\kappa\delta) \lambda, \mu \nu }\sigma _{\kappa \delta } + B_{\Sigma }^{\eta \beta \gamma\lambda ,\mu \nu }\nabla _{\eta }\omega _{\beta \gamma }\Big]. 
\label{deltaS}
\end{eqnarray}
 Different coefficients appearing on the right-hand side of Eq.~(\ref{deltaS}) are the kinetic coefficients for spin. They have tensor structures expressed in terms of $u^\mu,~g^{\mu\nu}$, and $\omega^{\mu\nu}$.  Explicit forms of these coefficients are as follows:
\begin{eqnarray}
B^{\lambda,\mu\nu}_{\Pi} &=&
B_{\Pi }^{(1)}u^{[\mu }\omega ^{\nu ]\lambda } + B_{\Pi }^{(2)}u^{\lambda }u^{\alpha }u^{[\mu }\omega ^{\nu ]}{}_{\alpha } +
B_{\Pi }^{(3)}\Delta ^{\lambda [\mu }u_{\alpha }\omega ^{\nu ]\alpha },
\label{eq:betaPi}\\
B_{\pi}^{\lambda \kappa \delta ,\mu \nu }&=&B_{\pi }^{(1)}\Delta^{[\mu \kappa }\Delta^{\lambda \delta }u_{\alpha }\omega ^{\nu ]\alpha }+B_{\pi }^{(2)}\Delta^{\lambda \delta }u^{[\mu }\omega ^{\nu ]\kappa }+B_{\pi }^{(3)}u^{[\mu }\Delta^{\nu ]\delta }\Delta^{\lambda}_{\alpha}\omega ^{\alpha \kappa } + B_{\pi }^{(4)}\Delta ^{\lambda [\mu }\omega ^{\rho \kappa } u_{\rho }\Delta^{\nu ]\delta },
\label{eq:betapi}\\
B_n^{\lambda \kappa ,}{}^{\mu \nu }&=&B_n^{(1)}\Delta ^{\lambda \kappa } \omega ^{\mu \nu }+B_n^{(2)}\Delta ^{\lambda \kappa }u^{\alpha }u^{[\mu }\omega ^{\nu ]}{}_{\alpha }+B_n^{(3)}\Delta ^{\lambda \alpha }\Delta ^{[\mu \kappa }\omega ^{\nu ]}{}_{\alpha }+B_n^{(4)}u^{[\mu }\Delta ^{\nu ]\kappa }u^{\rho }\omega ^{\lambda }{}_{\rho }+B_n^{(5)}\Delta ^{\lambda [\mu }\omega ^{\nu ]\kappa }\nn\\&&+B_n^{(6)}\Delta ^{\lambda [\mu }u^{\nu ]}u_{\alpha }\omega ^{\alpha \kappa},
\label{eq:betan}\\
B_{\Sigma }^{\eta \beta \gamma \lambda ,\mu \nu }&=&B_{\Sigma}^{(1)}\Delta^{\lambda \eta }g^{[\mu \beta }g^{\nu ]\gamma }+B_{\Sigma }^{(2)} u^{\gamma }\Delta ^{\lambda \eta }u^{[\mu }\Delta ^{\nu ]\beta } + B_{\Sigma}^{(3)}\left(\Delta ^{\lambda \eta }\Delta ^{\gamma [\mu }g^{\nu ]\beta }+\Delta ^{\lambda \gamma }\Delta ^{[\mu \eta }g^{\nu ]\beta }+\Delta ^{\gamma \eta }\Delta ^{\lambda [\mu }g^{\nu ]\beta }\right)\nn\\&&+B_{\Sigma}^{(4)}\Delta ^{\gamma \eta }\Delta ^{\lambda [\mu }\Delta^{\nu ]\beta}+B_{\Sigma}^{(5)}u^{\gamma }\Delta ^{\lambda \beta }u^{[\mu }\Delta ^{\nu ]\eta }\label{eq:betaSig},
\end{eqnarray}
where the scalar coefficients $B_{X}^{(i)}$ are explicitly defined in Appendix~\ref{Ac}. 

\end{widetext}

Equation~(\ref{deltaS}) is our main result. It shows that the dissipative spin effects are connected with the presence of expansion scalar, gradient of the ratio of chemical potential and temperature, the shear-flow tensor, and the gradient of the spin polarization tensor. All these quantities may be interpreted as ``thermodynamic forces'' that trigger dissipative currents. The first three among them are well known --- they lead to appearance of bulk pressure, diffusive current, and shear stress tensor. Interestingly, in the considered case, they also induce the dissipative part of the spin tensor. The fourth term in Eq.~(\ref{deltaS}) describes the induction of the disspative spin tensor by the gradient of the spin polarization tensor, hence, may be treated as a direct non-equilibrium interaction between spin degrees of freedom. 

Finally, we note that all the kinetic coefficients obtained from Eq.~(\ref{eq:col}) are proportional to the same relaxation time $\tau_{\rm eq}$. This implies that the equilibration times for momenta and spin degrees of freedom are the same. In phenomenological applications it is conceivable to vary the values of the relaxation times that appear in different kinetic coefficients, arguing that they describe independent physical phenomena. However, such modifications require further studies.


\section{Summary and Conclusions}

In this paper we have significantly extended the results obtained in Ref.~\cite{Bhadury:2020puc}. We used classical kinetic theory for particles with spin $\onehalf$ with Boltzmann statistics to obtain the structure of dissipative terms and the associated transport coefficients. We considered the relaxation time approximation for collision term in order to account for the interactions. This kinetic-theory framework was used to determine the structure of spin-dependent viscous and diffusive terms and explicitly evaluate a set of new kinetic coefficients that characterize dissipative spin dynamics.

Our main result is given by Eq.~(\ref{deltaS}), together with the explicit expressions for the kinetic coefficients $B$ given in the appendices. Equation~(\ref{deltaS}) shows that a non-equilibrium part of the spin tensor is produced by the thermodynamic forces such as expansion scalar, gradient of the ratio of chemical potential and temperature, the shear-flow tensor, and the gradient of the spin polarization tensor. Thus, the spin dissipative phenomena are connected with those leading to formation of bulk pressure, diffusion current, and the shear stress tensor. Probably, the most interesting term in Eq.~(\ref{deltaS}) is the last one, which describes induction of a non-equilibrium spin tensor by a gradient of the spin polarization tensor. In the future investigations, it would be interesting to analyze the role played by various coeficients appearing in Eq.~(\ref{deltaS}) and to find out which kind of corrections they imply for the spin tensor. The complicated tensor structure of the spin kinetic coefficients may lead to various interesting phenomena. 

\begin{acknowledgments}
W.F. and R.R. acknowledge the hospitality of National Institute of Science Education and Research where most of this work was done. S.B., A.J. and  A.K. would like to acknowledge the kind hospitality of Jagiellonian University and Institute of Nuclear Physics, Krakow, where part of this work was completed. A.J. was supported in part by the DST-INSPIRE faculty award under Grant No. DST/INSPIRE/04/2017/000038. A. K. was 
supported in part by the Department of Science and Technology, Government
of INDIA under the SERB National Post-Doctoral Fellowship Reference No.
PDF/2020/000648.
W.F. and R.R. were supported in part by the Polish National Science Centre Grants  No. 2016/23/B/ST2/00717 and No. 2018/30/E/ST2/00432.
\end{acknowledgments}

\onecolumngrid
\appendix

\section{Spin-space integrals}
\label{sec:appspin}

In this appendix several integrals over the spin space are explicitly done. The results obtained here are used throughout the paper in the calculations of the charge current, energy-momentum tensor, and the spin tensor.

\subsection{Normalization of spin integration measure}

We start with the calculation of the normalization of the spin integration measure~\cite{Florkowski:2018fap}. Since it is a Lorentz invariant quantity depending on the (external) momentum $p$, the calculations can be done in the particle rest frame (PRF) where $p^\mu = (m,0,0,0)$ and $s^\mu =(0,\sv_*)$,
\begin{equation}
    \int \mathrm{dS} = \frac{m}{\pi \mathfrak{s}} \int \mathrm{d^4}s\, \delta(s\cdot s + {\mathfrak{s}}^2)\, \delta(p\cdot s)
    = \frac{m}{\pi \mathfrak{s}} \int \mathrm{d}s_0 \int \mathrm{d|\sv_*|} |\sv_*|^2 \int \mathrm{d\Omega}\, \delta(|\sv_*|^2-{\mathfrak{s}}^2) \delta(m s_0).
\end{equation}
With the normalization $\int\mathrm{d\Omega}=\int\sin{\theta}\,\mathrm{d\theta}\int\mathrm{d\phi}=4\pi$ we obtain
\begin{align}
    \int \mathrm{dS} &= \frac{4\pi}{\pi{\mathfrak{s}}} \int \mathrm{d|\sv_*|} |\sv_*|^2\, \delta(|\sv_*|^2 -{\mathfrak{s}}^2) = 2. \label{A1}
\end{align}
The factor of 2 reflects here the two possibilities of the spin-$\onehalf$ projection.

\subsection{Spin average of \texorpdfstring{$s^{\mu\nu}$}{}}

While expanding the spin-dependent equilibrium distribution function in powers of $\omega$, we encounter the integrals of the form
\begin{equation}
    \int\mathrm{dS} \, s^{\mu\nu} = \frac{1}{m} \int \mathrm{dS}\, \epsilon^{\mu\nu\alpha\beta}\, p_\alpha s_\beta
    = \frac{1}{m}\, \epsilon^{\mu\nu\alpha\beta}\, p_\alpha \int \mathrm{dS}\, s_\beta. \label{A2}
\end{equation}
Since the last integral can be a function of momentum $p_\beta$ only, we can write
\begin{align}
    \int\mathrm{dS}\, s_\beta &= c\, p_\beta.
\end{align}
After contraction with $p$, this equation gives
\begin{align}
\int\mathrm{dS}\, (p\cdot s) &= c\, m^2,  \label{A3}
\end{align}
which implies that the constant $c$ equals zero, as $p\cdot s = 0$. Hence, throughout the paper we can use the property
\begin{align}
    \int\mathrm{dS}\,s^{\mu\nu} = 0. \label{A4}
\end{align}

\subsection{Spin average of \texorpdfstring{$s^{\mu\nu} s^{\alpha\beta}$}{} }

In the second order of expansions in $\omega$ we deal with the integrals of the form
\begin{equation}
    \int\mathrm{dS} \, s^{\mu\nu} s^{\alpha\beta} = \frac{1}{m^2} \int\mathrm{dS}\, \epsilon^{\mu\nu\rho\sigma}\, p_\rho s_\sigma\, \epsilon^{\alpha\beta\gamma\delta}\, p_\gamma\, s_\delta 
    = \frac{1}{m^2}\, \epsilon^{\mu\nu\rho\sigma}\, \epsilon^{\alpha\beta\gamma\delta}\, p_\rho\, p_\gamma \int \mathrm{dS}\, s_\sigma\,s_\delta.  \label{A5}
\end{equation}
Since the last integral can be a function of momenta and the metric tensor, we write
\begin{align}
    \int\mathrm{dS}\, s_\sigma\, s_\delta &= a\, g_{\sigma\delta} + b\, p_\sigma p_\delta, \label{A6}
\end{align}
where $a$ and $b$ are scalar coefficients. Multiplying \EQ{A6} by $p^\sigma p^\delta$ in the first case and contracting the indices in \EQ{A6} in the second case, we obtain two equations
\begin{align}
    \int \mathrm{dS}\, (p\cdot s)^2 &= a\, m^2 + b\, m^4 \label{A7}
\end{align}
and
\begin{align}
    \int \mathrm{dS}\,s^2 &= 4\, a + b\, m^2. \label{A8}
\end{align}
The left-hand sides of Eqs.~(\ref{A7}) and~(\ref{A8}) yield
\beq
    \int\! \mathrm{dS}\, (p\cdot s)^2 &=& 0, \nn
\eeq
\beq
    \int \mathrm{dS}\, s^2 &=& \frac{m}{\pi \mathfrak{s}} \int \mathrm{d^4}s\, (s \cdot s)\, \delta(s\cdot s + \mathfrak{s}^2)\, \delta(p \cdot s) \nonumber\\
    &=& -\frac{m}{\pi \mathfrak{s}} \int \mathrm{d}s_0 \int \mathrm{d|\sv_*|} |\sv_*|^4 \int \mathrm{d\Omega}\, \delta(|\sv_*|^2 - \mathfrak{s}^2)\, \delta(m s_0) \nonumber\\
    &=& - \frac{m}{\pi \mathfrak{s}}\! \int\! \mathrm{d}s_0\, \frac{\delta(s_0)}{m} \int \mathrm{d|\sv_*|} |\sv_*|^4\, \delta(|\sv_*|^2 - \mathfrak{s}^2)\, 4\pi \nonumber\\
    &=& - \frac{m}{\pi \mathfrak{s}} \frac{4\pi}{m} \frac{\mathfrak{s}^3}{2} = - 2\, \mathfrak{s}^2. 
\eeq
Thus, from Eqs.~(\ref{A7}) and~(\ref{A8}) we get
\begin{align}
    a\, m^2 + b\, m^4 &= 0,\\
    4\, a + b\, m^2 &= - 2\, \mathfrak{s}^2.
\end{align}
Solving these two equations we get $a=-2\,{\mathfrak{s}}^2/3$ and $b = 2\,{\mathfrak{s}}^2/(3 m^2)$. Hence we have
\begin{align}
    \int \mathrm{dS}\, s_\sigma\, s_\delta &= -\frac{2 \mathfrak{s}^2}{3} \bigg(g_{\sigma\delta} - \frac{p_\sigma p_\delta}{m^2}\bigg)
\end{align}
and
\begin{align}
    \int \mathrm{dS} \, s^{\mu\nu} s^{\alpha\beta} &= - \frac{2 \mathfrak{s}^2}{3m^2} \epsilon^{\mu\nu\rho\sigma}\, \epsilon^{\alpha\beta\gamma\delta}\, p_\rho\, p_\gamma \bigg(g_{\sigma\delta} - \frac{p_\sigma p_\delta}{m^2}\bigg). \label{A9}
\end{align}


\section{Thermodynamic integrals}
\label{sec:thermint}

Thermodynamic integrals considered in this work are given by the following expression
\beq
    I_{nq}^{(r)} &=& \frac{1}{(2q+1)!!} \int {\rm dP}\,(u\cdot p)^{n-2q-r} (\Delta_{\alpha\beta} p^{\alpha} p^{\beta})^q e^{-\beta \cdot p}.
\eeq
From the above formula, as the special cases, we obtain:
\beq
    I_{10}^{(0)} &=& \frac{T^3 z^2 }{2 \pi ^2} K_2(z),\\
    I_{20}^{(0)} &=& \frac{T^4 z^2}{2 \pi^2} \left[3 K_2(z) + zK_1(z)\right],\\
    I_{21}^{(0)} &=& - \frac{T^4 z^2}{2 \pi ^2} K_2(z),\\
    I_{30}^{(0)} &=& \frac{T^5 z^5 }{32\pi^2} \left[K_5(z)+K_3(z)-2K_1(z)\right],\\
    I_{31}^{(0)} &=& - \frac{T^5 z^5}{96\pi^2} \left[K_5(z)-3K_3(z)+2K_1(z)\right],\\
    I_{40}^{(0)} &=& \frac{T^6 z^6}{64 \pi^2} \left[K_{6}(z) + 2K_{4}(z) - K_{2}(z) - 2K_{0}(z)\right],\\
    I_{41}^{(0)} &=& - \frac{T^6 z^6}{192 \pi^2} \left[K_{6}(z) - 2K_{4}(z) - K_{2}(z) + 2K_{0}(z)\right],\\
    I_{42}^{(0)} &=& \frac{T^6 z^6}{960 \pi^2} \left[K_{6}(z) - 6K_{4}(z) + 15K_{2}(z) - 10K_{0}(z)\right].
\eeq
Here $K_n(z)$ denotes the modified Bessel functions of the second kind with the argument $z = m/T$. They are defined by the integral
\beq
    K_n(z) &=& \int_0^{\infty} \mathrm{d}x\, \cosh(nx)\, e^{- z \cosh x}.
\eeq
The other thermodynamic integrals which are needed in our calculation are given by the expressions
\beq
    I_{21}^{(1)} &=& - \frac{T^3 z^3}{6 \pi^2} \left[\frac{1}{4} K_3(z) - \frac{5}{4} K_1(z) + K_{i,1}(z)\right],\\
    I_{42}^{(1)} &=& \frac{T^5 z^5}{480 \pi^2} \Big[22K_{1}(z) - 7K_{3}(z) + K_5(z) - 16K_{i,1}(z)\Big],
\eeq
where
\beq
    K_{i,1}(z) = \int_0^{\infty } \mathrm{d}x \sech{x}\, e^{-z \cosh x}
    = \frac{\pi}{2} \Big[ 1 - z\, K_{0}(z)\, L_{-1}(z) - z\, K_{1}(z)\, L_{0}(z)\Big]
\eeq
is the first-order Bickley-Naylor function with $L_i$ being the modified Struve function.

Note that here we have not listed the functions $I_{20}^{(1)}$, $I_{30}^{(1)}$, $I_{31}^{(1)}$,  $I_{40}^{(1)}$, $I_{50}^{(1)}$, $I_{51}^{(1)}$ and $I_{52}^{(1)}$ as they all can be written in terms of the integrals listed above,  using the following recurrence relations
\bea
    I_{n,q}^{(r)} &=& I_{ n-1,q}^{(r-1)};~~~ n\geq2 q, \label{eq:ir1}\\
    I_{n,q}^{(0)} &=& \frac{1}{\beta} \left[(n - 2 q) I_{n-1,q}^{(0)} - I_{n-1, q-1}^{(0)}\right],\label{eq:ir2}\\
    \dot{I}_{n,q}^{(0)} &=& - \dot{\beta} I_{n+1, q}^{(0)}\,.\label{eq:ir3}
\eea


\section{Projection operators and entropy current}

\subsection{Properties of the projection operators}
\label{sec:projectors}

Herein we list useful relations involving the projection operators and the differential operator $\nabla^{\mu} \equiv  \Delta^{\mu \nu} \partial_\nu$:
\beq
    \Delta_{\mu \nu} \Delta^{\mu \nu } &=& 3, \qquad
    u_{\mu} \Delta^{\mu\nu} = u_{\mu} \Delta^{\nu \mu } = 0, \qquad
    \Delta^{\mu \nu} \Delta_{\nu}^{\lambda} = \Delta^{\mu \lambda}, \qquad
    u_{\mu} \nabla^{\mu}   = 0,\label{eq:identities1}\\
    \Delta_{\mu \nu}^{\alpha \beta} &=& \frac{1}{2} \left(\Delta_{\mu}^{\alpha} \Delta_{\nu}^{\beta} + \Delta_{\nu}^{\alpha} \Delta_{\mu}^{\beta} - \frac{2}{3} \Delta^{\alpha \beta} \Delta_{\mu \nu}\right),\label{eq:identities2a}\\
    u^{\mu } \Delta_{\mu \nu}^{\alpha \beta} &=& u^{\mu} \Delta_{\nu\mu }^{\alpha \beta} = u_{\alpha} \Delta_{\mu \nu}^{\alpha \beta} = u_{\alpha} \Delta_{\mu \nu}^{ \beta\alpha} =0,\label{eq:identities2b}\\
    \Delta^{\mu \nu} \Delta_{\mu \nu}^{\alpha \beta} &=& \Delta_{\mu \nu} \Delta^{\mu \nu}_{\alpha \beta} = 0,\label{eq:identities2c}\\
    \Delta^{\mu \lambda} \Delta_{\mu \nu}^{\alpha \beta} &=& \Delta_{\nu}^{\alpha \beta\lambda},\label{eq:identities2d}\\
    \Delta^{\alpha \beta \lambda \rho} \nabla_{\rho} u_{\lambda} &=& \frac{1}{2} \left(\Delta^{\alpha \lambda} \Delta^{\beta \rho} + \Delta^{\alpha \rho} \Delta^{\beta \lambda} - \frac{2}{3} \Delta^{\alpha \beta} \Delta^{\lambda \rho} \right) \nabla_{\rho} u_{\lambda} 
    = \frac{1}{2} \left(\nabla^{\beta} u^{\alpha} + \nabla^{\alpha} u^{\beta} - \frac{2}{3} \Delta^{\alpha \beta} \nabla^{\lambda} u_{\lambda}\right) \equiv \sigma^{\alpha \beta}.\label{eq:identities2e}
\eeq

\subsection{Entropy current conservation in equilibrium}
\label{sec:entropy}

In the following, we provide the proof of conservation of equilibrium entropy current defined in Eq.~\eqref{eq:H1}. We start with the charge current as defined in Eq.~\eqref{eq:Neq-sp0} 
\begin{equation}
N^\mu_{\rm eq} = \int dP   \int dS \, \, p^\mu \, \left[f^+_{s, \rm eq}(x,p,s)-f^-_{s, \rm eq}(x,p,s) \right].
\label{eq:Neq-sp0:A}    
\end{equation}
The distribution functions $f^\pm_{s, \rm eq}(x,p,s)$ in above equation are given by
\begin{equation}
f^\pm_{s, \rm eq}(x,p,s) =  f_{\rm eq}^{\pm}(x,p)\exp\left( \frac{1}{2} \omega_{\mu\nu} s^{\mu\nu} \right),
\label{eq:feqxps:A}
\end{equation}
where $f_{\rm eq}^{\pm}(x,p)=\exp\left[-p^\mu\beta_\mu(x)\pm\xi(x)\right]$ is the J\"uttner distribution function.
Using Eq.~(\ref{eq:feqxps:A}) in Eq.~(\ref{eq:Neq-sp0:A}) we get
\begin{equation}
N^\mu_{\rm eq} = 2 \sinh(\xi) \int dP \, \int dS \, p^\mu \, \exp\left(-p \cdot \beta+\frac{1}{2}  \omega_{\alpha \beta} s^{\alpha\beta} \right).
\label{eq:Neq-sp1:A}
\end{equation}
The above defined charge current is conserved in equilibrium, i.e., we must have
\begin{equation}
\partial_{\mu} N^\mu_{\rm eq} = 0.
\label{eq:Ncons1:A}
\end{equation}
Substituting Eq.~(\ref{eq:Neq-sp1:A}) in Eq.~(\ref{eq:Ncons1:A}) we obtain
\begin{equation}
\partial_{\mu} \left(2 \sinh(\xi) \int dP \, \int dS \, p^\mu \, \exp\left(-p \cdot \beta+\frac{1}{2}  \omega_{\alpha \beta} s^{\alpha\beta} \right)\right) = 0,
\label{eq:Ncons2:A}
\end{equation}
which leads to
\begin{align}
\int dP \, \int dS \, p^\mu \, \exp\left(-p \cdot \beta+\frac{1}{2}  \omega_{\alpha \beta} s^{\alpha\beta} \right)\left(-p^{\alpha}\partial_{\mu}\beta_{\alpha}+\frac{1}{2} s^{\alpha\beta}\partial_{\mu}\omega_{\alpha \beta}\right)\nonumber\\+\int dP \, \int dS \, p^\mu \, \exp\left(-p \cdot \beta+\frac{1}{2}  \omega_{\alpha \beta} s^{\alpha\beta} \right)\left(2\cosh\xi\partial_{\mu}\xi\right)=0
\label{eq:Ncons3:A}
\end{align}

Next, we note the definitions of the energy momentum tensor and spin tensor given in Eqs.~\eqref{eq:Teq-sp02} and \eqref{eq:Seq-sp01},
\begin{align}
T^{\mu \nu}_{\rm eq}=& \int dP   \int dS \, \, p^\mu p^\nu \, \left[f^+_{\rm eq}(x,p,s) + f^-_{\rm eq}(x,p,s) \right] \nonumber\\
=& 2 \cosh(\xi) \int dP \int dS \, p^\mu p^\nu \, \exp\left(- p \cdot \beta+\frac{1}{2}  \omega_{\alpha \beta} s^{\alpha\beta}\right),
\label{eq:Teq-sp02:A}
\end{align}
\begin{align}
S^{\lambda, \mu\nu}_{\rm eq}=& \int dP   \int dS \, \, p^\lambda \, s^{\mu \nu} 
\left[f^+_{\rm eq}(x,p,s) + f^-_{\rm eq}(x,p,s) \right] \nonumber \\
=& 2 \cosh(\xi) \int dP \int dS \, p^\lambda s^{\mu \nu} \, \exp\left( - p \cdot \beta+\frac{1}{2} \omega_{\alpha \beta} s^{\alpha\beta} \right).
\label{eq:Seq-sp01:A}
\end{align}
Using equations (\ref{eq:Teq-sp02:A}), (\ref{eq:Seq-sp01:A}), and (\ref{eq:Neq-sp1:A}) in (\ref{eq:Ncons3:A}) we get

\begin{align}
\left(-T^{\mu \nu}_{\rm eq}\partial_{\mu}\beta_{\alpha}+\frac{1}{2}S^{\mu, \alpha\beta}_{\rm eq}\left(\partial_{\mu}\omega_{\alpha \beta}\right)\right)\frac{\sinh\xi}{\cosh\xi}+N^\mu_{\rm eq}\left(\partial_{\mu}\xi\right)\frac{\cosh\xi}{\sinh\xi}=0
\label{eq:Ncons4:A}
\end{align}
Multiplying the above equation by $-\frac{\cosh\xi}{\sinh\xi}$ we get
\begin{align}
\left(T^{\mu \nu}_{\rm eq}\partial_{\mu}\beta_{\alpha}-\frac{1}{2}S^{\mu, \alpha\beta}_{\rm eq}\left(\partial_{\mu}\omega_{\alpha \beta}\right)\right)-N^\mu_{\rm eq}\left(\partial_{\mu}\xi\right)\frac{\cosh^2 \xi}{\sinh^2\xi}=0.
\label{eq:Ncons4.1:A}
\end{align}
The above equation can further be re-written as
\begin{align}
\left(T^{\mu \nu}_{\rm eq}\partial_{\mu}\beta_{\alpha}-\frac{1}{2}S^{\mu, \alpha\beta}_{\rm eq}\left(\partial_{\mu}\omega_{\alpha \beta}\right)\right)-N^\mu_{\rm eq}\left(\partial_{\mu}\xi\right)-N^\mu_{\rm eq}\left(\partial_{\mu}\xi\right)\frac{1}{\sinh^2\xi}=0.
\label{eq:Ncons4.2:A}
\end{align}
From discussions below Eq.~\eqref{eq:H2}, we have
\begin{equation}
P\beta^\mu = \frac{\cosh(\xi)}{\sinh(\xi)}  N^\mu_{\rm eq}.
\label{eq:NcalN1:A}
\end{equation}
From the above equation one finds
\begin{equation}
\partial_{\mu}(P\beta^\mu) = \frac{\cosh(\xi)}{\sinh(\xi)}  \partial_{\mu} N^\mu_{\rm eq}-N^\mu_{\rm eq}\left(\partial_{\mu}\xi\right)\frac{1}{\sinh^2{\xi}}.
\label{eq:NcalN2:A}
\end{equation}
The first term on the right hand side of the above equation vanishes because of conservation law of charge current (\ref{eq:Ncons1:A}). Therefore, from above equation we have
\begin{equation}
\partial_{\mu}(P\beta^\mu) =-N^\mu_{\rm eq}\left(\partial_{\mu}\xi\right)\frac{1}{\sinh^2{\xi}}
\label{eq:NcalN3:A}
\end{equation}
Using (\ref{eq:NcalN3:A}) in (\ref{eq:Ncons4.2:A}) we get 
\begin{align}
\left(T^{\mu \nu}_{\rm eq}\partial_{\mu}\beta_{\alpha}-\frac{1}{2}S^{\mu, \alpha\beta}_{\rm eq}\left(\partial_{\mu}\omega_{\alpha \beta}\right)-N^\mu_{\rm eq}\left(\partial_{\mu}\xi\right)+\partial_{\mu}(P\beta^\mu)\right)=0.
\label{eq:Ncons5:A}
\end{align}
Now from Eq.~\eqref{eq:H3} we obtain 
\begin{equation}
\partial_\mu H^\mu = \left( \partial_\mu \beta_\alpha \right) T^{\mu \alpha}_{\rm eq} 
-\frac{1}{2} \left( \partial_\mu \omega_{\alpha\beta} \right) S^{\mu, \alpha \beta}_{\rm eq}
- \left(\partial_\mu \xi \right)  N^\mu_{\rm eq} + \partial_{\mu}(P\beta^\mu).
\label{eq:H3:A}
\end{equation}
From Eq. (\ref{eq:Ncons5:A}) we can notice the right-hand side of Eq.~(\ref{eq:H3:A}) vanishes. Thus, we finally conclude that
\begin{equation}
\partial_\mu H^\mu =0
\end{equation}


\section{Calculation of the dissipative corrections \texorpdfstring{$\delta T^{\mu\nu}$}{}, \texorpdfstring{$\delta N^{\nu}$}{}, \texorpdfstring{$\delta S^{\lambda,\mu\nu}$}{}} \label{sec:appD}

\subsection{Dissipative corrections \texorpdfstring{$\delta N^{\nu}$}{} } \label{sec:appDN}

The dissipative part of the baryon current can be written as
\beq
    \delta N^\mu &=& \int\!  \mathrm{dP}\,\mathrm{dS}\,p^\mu \big(\delta f^+_s - \delta f^-_s\big). \label{delN}
\eeq
From Eq.~(\ref{del feq lim}) we obtain
\beq
    \delta f^+_s - \delta f^-_s &=& - \frac{\tau_{\rm eq}}{u\cdot p} e^{+ \xi - \beta \cdot p} \bigg[\left(1 + \frac{1}{2} s_{\alpha \beta} \omega^{\alpha \beta} \right) \left(p^{\mu}{\partial_\mu \xi} - p^{\lambda} p^{\mu} \partial_\mu \beta_{\lambda} \right) + \frac{1}{2} s_{\alpha \beta} p^{\mu} \partial_{\mu} \omega^{\alpha \beta} \bigg]\nn\\
    && -\frac{\tau_{\rm eq}}{u \cdot p} e^{-\xi - \beta \cdot p} \bigg[\left(1 + \frac{1}{2} s_{\alpha \beta} \omega^{\alpha \beta}\right) \left(p^{\mu}{\partial_\mu \xi} + p^{\lambda} p^{\mu} \partial_\mu \beta_{\lambda}\right) - \frac{1}{2} s_{\alpha \beta} p^{\mu} \partial_{\mu} \omega^{\alpha \beta} \bigg]\nn\\
    &=& - \frac{2 \tau_{\rm eq}}{u \cdot p} e^{-\beta \cdot p} \bigg[\left(1 + \frac{1}{2} s_{\alpha \beta} \omega^{\alpha \beta} \right)\! \left(\cosh\xi\; p^{\mu}\, \partial_\mu \xi - \sinh\xi\; p^{\lambda}\, p^{\mu}\, \partial_\mu \beta_{\lambda}\right) + \frac{1}{2} \sinh\xi\; s_{\alpha \beta}\, p^{\mu}\, \partial_{\mu} \omega^{\alpha \beta}\bigg].~\label{dfsub}
\eeq
Substituting Eq.~(\ref{dfsub}) into Eq.~(\ref{delN}) we get
\beq
    \delta N^\mu &=& - 2\tau_{\rm eq}\! \int\! \mathrm{dP}\,\mathrm{dS}\, \frac{p^\mu}{u\cdot p}\, e^{-\beta \cdot p} \Bigg[\!\left(\!1 + \frac{1}{2} s_{\alpha \beta} \omega^{\alpha \beta}\! \right)\! \left(\cosh\xi\; p^{\rho}\, \partial_\rho \xi - \sinh\xi\; p^{\lambda}\, p^{\rho}\, \partial_\rho \beta_{\lambda} \right)\! + \frac{1}{2} \sinh\xi\; s_{\alpha \beta}\, p^{\rho}\, \partial_{\rho} \omega^{\alpha\beta}\Bigg].~~\label{delN1}
\eeq
Now using Eqs.~(\ref{A1}) and (\ref{A4}) we can easily carry out the integration over spin variables, therefore,
\beq
    \delta N^\mu &=& - 4\, \tau_{\rm eq} \int \mathrm{dP}\, \frac{p^\mu}{u\cdot p} e^{-\beta \cdot p} \bigg[\cosh\xi\; p^{\rho}\, \partial_\rho \xi - \sinh\xi\; p^{\lambda}\, p^{\rho}\, \partial_\rho \beta_{\lambda}\bigg]\nn\\
    &=& - 4\, \tau_{\rm eq}\, \cosh\xi\; \partial_\rho \xi \int \mathrm{dP}\, \frac{p^\mu\, p^{\rho}}{u\cdot p}\, e^{-\beta \cdot p} + 4\,\tau_{\rm eq}\, \sinh\xi\; \partial_\rho \beta_{\lambda} \int \mathrm{dP}\, \frac{p^\mu\, p^\lambda\, p^\rho}{u\cdot p}\, e^{-\beta\cdot p}.~\label{delN2}
\eeq
Momentum integration can be carried out using the following useful integral formulas,
\bea
    I^{\mu_1 \mu_2\,\,\text{...}\,\,  \mu_n}_{(r)} &=& \int \frac{\mathrm{dP}}{(u\cdot p)^r} \left(p^{\mu_1} p^{\mu_2} \,\,\text{...}\,\,  p^{\mu_n} e^{- \beta \cdot p} \right) \nn\\
    &=& I_{ n{\rm 0}}^{(r)} u^{\mu_1} u^{\mu_2} \,\,\text{...}\,\,  u^{\mu_n} + I_{n {\rm 1}}^{(r)} \left(\Delta^{\mu_1 \mu_2} u^{\mu_3} u^{\mu_4} \,\,\text{...}\,\,  u^{\mu_n} + \text{permutations}\right) + \,\,\text{...}\,\,. \label{eq:imp_int}
\eea
Thus, in the cases which are of interest for us, the formula (\ref{eq:imp_int}) gives:
\beq
    I_{(r)}^{\mu\rho} &=& I_{20}^{(r)} u^{\mu} u^{\rho} + I_{21}^{(r)} \Delta^{\mu \rho}, \label{eq:I1}\\
    I_{(r)}^{\mu \nu \rho} &=& I_{30}^{(r)} u^{\mu} u^{\nu} u^{\rho} + I_{31}^{(r)} \left(\Delta^{\mu \nu} u^{\rho} + \Delta^{\mu \rho} u^{\nu} + \Delta^{\nu \rho} u^{\mu}\right), \label{eq:I2}\\
    I_{(r)}^{\mu \nu \lambda \rho} &=& I_{40}^{(r)} u^{\mu} u^{\nu} u^{\lambda} u^{\rho} + I_{41}^{(r)} \left(\Delta^{\mu \nu} u^{\lambda} u^{\rho} + \Delta^{\mu \lambda} u^{\nu} u^{\rho} + \Delta^{\nu \lambda} u^{\mu} u^{\rho} + \Delta^{\mu \rho} u^{\nu} u^{\lambda} + \Delta^{\nu \rho} u^{\mu} u^{\lambda} + \Delta^{\lambda \rho} u^{\mu} u^{\nu}\right)\nn\\
    &&+\, I_{42}^{(r)} \left(\Delta^{\mu \nu} \Delta^{\lambda \rho} + \Delta^{\mu \lambda} \Delta^{\nu \rho} + \Delta^{\mu \rho} \Delta^{\nu \lambda}\right), \label{eq:I3}\\
    I_{(r)}^{\mu \nu \lambda \rho \sigma} &=& I_{50}^{(r)} u^{\mu} u^{\nu} u^{\lambda} u^{\rho} u^{\sigma} + I_{51}^{(r)} \Big(\Delta^{\mu \nu} u^{\lambda} u^{\rho} u^{\sigma} + \Delta^{\mu \lambda} u^{\nu} u^{\rho} u^{\sigma} + \Delta^{\nu \lambda} u^{\mu} u^{\rho} u^{\sigma} + \Delta^{\mu \rho} u^{\nu} u^{\lambda} u^{\sigma} + \Delta^{\nu \rho} u^{\mu} u^{\lambda} u^{\sigma} \nn\\
    &&+\, \Delta^{\mu \sigma} u^{\nu} u^{\lambda} u^{\rho} + \Delta^{\nu \sigma} u^{\mu} u^{\lambda} u^{\rho} + \Delta^{\lambda \rho} u^{\mu} u^{\nu} u^{\sigma} + \Delta^{\lambda \sigma} u^{\mu} u^{\nu} u^{\rho} + \Delta^{\rho \sigma} u^{\mu} u^{\nu} u^{\lambda} \Big)\nn\\
    &&+\, I_{52}^{(r)}\Big[u^{\mu} \left(\Delta^{\nu \lambda} \Delta^{\rho \sigma} + \Delta^{\nu \rho} \Delta^{\lambda \sigma} + \Delta^{\nu \sigma} \Delta^{\lambda \rho} \right) + u^{\nu} \left(\Delta^{\mu \lambda} \Delta^{\rho \sigma} + \Delta^{\mu \rho} \Delta^{\lambda \sigma} + \Delta^{\mu \sigma} \Delta^{\lambda \rho} \right)\nn\\
    &&+\, u^{\lambda} \left(\Delta^{\mu \nu} \Delta^{\rho \sigma} + \Delta^{\mu \rho} \Delta^{\nu \sigma} + \Delta^{\mu \sigma} \Delta^{\nu \rho}\right) + u^{\rho} \left(\Delta^{\mu \nu} \Delta^{\lambda \sigma} + \Delta^{\mu \lambda} \Delta^{\nu \sigma} + \Delta^{\mu \sigma} \Delta^{\nu \lambda}\right) \nn\\
    &&+\, u^{\sigma} \left(\Delta^{\mu \nu} \Delta^{\lambda \rho}  + \Delta^{\mu \lambda} \Delta^{\nu \rho} + \Delta^{\mu \rho} \Delta^{\nu \lambda}\right) \Big].\label{eq:I4}
\eeq
Using the integral formula (\ref{eq:imp_int}) in Eq.~(\ref{delN2}) we get
\beq
    \delta N^\mu &=& - 4 \tau_{\rm eq} \cosh\xi\, I_{(1)}^{\mu \rho}\, \partial_\rho \xi + 4 \tau_{\rm eq}\, \sinh\xi\, I_{(1)}^{\mu \lambda \rho}\, \partial_\rho \beta_{\lambda}, \nn\\
     &=& - 4 \tau_{\rm eq} \cosh\xi \left(I_{20}^{(1)} u^{\mu} u^{\rho}  + I_{21}^{(1)} \Delta^{\mu \rho}\right) \partial_\rho \xi + 4 \tau_{\rm eq} \sinh\xi \left[I_{30}^{(1)} u^{\mu} u^{\lambda} u^{\rho} + I_{31}^{(1)} \left(\Delta^{\mu \lambda} u^{\rho}+ \Delta^{\mu \rho} u^{\lambda} + \Delta^{\lambda \rho} u^{\mu} \right)\right] \partial_\rho \beta_{\lambda}.\nn\\\label{delN3}
 \eeq 
One can express the space-like (transverse) derivative operator as
\beq
    \nabla_{\rho} = \Delta_{\rho}^{\alpha} \partial_{\alpha} =(g_{\rho}^{\alpha} - u_\rho u^\alpha) \partial_{\alpha} = \partial_{\rho} - u_\rho D. \label{B8}
\eeq
Using Eq.~(\ref{B8}) we can write
\beq
    \partial_\rho \xi &=& \left(\nabla_\rho + u_\rho D \right) \xi = \nabla_{\rho} \xi + u_{\rho}\dot{\xi}, \label{B9}\\
    \partial_\rho \omega^{\mu\nu} &=& \left(\nabla_\rho + u_\rho D \right) \omega^{\mu\nu} = \nabla_{\rho} \omega^{\mu\nu} + u_{\rho} \dot{\omega}^{\mu\nu}, \label{B9.1}\\
    \partial_\rho \beta_\lambda &=& \partial_\rho (\beta u_\lambda) = \beta \partial_\rho u_\lambda + u_\lambda \partial_\rho\beta. \label{B10}
\eeq
Again using Eq.~(\ref{B8}) in Eq.~(\ref{B10}) we get 
\beq
    \partial_\rho\beta_\lambda = \beta\left(\nabla_\rho + u_\rho D \right) u_\lambda + u_\lambda \left( \nabla_{\rho} + u_{\rho} D \right)\beta\
    = \beta\nabla_\rho u_\lambda + \beta u_\rho \dot{u}_\lambda + u_\lambda \nabla_\rho \beta + u_\lambda u_\rho \dot{\beta}. \label{B11}
\eeq
Using Eqs.~(\ref{B9}) and  (\ref{B11}) in Eq.~(\ref{delN3}) one gets
\beq
    \delta N^\mu &=& - 4 \tau_{\rm eq} \cosh\xi \left(I_{20}^{(1)} u^{\mu} u^{\rho} + I_{21}^{(1)} \Delta^{\mu \rho}\right) \left(\nabla_\rho \xi + u_\rho \dot{\xi} \right) + 4 \tau_{\rm eq} \sinh\xi \Big[I_{30}^{(1)} u^{\mu} u^{\lambda} u^{\rho} \nn\\
    &&+\, I_{31}^{(1)} \left(\Delta^{\mu \lambda} u^{\rho} + \Delta^{\mu \rho} u^{\lambda} + \Delta^{\lambda \rho} u^{\mu}\right)\Big] \left(\beta\nabla_{\rho} u_\lambda + \beta u_{\rho} \dot{u}_{\lambda} + u_\lambda \nabla_{\rho} \beta + u_\lambda u_{\rho} \dot{\beta} \right). \label{delN4}
 \eeq 

\subsection{Dissipative corrections \texorpdfstring{$\delta T^{\mu\nu}$}{} } \label{sec:appDT}

The dissipative part of the energy-momentum tensor can be written as
\beq
    \delta T^{\mu\nu} &=& \int \mathrm{dP}\,\mathrm{dS}\, p^\mu p^\nu \big(\delta f^+_s + \delta f^-_s\big). \label{delT}
\eeq
From Eq.~\eqref{del feq lim} we find the sum of the out-of-equilibrium corrections to the distribution functions for particles and antiparticles
\beq
    \delta f^+_s + \delta f^-_s &=& - \frac{\tau _{\rm eq}}{u\cdot p} e^{+\xi -\beta \cdot p} \bigg[\left(1 + \frac{1}{2} s_{\alpha \beta} \omega^{\alpha \beta}\right) \left(p^{\rho} \partial_\rho \xi - p^{\lambda} p^{\rho} \partial_\rho \beta_{\lambda} \right) + \frac{1}{2} s_{\alpha \beta} p^{\rho} \partial_{\rho} \omega^{\alpha \beta}\bigg]\nn\\
    &&+\, \frac{\tau_{\rm eq}}{u\cdot p} e^{- \xi - \beta\cdot p} \bigg[\left(1 + \frac{1}{2} s_{\alpha \beta} \omega^{\alpha \beta}\right) \left(p^{\rho}\partial_\rho \xi + p^{\lambda } p^{\rho} \partial_\rho \beta_{\lambda} \right) - \frac{1}{2} s_{\alpha \beta} p^{\rho} \partial_{\rho} \omega^{\alpha \beta} \bigg]\nn\\
    &=& - \frac{2 \tau_{\rm eq}}{u\cdot p} e^{-\beta\cdot p} \bigg[\left(1 + \frac{1}{2} s_{\alpha \beta} \omega^{\alpha \beta}\right) \left(\sinh \xi\, p^{\rho}\, \partial_\rho \xi - \cosh \xi\, p^{\lambda}\, p^{\rho}\, \partial_\rho \beta_{\lambda}\right) + \frac{1}{2} \cosh \xi\, s_{\alpha \beta}\, p^{\rho}\, \partial_{\rho} \omega^{\alpha \beta} \bigg]. \label{dfsum}
\eeq
Substituting Eq.~\eqref{dfsum} in Eq.~\eqref{delT}, we obtain
\beq
    \delta T^{\mu\nu} &=& - 2\, \tau_{\rm eq} \!\int\! \mathrm{dP}\,\mathrm{dS}\, \frac{p^\mu p^\nu}{u\cdot p}e^{-\beta\cdot p} \Bigg[\!\!\left(\!1+\frac{1}{2} s_{\alpha \beta} \omega ^{\alpha \beta}\!\right)\!\left(\sinh\xi\, p^{\rho} \partial_\rho \xi - \cosh\xi\, p^{\lambda} p^{\rho} \partial_\rho \beta_{\lambda}\right) + \frac{1}{2} \cosh\xi s_{\alpha \beta} p^{\rho} \partial_{\rho} \omega^{\alpha \beta} \Bigg].~~~~~\label{delT1}
\eeq
Integration over spin variables and using Eqs.~\eqref{A1} and~\eqref{A4} leads to
\beq
    \delta T^{\mu\nu} &=& - 4 \tau_{\rm eq}\int \mathrm{dP}\, \frac{p^\mu p^\nu}{u\cdot p}e^{-\beta \cdot p} \left(\sinh \xi p^{\rho}\, \partial_\rho \xi - \cosh \xi p^{\lambda} p^{\rho}\, \partial_\rho \beta_{\lambda}\right) \nn\\
    &=& - 4 \tau _{\rm eq} \sinh \xi\, \partial_\rho \xi \int \mathrm{dP}\, \frac{p^\mu\, p^\nu\, p^{\rho}}{u\cdot p} e^{-\beta\cdot p} + 4 \tau_{\rm eq} \cosh \xi\, \partial_\rho \beta_\lambda \int \mathrm{dP}\, \frac{p^\mu p^\nu p^\lambda p^{\rho} }{u\cdot p} e^{-\beta\cdot p}. \label{delT2}
\eeq
Using the integral formula \eqref{eq:imp_int} in the above equation we obtain
\beq
    \delta T^{\mu\nu} &=& -4\, \tau_{\rm eq} \sinh \xi\, I_{(1)}^{\mu \nu \rho}\, \partial_\rho \xi + 4\, \tau_{\rm eq} \cosh \xi\, I_{(1)}^{\mu \nu \lambda \rho }\, \partial_\rho \beta_\lambda . \label{delT3.1}
\eeq
Subsequently, using Eqs.~\eqref{eq:I2} and \eqref{eq:I3}, we find
\beq
    \delta T^{\mu\nu} &=& - 4 \tau_{\rm eq} \sinh \xi\, \partial_\rho \xi \left[I_{30}^{(1)} u^{\mu} u^{\nu} u^{\rho} + I_{31}^{(1)} \left(\Delta^{\mu \nu} u^{\rho} + \Delta^{\mu \rho} u^{\nu} + \Delta^{\nu \rho} u^{\mu} \right)\right] + 4 \tau_{\rm eq} \cosh \xi\, \partial_\rho \beta_\lambda \bigg[ I_{40}^{(1)} u^{\mu} u^{\nu} u^{\lambda} u^{\rho} \nn\\
    &&+\, I_{41}^{(1)} \left(\Delta^{\mu \nu} u^{\lambda} u^{\rho} + \Delta^{\mu \lambda} u^{\nu} u^{\rho} + \Delta^{\nu \lambda} u^{\mu} u^{\rho} + \Delta^{\mu \rho} u^{\lambda} u^{\nu} + \Delta^{\nu \rho} u^{\lambda} u^{\mu} + \Delta^{\lambda \rho} u^{\mu} u^{\nu}\right)\nn\\
    &&+\, I_{42}^{(1)} \left(\Delta^{\mu \nu} \Delta^{\lambda \rho} + \Delta^{\mu \lambda} \Delta^{\nu \rho} + \Delta^{\mu \rho} \Delta^{\nu \lambda} \right)\bigg]. \label{delT3.2}
\eeq
Furthermore, using Eqs.~\eqref{B9} and \eqref{B11} in the above equation we get
\beq
    \delta T^{\mu\nu} &=& - 4 \tau_{\rm eq} \sinh \xi \left[ I_{30}^{(1)} u^{\mu} u^{\nu} u^{\rho} + I_{31}^{(1)} \left(\Delta^{\mu \nu} u^{\rho} + \Delta^{\mu \rho} u^{\nu} + \Delta^{\nu \rho} u^{\mu}\right)\right] \left(\nabla_{\rho} \xi + u_{\rho} \dot{\xi} \right) + 4 \tau_{\rm eq} \cosh \xi \bigg[I_{40}^{(1)} u^{\mu} u^{\nu} u^{\lambda} u^{\rho} \nn\\
    &&+\, I_{41}^{(1)}\left(\Delta^{\mu \nu} u^{\lambda} u^{\rho} + \Delta^{\mu \lambda} u^{\nu} u^{\rho} + \Delta^{\nu \lambda} u^{\mu} u^{\rho} + \Delta^{\mu \rho} u^{\lambda} u^{\nu} + \Delta^{\nu \rho} u^{\lambda} u^{\mu} + \Delta^{\lambda \rho} u^{\mu} u^{\nu} \right)\nn\\
    &&+\, I_{42}^{(1)} \left(\Delta^{\mu \nu} \Delta^{\lambda \rho} + \Delta^{\mu \lambda} \Delta^{\nu \rho} + \Delta^{\mu \rho} \Delta^{\nu \lambda}\right)\bigg] \left(\beta\nabla_{\rho} u_\lambda + \beta u_{\rho} \dot{u}_{\lambda} + u_\lambda \nabla_{\rho} \beta + u_\lambda u_{\rho} \dot{\beta}\right).\label{delT3}
\eeq

\subsection{Dissipative corrections  \texorpdfstring{$\delta S^{\lambda,\mu\nu}$}{}} \label{sec:appDS}
Dissipative part of the spin-tensor is given by the formula
\beq
    \delta S^{\lambda,\mu\nu} &=& \int \mathrm{dP\,dS}\, p^{\lambda}\, s^{\mu\nu} (\delta f^+_s + \delta f^-_s). \label{eq_dspin1}
\eeq
Using Eq.~\eqref{dfsum} in Eq.~\eqref{eq_dspin1} one gets
\beq
    \delta S^{\lambda,\mu\nu}\!&=&\!-2 \tau_{\rm eq} \!\int\! \mathrm{dP\,dS}\, \frac{p^{\lambda}\, s^{\mu\nu}}{u\cdot p} e^{-\beta\cdot p} \Bigg[\!\! \left(\!1\!+\!\frac{1}{2} s_{\alpha \beta}\, \omega^{\alpha \beta} \!\right)\!\! \big(\sinh \xi\, p^{\rho}\, \partial_\rho \xi - \cosh \xi\, p^{\kappa} p^{\rho}\, \partial_\rho \beta_{\kappa}\big) + \frac{1}{2} \cosh \xi\, s_{\alpha \beta}\, p^{\rho}\, \partial_{\rho} \omega^{\alpha \beta} \!\Bigg]. ~~~~~~~~\label{eq_dspin1.11}
\eeq
With the help of Eqs.~\eqref{A4} and \eqref{A9} the integration over the spin degrees of freedom in the above equation can be easily performed giving
\beq
    \delta S^{\lambda,\mu\nu} &=& - \frac{4 \mathfrak{s}^2 \tau_{\rm eq}}{3\, m^2} \!\int\! \mathrm{dP}\, \frac{p^{\lambda}}{u\cdot p} e^{-\beta\cdot p} \bigg(\!\sinh \xi\, p^{\rho}\, \partial_\rho \xi - \cosh \xi\, p^{\kappa}\, p^{\rho}\, \partial_\rho \beta_{\kappa} + \cosh \xi\, p^{\rho}\, \partial_{\rho} \bigg) \big(m^2 \omega ^{\mu \nu} + 2\, p^{\alpha}\, p^{[\mu}\, \omega^{\nu]}{}_{\alpha}\big).~~~~ \label{eq_dspin2}
\eeq
The above equation can further be written as a sum of six terms
\beq
    \delta S^{\lambda,\mu\nu} &=& - \underbrace{\frac{4\, \mathfrak{s}^2}{3} \tau_{\rm eq} \sinh \xi\, \partial_\rho \xi \!\int\! \mathrm{dP}\, \frac{p^\lambda\, p^{\rho}\, \omega^{\mu \nu}}{u\cdot p} e^{-\beta\cdot p}}_{I} + \underbrace{\frac{4\, \mathfrak{s}^2}{3} \tau_{\rm eq} \cosh \xi\, \partial_\rho \beta_\kappa \!\int\! \mathrm{dP}\, \frac{p^\lambda\, p^\kappa\, p^{\rho}\, \omega^{\mu \nu}}{u\cdot p} e^{-\beta \cdot p}}_{II} \nn\\
    &&-\, \underbrace{\frac{4\, \mathfrak{s}^2}{3} \tau_{\rm eq} \cosh \xi\, \partial_\rho \omega^{\mu\nu} \!\int\! \mathrm{dP}\, \frac{p^\lambda p^{\rho}}{u\cdot p} e^{-\beta\cdot p}}_{III} - \underbrace{\frac{8\, \mathfrak{s}^2}{3\,m^2} \tau_{\rm eq} \sinh \xi\, \partial_\rho \xi \!\int\! \mathrm{dP}\, \frac{p^\lambda\, p^{\rho}\, p^\alpha\, p^{[\mu}\, \omega^{\nu]}_{~~{\alpha}}}{u\cdot p} e^{-\beta\cdot p}}_{IV} \nn\\
    &&+\, \underbrace{\frac{8\, \mathfrak{s}^2}{3\, m^2} \tau_{\rm eq} \cosh \xi\, \partial_\rho \beta_\kappa \!\int\! \mathrm{dP}\, \frac{p^\lambda\, p^\kappa\, p^{\rho}\, p^\alpha\, p^{[\mu}\, \omega^{\nu]}_{~~{\alpha}}}{u\cdot p} e^{-\beta\cdot p}}_{V}
    - \underbrace{\frac{8\, \mathfrak{s}^2}{3\,m^2} \tau_{\rm eq} \cosh \xi \!\int\! \mathrm{dP}\, \frac{p^\lambda\, p^{\rho}\, p^\alpha\, p^{[\mu}\, \partial_{\rho}{\omega}^{\nu]}_{~~{\alpha}}}{u\cdot p} e^{-\beta\cdot p}}_{VI}. ~~~~~~\label{eq_dspin3}
\eeq
Now we evaluate one by one each of the terms appearing in this expression.

\medskip
\noindent\textbf{Term I:}
\medskip
\beq
   I &=& \frac{4\, \mathfrak{s}^2}{3} \tau_{\rm eq} \sinh \xi\, \partial_\rho \xi\, \omega^{\mu \nu} \!\int\! \mathrm{dP}\, \frac{p^\lambda\, p^{\rho}}{u\cdot p} e^{-\beta\cdot p}.
\eeq
Using Eqs.~(\ref{B9}), (\ref{eq:imp_int}) and (\ref{eq:I1}), we can get
\beq
    I &=& \frac{4\, \mathfrak{s}^2}{3} \tau_{\rm eq} \sinh \xi\, (\nabla_{\rho} \xi + u_{\rho} \dot{\xi}) \omega^{\mu \nu} I_{(1)}^{\lambda \rho} \nn\\
    &=& \frac{4\, \mathfrak{s}^2}{3} \tau_{\rm eq} \sinh \xi (\nabla_{\rho} \xi +\dot{\xi} u_{\rho}) \omega^{\mu \nu} \left(I_{20}^{(1)} u^{\lambda} u^{\rho} + I_{21}^{(1)} \Delta^{\lambda \rho}\right) \nn\\
    &=&\frac{4\, \mathfrak{s}^2}{3} \tau_{\rm eq} \sinh \xi \omega^{\mu \nu} \left(I_{21}^{(1)}\, \nabla^{\lambda} \xi + \dot{\xi} u^{\lambda}\, I_{20}^{(1)}\right). \label{T1}
\eeq
\medskip
\noindent\textbf{Term II:}
%
\beq
    II&=&\frac{4\, \mathfrak{s}^2}{3} \tau_{\rm eq} \cosh \xi\, \partial_\rho \beta_\kappa \!\int\! \mathrm{dP}\, \frac{p^\lambda\, p^\kappa\, p^{\rho}\, \omega^{\mu \nu}}{u\cdot p} e^{-\beta\cdot p}.
\eeq
Using Eqs.~(\ref{B11}), (\ref{eq:imp_int}), and (\ref{eq:I2}) this term can be written as 
\beq
    II&=& \frac{4\, \mathfrak{s}^2}{3} \tau_{\rm eq} \cosh \xi\, \omega^{\mu\nu}\! \left(\beta\nabla_{\rho} u_\kappa\!+\! \beta u_{\rho} \dot{u}_{\kappa}\!+\! u_\kappa \nabla_{\rho} \beta\!+\! u_\kappa u_{\rho} \dot{\beta}\right)\! \left[I_{30}^{(1)} u^{\lambda} u^{\kappa} u^{\rho}\!+\!I_{31}^{(1)} \left(\Delta^{\lambda \kappa} u^{\rho} + \Delta^{\lambda \rho} u^{\kappa} + \Delta^{\kappa \rho} u^{\lambda}\right) \right].\!~~~~~
\eeq
This expression simplifies to
\beq
    II&=& \frac{4\, \mathfrak{s}^2}{3} \tau_{\rm eq} \cosh \xi\, \omega^{\mu\nu} \left[I_{30}^{(1)} \dot{\beta} u^{\lambda} + I_{31}^{(1)} \left(\beta u^{\lambda} \theta + \beta \dot{u}^{\lambda} + \nabla^{\lambda} \beta\right) \right].\label{T2}
\eeq
\medskip
\noindent\textbf{Term III:}
\medskip
\beq
    III&=& \frac{4\, \mathfrak{s}^2}{3} \tau_{\rm eq} \cosh \xi\, \partial_\rho \omega^{\mu\nu} \!\int\! \mathrm{dP}\, \frac{p^\lambda\, p^{\rho}}{u\cdot p} e^{-\beta\cdot p}.
\eeq
Using Eqs.~(\ref{B9.1}), (\ref{eq:imp_int}), and (\ref{eq:I1}) we can write
\beq
    III&=& \frac{4\, \mathfrak{s}^2}{3} \tau_{\rm eq} \cosh \xi\, \left(\nabla_{\rho} \omega^{\mu\nu} + u_{\rho} \dot{\omega}^{\mu\nu}\right) I_{(1)}^{\lambda\rho}\nn\\
    &=&\frac{4\, \mathfrak{s}^2}{3} \tau_{\rm eq} \cosh \xi\, \left(\nabla_{\rho} \omega^{\mu\nu} + u_{\rho} \dot{\omega}^{\mu\nu}\right) \left(I_{20}^{(1)} u^{\lambda} u^{\rho} + I_{21}^{(1)} \Delta^{\lambda \rho}\right)\nn\\
    &=& \frac{4\, \mathfrak{s}^2}{3} \tau_{\rm eq} \cosh \xi\, \left(I_{21}^{(1)}\, \nabla^{\lambda} \omega^{\mu\nu} + I_{20}^{(1)}\, u^{\lambda} \dot{\omega}^{\mu\nu} \right). \label{T3}
\eeq
\medskip
\noindent\textbf{Term IV:}
\medskip
\beq
    IV=\frac{8\, \mathfrak{s}^2}{3\,m^2} \tau_{\rm eq} \sinh \xi\, \partial_\rho \xi \!\int\! \mathrm{dP}\, \frac{p^\lambda\, p^{\rho}\, p^\alpha\, p^{[\mu}\, \omega^{\nu]}_{~~{\alpha}}}{u\cdot p} e^{-\beta\cdot p}.
\eeq
Using Eqs.~(\ref{B9}), (\ref{eq:imp_int}), and (\ref{eq:I3}) we find
\beq
    IV&=&\frac{8\, \mathfrak{s}^2}{3\,m^2} \tau_{\rm eq} \sinh \xi\, (\nabla_{\rho} \xi + u_{\rho} \dot{\xi}) I_{(1)}^{\lambda \rho \alpha [\mu}{\omega}^{\nu]}_{~~{ \alpha}}\nn\\
    &=& \frac{8\, \mathfrak{s}^2}{3\,m^2} \tau_{\rm eq} \sinh \xi\, (\nabla_{\rho} \xi + u_{\rho} \dot{\xi}) \Bigg[ I_{40}^{(1)} u^{\lambda} u^{\rho} u^{\alpha} u^{[\mu} + I_{41}^{(1)} \!\Big(\! \Delta^{\lambda \rho} u^{\alpha} u^{[\mu} + \Delta^{\lambda \alpha} u^{\rho} u^{[\mu} + \Delta^{\lambda [\mu} u^{\rho} u^{\alpha} \nn\\
    &&+\, \Delta^{\rho \alpha} u^{\lambda} u^{[\mu} + \Delta^{\rho [\mu} u^{\lambda} u^{\alpha} + \Delta^{\alpha [\mu} u^{\lambda} u^{\rho}\Big)
    + I_{42}^{(1)}\left(\Delta^{\lambda \rho} \Delta^{\alpha [\mu} + \Delta^{\lambda \alpha} \Delta^{\rho [\mu} + \Delta^{\lambda [\mu} \Delta^{\rho \alpha}\right)\!\Bigg] {\omega}^{\nu]}_{~~{\alpha}}\nn\\
    &=& \frac{8\, \mathfrak{s}^2}{3\,m^2} \tau_{\rm eq} \sinh \xi \Bigg[I_{41}^{(1)} \left(u^{\alpha} u^{[\mu} \omega^{\nu]}{}_{\alpha}\, \nabla^{\lambda} \xi + u^{\lambda} u^{[\mu} \omega^{\nu]}{}_{\alpha} \nabla^{\alpha} \xi + u^{\lambda} u^{\alpha} \omega^{[\nu}{}_{\alpha} \nabla^{\mu]} \xi\right)\nn\\
    &&+\, I_{42}^{(1)}\left(\Delta^{\alpha [\mu} \omega^{\nu ]}{}_{\alpha} \nabla^{\lambda}\xi + \Delta^{\lambda \alpha} \omega^{[\nu}{}_{\alpha} \nabla^{\mu]} \xi + \Delta^{\lambda [\mu} \omega^{\nu]}{}_{\alpha} \nabla^{\alpha} \xi \right)\nn\\
    &&+\, \dot{\xi} \left\{I_{40}^{(1)} u^{\lambda} u^{\alpha} u^{[\mu} \omega^{\nu]}{}_{\alpha} + I_{41}^{(1)} \left(\Delta^{\lambda \alpha} u^{[\mu} \omega^{\nu]}{}_{\alpha} + \Delta^{\lambda [\mu} u^{\alpha} \omega^{\nu]}{}_{\alpha} + \Delta^{\alpha [\mu} u^{\lambda} \omega^{\nu]}{}_{\alpha}\right)\right\}\Bigg] \label{T4}.
\eeq
\medskip
\noindent\textbf{Term V:}
\medskip
\beq
    V&=& \frac{8\, \mathfrak{s}^2}{3\, m^2} \tau_{\rm eq} \cosh \xi\, \partial_\rho \beta_\kappa \!\int\! \mathrm{dP}\, \frac{p^\lambda\, p^\kappa\, p^{\rho}\, p^\alpha\, p^{[\mu} \omega^{\nu]}_{~~{\alpha}}}{u\cdot p}\, e^{-\beta\cdot p}.
\eeq
Using Eqs.~(\ref{B11}), (\ref{eq:imp_int}) and (\ref{eq:I4}) this term can be written as
\beq
    V&=&\frac{8\, \mathfrak{s}^2}{3\, m^2} \tau_{\rm eq} \cosh \xi \left(\beta \nabla_{\rho} u_\kappa + \beta u_{\rho} \dot{u}_{\kappa} + u_\kappa \nabla_{\rho} \beta + u_\kappa u_{\rho} \dot{\beta}\right) I_{(1)}^{\lambda\kappa\rho \alpha [\mu}{\omega}^{\nu]}_{~~{\alpha}} \nn\\
    &=&\frac{8\, \mathfrak{s}^2}{3\, m^2} \tau_{\rm eq} \cosh \xi \left(\beta \nabla_{\rho} u_\kappa + \beta u_{\rho} \dot{u}_{\kappa} + u_\kappa \nabla_{\rho} \beta + u_\kappa u_{\rho} \dot{\beta} \right)\! \Bigg[I_{50}^{(1)} u^{\lambda} u^{\kappa} u^{\rho} u^{\alpha} u^{[\mu} + I_{51}^{(1)}\Big(\Delta^{\lambda \kappa} u^{\rho} u^{\alpha} u^{[\mu} + \Delta^{\lambda \rho} u^{\kappa} u^{\alpha} u^{[\mu} \nn\\
    &&+\, \Delta^{\lambda \alpha} u^{\kappa} u^{\rho} u^{[\mu} + \Delta^{\lambda [\mu} u^{\kappa} u^{\rho} u^{\alpha} + \Delta^{\kappa \rho} u^{\lambda} u^{\alpha} u^{[\mu} + \Delta^{\kappa \alpha} u^{\lambda} u^{\rho} u^{[\mu} + \Delta^{\kappa [\mu} u^{\lambda} u^{\rho} u^{\alpha} + \Delta^{\rho \alpha} u^{\lambda} u^{\kappa} u^{[\mu} + \Delta^{\rho [\mu} u^{\lambda} u^{\kappa} u^{\alpha} \nn\\
    &&+\, \Delta^{\alpha [\mu} u^{\lambda} u^{\kappa} u^{\rho} \Big) + I_{52}^{(1)} \bigg\{ u^{\lambda} \left(\Delta^{\kappa \rho} \Delta^{\alpha [\mu} + \Delta^{\kappa \alpha} \Delta^{\rho [\mu} + \Delta^{\kappa [\mu} \Delta^{\rho \alpha} \right) + u^{\kappa} \left(\Delta^{\lambda \rho} \Delta^{\alpha [\mu} + \Delta^{\lambda \alpha} \Delta^{\rho [\mu} + \Delta^{\lambda [\mu} \Delta^{\rho \alpha} \right) \nn\\
    &&+ u^{\rho} \!\left(\!\Delta^{\lambda \kappa} \Delta^{\alpha [\mu} + \Delta^{\lambda \alpha} \Delta^{\kappa [\mu} + \Delta^{\lambda [\mu} \Delta^{\kappa \alpha} \!\right)\! + u^{\alpha} \!\left(\!\Delta^{\lambda \kappa} \Delta^{\rho [\mu} + \Delta^{\lambda \rho} \Delta^{\kappa [\mu} + \Delta^{\lambda [\mu} \Delta^{\kappa \rho} \!\right) \nn\\ 
    &&+\, u^{[\mu} \!\left(\Delta^{\lambda \kappa} \Delta^{\rho \alpha} + \Delta^{\lambda \rho} \Delta^{\kappa \alpha} + \Delta^{\lambda \alpha} \Delta^{\kappa \rho} \right)\!\bigg\}\! \Bigg] {\omega}^{\nu]}_{~~{\alpha}}\nn\\
    &=& \frac{8\, \mathfrak{s}^2}{3\, m^2}\tau _{\rm eq} \cosh \xi\, \Bigg[I_{50}^{(1)} \dot{\beta} u^{\lambda} u^{\alpha} u^{[\mu} \omega^{\nu]}{}_{\alpha} + I_{51}^{(1)} \bigg(\beta \theta u^{\lambda} u^{\alpha} u^{[\mu} \omega^{\nu]}{}_{\alpha} + \beta \dot{u}^{\lambda} u^{\alpha} u^{[\mu} \omega^{\nu]}{}_{\alpha} + \beta \dot{u}^{\alpha} u^{\lambda} u^{[\mu} \omega^{\nu]}{}_{\alpha} + \beta \dot{u}^{[\mu}u^{\lambda} u^{\alpha} \omega^{\nu]}{}_{\alpha} \nn\\
    &&+\, u^{\alpha} u^{[\mu} \omega^{\nu]}{}_{\alpha}\, \nabla^{\lambda}\beta + u^{\lambda} u^{[\mu} \omega^{\nu]}{}_{\alpha}\, \nabla^{\alpha} \beta + u^{\lambda} u^{\alpha} \omega^{[\nu}{}_{\alpha} \nabla^{\mu]}\beta + \dot{\beta} \Delta^{\lambda \alpha} u^{[\mu} \omega^{\nu]}{}_{\alpha} + \dot{\beta} \Delta^{\lambda [\mu} u^{\alpha} \omega^{\nu]}{}_{\alpha} + \dot{\beta} \Delta^{\alpha [\mu} u^{\lambda} \omega^{\nu]}{}_{\alpha}\bigg)\nn\\
    &&+\, I_{52}^{(1)} \bigg(\!\beta\theta u^{\lambda} \Delta^{\alpha [\mu} \omega^{\nu ]}{}_{\alpha} + \beta u^{\lambda} \omega^{[\nu}{}_{\alpha} \nabla^{\mu]}u^{\alpha} + \beta u^{\lambda} \omega^{[\nu}{}_{\alpha} \nabla^{\alpha} u^{\mu]} + \beta u^{\alpha} \omega^{[\nu}{}_{\alpha} \nabla^{\mu]} u^{\lambda} + \beta u^{\alpha} \omega^{[\nu}{}_{\alpha} \nabla^{\lambda} u^{\mu]} + u^{\alpha} \Delta^{\lambda [\mu} \omega^{\nu]}{}_{\alpha}\, \beta\, \theta\nn\\
    &&+\, \beta u^{[\mu} \omega^{\nu]}{}_{\alpha} \nabla^{\alpha} u^{\lambda} + \beta u^{[\mu} \omega^{\nu]}{}_{\alpha} \nabla^{\lambda} u^{\alpha} + \beta \theta u^{[\mu} \Delta^{\lambda \alpha} \omega^{\nu]}{}_{\alpha} + \beta \dot{u}^{\lambda} \Delta^{\alpha [\mu} \omega^{\nu]}{}_{\alpha} + \beta \Delta^{\lambda \alpha} \dot{u}^{[\mu} \omega^{\nu]}{}_{\alpha} + \beta \dot{u}^{\alpha} \Delta^{\lambda [\mu} \omega^{\nu]}{}_{\alpha}\nn\\
    &&+\, \Delta^{\alpha [\mu} \omega^{\nu]}{}_{\alpha} \nabla^{\lambda} \beta + \Delta^{\lambda \alpha} \omega^{[\nu}{}_{\alpha} \nabla^{\mu]}\beta + \Delta^{\lambda [\mu} \omega^{\nu]}{}_{\alpha} \nabla^{\alpha} \beta \bigg)\Bigg] \label{T5}.
\eeq
\medskip
\noindent\textbf{Term VI:}
%
\beq
    VI&=& \frac{8\, \mathfrak{s}^2}{3\, m^2} \tau_{\rm eq} \cosh \xi \!\int\! \mathrm{dP}\, \frac{p^\lambda\, p^{\rho}\, p^\alpha\, p^{[\mu}\, \partial_{\rho}\, \omega^{\nu]}_{~~\alpha}}{u\cdot p}\, e^{-\beta\cdot p}.
\eeq
Using Eqs.~(\ref{B9.1}), (\ref{eq:imp_int}) and (\ref{eq:I3}) we obtain
\beq
    VI&=& \frac{8\, \mathfrak{s}^2}{3\, m^2} \tau_{\rm eq} \cosh \xi\, I_{(1)}^{\lambda \rho \alpha[\mu} \partial_\rho \omega^{\nu]}{}_{\alpha}\nn\\
    &=&\frac{8\, \mathfrak{s}^2}{3\, m^2} \tau_{\rm eq} \cosh \xi \bigg[ I_{40}^{(1)} u^{\lambda} u^{\rho} u^{\alpha} u^{[\mu} + I_{41}^{(1)} \Big(\Delta^{\lambda \rho} u^{\alpha} u^{[\mu} + \Delta^{\lambda \alpha} u^{\rho} u^{[\mu} + \Delta^{\lambda [\mu} u^{\rho} u^{\alpha} + \Delta^{\rho \alpha} u^{\lambda} u^{[\mu} + \Delta^{\rho [\mu} u^{\lambda} u^{\alpha} \nn\\
    &&+\, \Delta^{\alpha [\mu} u^{\lambda} u^{\rho} \Big)+ I_{42}^{(1)}\left(\Delta ^{\lambda \rho }\Delta ^{\alpha [\mu }+\Delta ^{\lambda \alpha }\Delta ^{\rho [\mu }+\Delta ^{\lambda [\mu }\Delta ^{\rho \alpha }\right)\bigg]\left(\nabla _{\rho }\omega ^{\nu ]}{}_{\alpha }+u_{\rho }\dot{\omega }^{\nu ]}{}_{\alpha }\right)\nn\\
    &=& \frac{8\, \mathfrak{s}^2}{3\, m^2} \tau_{\rm eq} \cosh \xi \bigg[ I_{40}^{(1)} u^{\lambda} u^{\alpha} u^{[\mu} \dot{\omega}^{\nu]}{}_{\alpha} + I_{41}^{(1)} \Big(u^{\alpha} u^{[\mu} \nabla^{\lambda} \omega^{\nu]}{}_{\alpha} + u^{\lambda} u^{[\mu} \nabla^{\alpha} \omega^{\nu]}{}_{\alpha} + u^{\lambda} u^{\alpha} \nabla^{[\mu} \omega^{\nu]}{}_{\alpha} + \Delta^{\lambda \alpha} u^{[\mu} \dot{\omega}^{\nu]}{}_{\alpha} \nn\\
    &&+\, \Delta^{\lambda [\mu} u^{\alpha} \dot{\omega}^{\nu]}{}_{\alpha}
    + \Delta^{\alpha [\mu} u^{\lambda} \dot{\omega}^{\nu]}{}_{\alpha} \Big) + I_{42}^{(1)} \left(\Delta^{\alpha [\mu}\, \nabla^{\lambda} \omega^{\nu]}{}_{\alpha} + \Delta^{\lambda \alpha} \nabla^{[\mu} \omega^{\nu]}{}_{\alpha} + \Delta^{\lambda [\mu} \nabla^{\alpha} \omega^{\nu]}{}_{\alpha} \right)\bigg] \label{T6}.
\eeq
Now substituting Eqs.~(\ref{T1}), (\ref{T2}), (\ref{T3}), (\ref{T4}), (\ref{T5}), and (\ref{T6}) into Eq.~(\ref{eq_dspin3}) one can obtain the following expression for the dissipative correction to the spin tensor
\beq
    \delta S^{\lambda,\mu\nu} &=& \frac{4\, \mathfrak{s}^2}{3} \tau_{\rm eq} \Bigg[ - \sinh \xi \bigg\{I_{21}^{(1)} \omega^{\mu \nu}\, \nabla^{\lambda} \xi + I_{20}^{(1)}\, \dot{\xi}\, u^{\lambda}\, \omega^{\mu \nu} + \frac{2}{m^2}\bigg(I_{41}^{(1)} \left( u^{\alpha} u^{[\mu} \omega^{\nu]}{}_{\alpha}\, \nabla^{\lambda} \xi + u^{\lambda} u^{[\mu} \omega^{\nu]}{}_{\alpha}\, \nabla^{\alpha} \xi + u^{\lambda} u^{\alpha} \omega^{[\nu}{}_{\alpha}\, \nabla^{\mu]} \xi\right)\nn\\
    &+& I_{42}^{(1)} \left(\Delta^{\alpha [\mu} \omega^{\nu]}{}_{\alpha}\, \nabla^{\lambda} \xi + \Delta^{\lambda \alpha} \omega^{[\nu}{}_{\alpha}\, \nabla^{\mu]} \xi + \Delta^{\lambda [\mu} \omega^{\nu]}{}_{\alpha}\, \nabla^{\alpha} \xi\right) + \dot{\xi}\, I_{40}^{(1)} u^{\lambda} u^{\alpha} u^{[\mu} \omega^{\nu]}{}_{\alpha}\nn\\
    &+& \dot{\xi}\, I_{41}^{(1)} \left(\Delta^{\lambda \alpha} u^{[\mu} \omega^{\nu]}{}_{\alpha} + \Delta^{\lambda [\mu} u^{\alpha} \omega^{\nu]}{}_{\alpha} + \Delta^{\alpha [\mu} u^{\lambda} \omega^{\nu]}{}_{\alpha}\right)\bigg)\bigg\} \nn\\
    &+& \cosh \xi \bigg\{I_{30}^{(1)}\, \dot{\beta}\, u^{\lambda} \omega^{\mu \nu} + I_{31}^{(1)} \left(\beta\,  \theta\, u^{\lambda} + \beta\, \dot{u}^{\lambda} + \nabla^{\lambda} \beta \right) \omega^{\mu \nu} + \frac{2}{m^2} \dot{\beta}\, I_{50}^{(1)} u^{\lambda} u^{\alpha} u^{[\mu} \omega^{\nu]}{}_{\alpha}\nn\\
    &+& \frac{2}{m^2} I_{51}^{(1)} \bigg(\beta\, \theta\, u^{\lambda} u^{\alpha} u^{[\mu} \omega^{\nu]}{}_{\alpha} + \left(\beta\, \dot{u}^{\lambda} + \nabla^{\lambda} \beta \right) u^{\alpha} u^{[\mu} \omega^{\nu]}{}_{\alpha} + \left(\beta\, \dot{u}^{\alpha} + \nabla^{\alpha} \beta \right) u^{\lambda} u^{[\mu} \omega^{\nu]}{}_{\alpha}\nn\\
    &+&\left(\beta\, \dot{u}^{[\mu} + \nabla^{[\mu} \beta \right) \omega^{\nu]}{}_{\alpha} u^{\lambda} u^{\alpha} + \dot{\beta} \left(\Delta^{\lambda \alpha} u^{[\mu} \omega^{\nu]}{}_{\alpha} + \Delta^{\lambda [\mu} u^{\alpha} \omega^{\nu]}{}_{\alpha} + \Delta^{\alpha [\mu} u^{\lambda} \omega^{\nu]}{}_{\alpha}\right)\bigg)\nn\\
    &+& \frac{2}{m^2} I_{52}^{(1)}\bigg(\beta\, \theta\, u^{\lambda} \Delta^{\alpha [\mu} \omega^{\nu]}{}_{\alpha} + \beta u^{\lambda} \omega^{[\nu}{}_{\alpha} \nabla^{\mu]} u^{\alpha} + \beta u^{\lambda} \omega^{[\nu}{}_{\alpha} \nabla^{\alpha} u^{\mu]} + \beta\, u^{\alpha} \omega^{[\nu}{}_{\alpha} \nabla^{\mu]} u^{\lambda} + \beta\, u^{\alpha} \omega^{[\nu}{}_{\alpha} \nabla^{\lambda} u^{\mu]} + u^{\alpha} \Delta^{\lambda [\mu} \omega^{\nu]}{}_{\alpha} \beta\, \theta\nn\\
    &+& \beta\, u^{[\mu} \omega^{\nu]}{}_{\alpha} \nabla^{\alpha} u^{\lambda} + \beta\, u^{[\mu} \omega^{\nu]}{}_{\alpha} \nabla^{\lambda} u^{\alpha} + \beta \theta u^{[\mu} \Delta^{\lambda\alpha} \omega^{\nu]}{}_{\alpha} + \Delta ^{\alpha [\mu }\left(\beta\, \dot{u}^{\lambda} + \nabla^{\lambda} \beta\right) \omega^{\nu]}{}_{\alpha} + \Delta^{\lambda \alpha} \left(\beta \dot{u}^{[\mu} + \nabla^{[\mu}\beta\right) \omega^{\nu]}{}_{\alpha}\nn\\
    &+& \Delta^{\lambda [\mu} \left(\beta\, \dot{u}^{\alpha} + \nabla^{\alpha} \beta\right) \omega^{\nu]}{}_{\alpha}\bigg) - \left(I_{21}^{(1)} \nabla^{\lambda} \omega^{\mu \nu} + I_{20}^{(1)}\, u^{\lambda}\, \dot{\omega}^{\mu \nu} \right) - \frac{2}{m^2} \bigg(I_{40}^{(1)} u^{\lambda} u^{\alpha} u^{[\mu} \dot{\omega}^{\nu]}{}_{\alpha}\nn\\
    &+& I_{41}^{(1)} \left(u^{\alpha} u^{[\mu} \nabla^{\lambda} \omega^{\nu]}{}_{\alpha} + u^{\lambda} u^{[\mu} \nabla^{\alpha} \omega^{\nu]}{}_{\alpha} + u^{\lambda} u^{\alpha} \nabla^{[\mu} \omega^{\nu]}{}_{\alpha} + \Delta^{\lambda \alpha} u^{[\mu} \dot{\omega}^{\nu]}{}_{\alpha} + \Delta^{\lambda [\mu} u^{\alpha} \dot{\omega}^{\nu]}{}_{\alpha} + \Delta^{\alpha [\mu} u^{\lambda} \dot{\omega}^{\nu]}{}_{\alpha} \right)\nn\\
    &+& I_{42}^{(1)}\left(\Delta ^{\alpha [\mu }\nabla ^{\lambda }\omega ^{\nu ]}{}_{\alpha }+\Delta ^{\lambda \alpha }\nabla ^{[\mu }\omega ^{\nu ]}{}_{\alpha }+\Delta ^{\lambda [\mu }\nabla ^{\alpha }\omega ^{\nu ]}{}_{\alpha }\right)\bigg)\bigg\}\Bigg]. \label{eq_dspin4}
\eeq


\section{Eliminating \texorpdfstring{$\dot{\xi}$}{}, \texorpdfstring{$\dot{\beta}$}{}, \texorpdfstring{$\dot{u}^{\mu}$}{} and  \texorpdfstring{$\dot{\omega }^{\mu \nu }$}{} from \texorpdfstring{$\delta S^{\lambda,\mu\nu}$}{}}
\label{Ac}

Note that the derivation of equations that specify the convective derivatives $\dot{\xi}$, $\dot{\beta}$, and $\dot{u}^{\mu}$ has already been done in Sec.~\ref{Sbc} and our results are reported in Eqs. (\ref{eq:xidot1}), (\ref{eq:betadot1}), and (\ref{eq:udot1}). Here we present important steps needed for derivation of the dynamical equation for $\dot{\omega}^{\mu\nu}$. Substituting Eq.~(\ref{eq:Smunulambda_de_Groot2.1}) in Eq.~(\ref{eq:con_S1}) we can get
\beq
    \dot{\omega}^{\mu \nu} &=& - \frac{1}{\left(I_{10}^{(0)}-\frac{2}{m^2} I_{31}^{(0)}\right)} \left(I_{10}^{(0)} \theta\,  \omega^{\mu \nu} + I_{10}^{(0)} \dot{\xi} \omega^{\mu \nu} \tanh \xi + \dot{I}_{10}^{(0)} \omega^{\mu \nu} \right)\nn\\
    &-& \frac{2}{\left(m^2\, I_{10}^{(0)} - 2\, I_{31}^{(0)} \right)} \Bigg[\tanh \xi \left(\left(I_{30}^{(0)} - 3 I_{31}^{(0)}\right) \dot{\xi} u^{\alpha} u^{[\mu} \omega^{\nu]}{}_{\alpha} + I_{31}^{(0)} \partial_\lambda \xi \left(u^{[\mu} \omega^{\nu]\lambda} - \omega^{\mu \nu} u^{\lambda} + u^{\alpha} g^{\lambda [\mu} \omega^{\nu]}{}_{\alpha} \right)\right)\nn\\
    &+& \left(\dot{I}_{30}^{(0)} - 3 \dot{I}_{31}^{(0)}\right) u^{\alpha} u^{[\mu} \omega^{\nu]}{}_{\alpha} + \left( u^{[\mu} \omega^{\nu]\lambda}\, \partial_{\lambda} I_{31}^{(0)} - \dot{I}_{31}^{(0)} \omega^{\mu \nu} + u^{\alpha} g^{\lambda[\mu} \omega^{\nu]}{}_{\alpha}\, \partial_\lambda I_{31}^{(0)}\right)\nn\\
    &+& \left(I_{30}^{(0)} - 3 I_{31}^{(0)}\right) \theta\, u^{\alpha} u^{[\mu} \omega^{\nu]}{}_{\alpha} + \left(I_{30}^{(0)} - 3 I_{31}^{(0)}\right) \dot{u}^{\alpha} u^{[\mu} \omega^{\nu]}{}_{\alpha} + \left(I_{30}^{(0)} - 3 I_{31}^{(0)}\right) u^{\alpha} \dot{u}^{[\mu} \omega^{\nu]}{}_{\alpha} \nn\\
    &+& \left(I_{30}^{(0)} - 3 I_{31}^{(0)}\right) u^{\alpha} u^{[\mu} \dot{\omega}^{\nu]}{}_{\alpha} + I_{31}^{(0)} \left( \omega^{[\nu\lambda} \partial_{\lambda} u^{\mu]} + u^{[\mu} \partial_{\lambda} \omega^{\nu] \lambda} - \theta\, \omega^{\mu \nu} + g^{\lambda [\mu} \omega^{\nu]}{}_{\alpha} \partial_\lambda u^{\alpha} + u^{\alpha} \partial^{[\mu} \omega^{\nu]}{}_{\alpha} \!\right)\!\Bigg].
\eeq
Using the relations  $\dot{I}_{10}^{(0)}=-\dot{\beta}I_{20}^{(0)}$,  $\dot{I}_{30}^{(0)}=-\dot{\beta } I_{40}^0$, $\dot{I}_{31}^{(0)}=-\dot{\beta } I_{41}^{(0)} $, ${\partial_{\lambda} I_{30}^{(0)}}=-\left(\partial_{\lambda} \beta\right)I_{40}^{(0)}$, ${\partial_{\lambda} I_{31}^{(0)}}=-{\left(\partial_{\lambda} \beta \right)}I_{41}^{(0)}$ and substituting $\partial_{\lambda}=\nabla_{\lambda}+u_{\lambda}D$ in the above equation we obtain
\beq
    \dot{\omega}^{\mu \nu} &=& - \frac{2}{\left(m^2\, I_{10}^{(0)} - 2\, I_{31}^{(0)}\right)} \left(I_{30}^{(0)} - I_{31}^{(0)}\right) u^{\alpha} u^{[\mu} \dot{\omega}^{\nu]}{}_{\alpha} - \frac{1}{\left(I_{10}^{(0)} - \frac{2}{m^2} I_{31}^{(0)}\right)} \left(I_{10}^{(0)} \theta\, \omega^{\mu \nu} + I_{10}^{(0)} \dot{\xi} \omega^{\mu \nu} \tanh \xi - \dot{\beta} I_{20}^{(0)} \omega^{\mu \nu} \right)\nn\\
    &-& \frac{2}{\left(m^2\, I_{10}^{(0)} - 2\, I_{31}^{(0)} \right)} \Bigg[ \tanh \xi \left(I_{30}^{(0)} - 3 I_{31}^{(0)}\right) \dot{\xi}\, u^{\alpha} u^{[\mu} \omega^{\nu]}{}_{\alpha} + \tanh \xi\, I_{31}^{(0)} \left(\nabla_{\lambda} \xi + \dot{\xi} u_{\lambda} \right) \left(u^{[\mu} \omega^{\nu] \lambda} - \omega^{\mu \nu} u^{\lambda} + u^{\alpha} g^{\lambda [\mu} \omega^{\nu]}{}_{\alpha} \right)\nn\\
    &-& \dot{\beta} \left(I_{40}^{(0)} - 3 I_{41}^{(0)}\right) u^{\alpha} u^{[\mu} \omega^{\nu]}{}_{\alpha} - \left(\left(\nabla_{\lambda} \beta + \dot{\beta} u_{\lambda} \right) I_{41}^{(0)} u^{[\mu} \omega^{\nu]\lambda} - \dot{\beta}{I}_{41}^{(0)} \omega^{\mu \nu} + \left(\nabla_{\lambda} \beta + \dot{\beta } u_{\lambda} \right) I_{41}^{(0)} u^{\alpha} g^{\lambda [\mu} \omega^{\nu]}{}_{\alpha} \right)\nn\\
    &+& \left(I_{30}^{(0)} - 3 I_{31}^{(0)}\right) \theta\, u^{\alpha} u^{[\mu} \omega^{\nu]}{}_{\alpha} + \left(I_{30}^{(0)} - 2 I_{31}^{(0)} \right) \dot{u}^{\alpha} u^{[\mu} \omega^{\nu]}{}_{\alpha} + \left(I_{30}^{(0)} - 2 I_{31}^{(0)}\right) u^{\alpha} \dot{u}^{[\mu} \omega^{\nu]}{}_{\alpha}\nn\\
    &+& {I}_{31}^{(0)} \left(\omega^{[\nu \lambda} \nabla_{\lambda} u^{\mu]} + u^{[\mu} \nabla_{\lambda} \omega^{\nu]\lambda} - \theta\, \omega^{\mu \nu} + \omega^{[\nu}{}_{\alpha} \nabla^{\mu]} u^{\alpha} + u^{\alpha} \nabla^{[\mu} \omega^{\nu]}{}_{\alpha} \right)\Bigg]. \label{eq:omunudot1}
\eeq
We first eliminate $u^{\alpha}\left(u^{[\mu }\dot{\omega }^{\nu ]}{}_{\alpha }\right)$ from the above expression. Contracting the resulting  equation with $u_{\nu}$ and using $I_{30}^{(0)}-2 I_{31}^{(0)}=-\beta I_{41}^{(0)} $, $I_{30}^{(0)}-I_{31}^{(0)}=I_{31}^{(0)}-\beta I_{41}^{(0)}$ at appropriate places we obtain
\beq
    u_{\nu} \dot{\omega}^{\mu \nu} &=& - \frac{m^2}{m^2 I_{10}^{(0)} - \left(I_{30}^{(0)} + I_{31}^{(0)}\right)} \left(I_{10}^{(0)} \theta\, \omega^{\mu \nu} u_{\nu} + I_{10}^{(0)} \dot{\xi}\, \omega^{\mu \nu}\, \tanh \xi\, u_{\nu} - I_{20}^{(0)} \dot{\beta}\, \omega^{\mu \nu}\, u_{\nu}\right)\nn\\
    &-& \frac{1}{m^2 I_{10}^{(0)} - \left(I_{30}^{(0)} + I_{31}^{(0)}\right)} \bigg[\! - \tanh \xi \left(I_{30}^{(0)} + I_{31}^{(0)}\right) \dot{\xi}\, \omega^{\mu \nu} u_{\nu} - \tanh \xi I_{31}^{(0)} \Delta^{\mu}_{\nu} \omega^{\nu}{}_{\lambda} \nabla^{\lambda} \xi\nn\\
    &+& \dot{\beta} \left(I_{40}^{(0)} + I_{41}^{(0)} \right) \omega^{\mu \nu}\, u_{\nu} + \left(\beta\, \dot{u}^{\alpha} + \nabla^{\alpha} \beta \right) I_{41}^{(0)} \Delta^{\mu}{}_{\nu} \omega^{\nu}{}_{\alpha}\nn\\
    &+& \left(\left(I_{41}^{(0)} \beta -I_{31}^{(0)}\right) \theta\, \omega^{\mu \nu} u_{\nu} + I_{31}^{(0)} \left(\omega^{\nu \lambda} u_{\nu} \nabla_{\lambda} u^{\mu} - \Delta^{\mu}{}_{\nu} \nabla_{\lambda} \omega^{\nu \lambda} + u_{\nu} \omega^{\nu}{}_{\alpha} \nabla^{\mu} u^{\alpha} + u^{\alpha} u_{\nu} \nabla^{\mu} \omega^{\nu}{}_{\alpha} \right)\right)\bigg].
\eeq
Now eliminating $\dot{\xi}$, $\dot{\beta}$, and $\dot{u}^{\mu}$ (with the help of Eqs.~(\ref{eq:xidot1}), (\ref{eq:betadot1}) and (\ref{eq:udot1})) the above equation  can be written as
\beq
    u_{\nu} \dot{\omega}^{\mu \nu} &=& C_{\Pi}^{\mu} \theta + C_{n \lambda}^{\mu} \nabla^{\lambda} \xi + C_{\pi \alpha}\, \sigma^{\alpha \mu} + C_{\Sigma \nu}^{\mu} \nabla_{\lambda} \omega^{\nu \lambda}. \label{eq:unuomunudot1}
\eeq
Various $C-$coefficients appearing in the above equation are as follows:
\beq
    C_{\Pi}^{\mu} &=& C_{\Pi}\, u_{\nu}\, \omega^{\mu \nu},\label{eq:CPi1}\\
	C_{n \lambda}^{\mu} &=& C_{n}\, \Delta^{\mu}_{\nu}\, \omega^{\nu}{}_{\lambda},\label{eq:Cn1}\\
	C_{\pi\alpha} &=& C_{\pi}\, u_{\nu}\, \omega^{\nu}{}_{\alpha}, \label{eq:Cpia1}\\
    C_{\Sigma \nu}^{\mu} &=& C_{\Sigma}\, \Delta^{\mu}_{\nu}, \label{eq:Csig1}
\eeq	
where
\beq
    C_{\Pi} &=& - \frac{1}{m^2 I_{10}^{(0)} - \left(I_{30}^{(0)} + I_{31}^{(0)}\right)} \Bigg[m^2 \xi_{\theta} \tanh \xi I_{10}^{(0)} - m^2\, \beta_\theta\, I_{20}^{(0)} + m^2\, I_{10}^{(0)} - \tanh \xi \left(I_{30}^{(0)} + I_{31}^{(0)}\right) \xi_{\theta} \nn\\
    && \qquad\qquad\qquad\qquad\qquad\qquad + \beta_{\theta} \left(I_{40}^{(0)} + I_{41}^{(0)}\right) + \beta\, I_{41}^{(0)} - \frac{5}{3} I_{31}^{(0)}\Bigg], \label{eq:CPI}\\
    C_{n} &=& \frac{\tanh \xi}{m^2\, I_{10}^{(0)} - \left(I_{30}^{(0)} + I_{31}^{(0)}\right)} \left(I_{31}^{(0)} - \frac{n_0 I_{41}^{(0)}}{\varepsilon_0 + P_0}\right),\label{eq:CN}\\
    C_{\pi} &=& - \frac{2 I_{31}^{(0)}}{m^2\, I_{10}^{(0)} - \left(I_{30}^{(0)} + I_{31}^{(0)}\right)}, \label{eq:Cpi}\\
    C_{\Sigma} &=& \frac{I_{31}^{(0)}}{m^2\, I_{10}^{(0)} - \left(I_{30}^{(0)} + I_{31}^{(0)}\right)}.\label{CSIG}
\eeq
Using Eq.~(\ref{eq:unuomunudot1}) and the recurrence relation $I_{30}^{(0)}-2 I_{31}^{(0)}=-\beta I_{41}^{(0)}$ in (\ref{eq:omunudot1}) and then eliminating $\dot{\xi}$, $\dot{\beta}$, and $\dot{u}^{\mu}$ (using Eqs.~(\ref{eq:xidot1}), (\ref{eq:betadot1}), and (\ref{eq:udot1})) we obtain
\beq
    \dot{\omega}^{\mu \nu} = D_{\Pi}^{\mu \nu} \theta + D_n^{[\mu \nu]}{}_{\alpha} \left(\nabla^{\alpha} \xi \right) + D_{\pi}^{[\nu}{}_{\lambda} \sigma^{\lambda\mu]} + D_{\Sigma 1}^{\alpha} \nabla^{[\mu} \omega^{\nu]}{}_{\alpha} + D_{\Sigma 2}^{[\mu \nu] \alpha}\, \nabla^{\lambda} \omega_{\alpha \lambda}. \label{eq:dotomega1}
\eeq
The various $D$-coefficients appearing in the above equation are given by the following expressions
\beq
    D_{\Pi}^{\mu \nu} &=& D_{\Pi 1} \omega^{\mu \nu} + D_{\Pi 2} u^{\alpha} u^{[\mu} \omega^{\nu]}{}_{\alpha},\label{eq:DPi1}\\
    D_n^{[\mu \nu]}{}_{\alpha} &=& - D_{n1}\left(u^{[\mu} \omega^{\nu]}{}_{\alpha} + g^{[\mu}{}_{\alpha} u^{\kappa} \omega^{\nu]}{}_{\kappa}\right) - D_{n2} u^{[\mu} \Delta^{\nu]}{}_{\rho} \omega^{\rho}{}_{\alpha},\label{eq:Dnmunua1.1}\\
    D_{\pi}^{[\mu}{}_{\lambda} &=& - \omega^{[\mu}{}_{\lambda} \frac{4 I_{31}^{(0)}}{(m^2\, I_{10}^{(0)} - 2 I_{31}^{(0)})} - u^{[\mu} u^{\alpha} \omega_{\alpha \lambda} \frac{4 (I_{30}^{(0)} - I_{31}^{(0)}) I_{31}^{(0)}}{(m^2\, I_{10}^{(0)} - 2\, I_{31}^{(0)}) \big[m^2\, I_{10}^{(0)} - (I_{30}^{(0)} + I_{31}^{(0)})\big]},\label{eq:Dnulam1}\\
    D_{\Sigma 1}^{\alpha} &=& - u^{\alpha} \frac{2\, I_{31}^{(0)}}{(m^2\, I_{10}^{(0)} - 2\, I_{31}^{(0)})}, \label{eq:Dsig1.1}\\
    D_{\Sigma 2}^{[\mu \nu] \alpha} &=& - u^{[\mu} g^{\nu]\alpha} \frac{2\, I_{31}^{(0)}}{ \left(m^2\, I_{10}^{(0)} - 2\, I_{31}^{(0)}\right)} - u^{[\mu} \Delta^{\nu]\alpha} \frac{2\, (I_{30}^{(0)} - I_{31}^{(0)}) I_{31}^{(0)}}{\left( m^2\, I_{10}^{(0)} - 2\, I_{31}^{(0)}\right) \left[m^2\, I_{10}^{(0)} - \left( I_{30}^{(0)} + I_{31}^{(0)}\right)\right]},\label{eq:Dsig2.1}
\eeq
where
\beq
    D_{\Pi 1} &=& - \frac{1}{\left(I_{10}^{(0)} - \frac{2}{m^2} I_{31}^{(0)}\right)} \left(\xi_{\theta} \tanh \xi\, I_{10}^{(0)} - \beta_{\theta} I_{20}^{(0)} + I_{10}^{(0)} - \frac{2}{m^2} \xi _{\theta}\, \tanh \xi\, I_{31}^{(0)} + \frac{2\, \beta_{\theta}\, I_{41}^{(0)}}{m^2} - \frac{10 I_{31}^{(0)}}{3\, m^2}\right),\label{eq:a1.1}\\
    D_{\Pi 2} &=& \frac{2}{m^2\, I_{10}^{(0)} - 2\, I_{31}^{(0)}} \Bigg[\beta_{\theta} \!\left(I_{40}^{(0)} - I_{41}^{(0)}\right) - \xi_{\theta}\! \left(I_{30}^{(0)} - I_{31}^{(0)}\right) \tanh \xi - \left(I_{30}^{(0)} - \frac{11}{3} I_{31}^{(0)}\right) + \frac{\left(I_{30}^{(0)} - I_{31}^{(0)}\right)}{m^2\, I_{10}^{(0)} - I_{30}^{(0)} - I_{31}^{(0)}}\nn\\
    && \!\!\!\!\!\times\bigg(\!m^2\, \xi_{\theta} \tanh\xi\, I_{10}^{(0)} - m^2 \beta_{\theta}\, I_{20}^{(0)} + m^2 I_{10}^{(0)} - \xi_{\theta}\! \left(\!I_{30}^{(0)} + I_{31}^{(0)}\!\right)\! \tanh \xi + \beta_\theta\! \left(\!I_{40}^{(0)} + I_{41}^{(0)}\!\right)\! +\beta I_{41}^{(0)} - \frac{5}{3} I_{31}^{(0)}\bigg)\! \Bigg],\label{eq:a2.1}\\
    D_{n1} &=& \frac{2 \tanh \xi}{\left(m^2\, I_{10}^{(0)} - 2\, I_{31}^{(0)}\right)} \left(I_{31}^{(0)} - \frac{n_0\, I_{41}^{(0)}}{\varepsilon_0 + P_0}\right), \label{eq:R_1.1}\\
    D_{n2} &=& \frac{\tanh \xi}{m^2I_{10}^{(0)} - \left(I_{30}^{(0)} + I_{31}^{(0)}\right)} \left(I_{31}^{(0)} - \frac{n_0 I_{41}^{(0)}}{\varepsilon_0 + P_0}\right) \frac{2 \left(I_{30}^{(0)} - I_{31}^{(0)}\right)}{\left(m^2\, I_{10}^{(0)} - 2\, I_{31}^{(0)}\right)}. \label{eq:R_3.1}
\eeq
Using Eqs. (\ref{eq:xidot1}), (\ref{eq:betadot1}), (\ref{eq:udot1}), (\ref{eq:unuomunudot1}), (\ref{eq:dotomega1}), and (\ref{eq_dspin4}), we finally obtain
\begin{eqnarray}
    \delta S^{\lambda,\mu\nu} &=& \tau_{\rm eq} \Big[B^{\lambda,\mu\nu}_{\Pi}\, \theta + B^{\lambda\kappa,\mu\nu}_{n}\, (\nabla_\kappa \xi) + B_{\pi}^{\lambda \kappa \delta, \mu \nu} \sigma_{\kappa \delta} + B_{\Sigma}^{\eta \beta \gamma \lambda, \mu \nu} \nabla_{\eta} \omega_{\beta \gamma}\Big], \label{deltaS1}
\end{eqnarray}
where different coefficients appearing on the right-hand side of Eq.~(\ref{deltaS1}) are the kinetic coefficients for spin-related phenomena.  These coefficients are listed in Eqs. (\ref{eq:betaPi}), (\ref{eq:betapi}), (\ref{eq:betan}), and (\ref{eq:betaSig}) where:
\begin{eqnarray}
    B_{\Pi}^{(1)} &=& \frac{4\, \mathfrak{s}^2}{3}\Bigg(\!-\frac{2}{m^2} \xi_{\theta}\, \sinh \xi\, I_{41}^{(1)} + \frac{2}{m^2} I_{51}^{(1)} \beta_{\theta}\, \cosh \xi + \frac{10}{3m^2} I_{52}^{(1)}  \beta\, \cosh \xi - \frac{2}{m^2} I_{41}^{(1)} \cosh \xi\, D_{\Pi 1}\Bigg),\label{eq:betaPi1}\\
    B_{\Pi}^{(2)} &=& \frac{4\, \mathfrak{s}^2}{3} \Bigg[\!-\frac{2}{m^2} \xi_{\theta}\, \sinh \xi I_{40}^{(1)} + \frac{4}{m^2} \xi_{\theta}\, \sinh \xi\, I_{41}^{(1)} + \frac{2}{m^2} I_{50}^{(1)} \beta_{\theta}\, \cosh \xi + \frac{2}{m^2} I_{51}^{(1)} \beta\, \cosh \xi - \frac{4}{m^2} I_{51}^{(1)} \beta_\theta\, \cosh \xi \nn\\
    &-& \frac{20}{3m^2} I_{52}^{(1)} \beta\, \cosh \xi - \left(I_{20}^{(1)} - \frac{3}{m^2} I_{41}^{(1)}\right) \cosh \xi\, D_{\Pi 2} - \frac{2}{m^2} \left(I_{40}^{(1)} - 2\, I_{41}^{(1)}\right) \cosh \xi\, C_\Pi\Bigg], \label{eq:betaPi2}\\
    B_{\Pi }^{(3)} &=& \frac{4\, \mathfrak{s}^2}{3}\Bigg( - \frac{2}{m^2} \xi_{\theta}\, \sinh \xi I_{41}^{(1)} + \frac{2}{m^2} I_{51}^{(1)} \beta_{\theta}\, \cosh \xi + \frac{10}{3m^2} I_{52}^{(1)} \beta\, \cosh \xi - \frac{2}{m^2} I_{41}^{(1)} \cosh \xi\, C_\Pi\Bigg), \label{eq:betaPi3}
\end{eqnarray}
\begin{eqnarray}
	B_{\pi}^{(1)} &=& \frac{16\, \mathfrak{s}^2}{3\, m^2} \beta\, \cosh \xi\, I_{42}^{(0)},
	\label{eq:betapi1}\\
	B_{\pi}^{(2)} &=&\frac{16\, \mathfrak{s}^2}{3\, m^2} \cosh \xi \left(\beta\, I_{42}^{(0)} - \frac{ I_{41}^{(1)}\, I_{31}^{(0)}}{m^2\, I_{10}^{(0)} - 2\, I_{31}^{(0)}}\right),
	\label{eq:betapi2}\\
	B_{\pi}^{(3)} &=& \frac{16\, \mathfrak{s}^2}{3\, m^2} \cosh \xi \left(\frac{I_{41}^{(1)}\, I_{31}^{(0)}}{m^2\, I_{10}^{(0)} - 2\, I_{31}^{(0)}}\right),
	\label{eq:betapi3}\\
	B_{\pi}^{(4)} &=& \frac{16\, \mathfrak{s}^2}{3\, m^2} \cosh \xi \left(\frac{I_{41}^{(1)}\, I_{31}^{(0)}}{m^2 I_{10}^{(0)}-\left(I^{(0)}_{30}+I_{31}^{(0)}\right)}\right),
	\label{eq:betapi4}
\end{eqnarray}
\begin{eqnarray}
	B_{n}^{(1)} &=& \frac{4 \mathfrak{s}^2}{3\, m^2} \cosh \xi \left[-\tanh \xi \left(m^2\, I_{21}^{(1)} - 2\, I_{42}^{(1)}\right) + \bigg(\frac{n_0 \tanh(\xi)}{\varepsilon_0 + P_0} \bigg) \left(m^2\, I_{31}^{(1)} - 2\, I_{52}^{(1)}\right)\right],\label{eq:betan1}\\
	B_{n}^{(2)} &=& \frac{8 \mathfrak{s}^2}{3\, m^2} \cosh \xi  \Bigg[\!-\tanh \xi \!\left(I_{41}^{(1)} - I_{42}^{(1)}\right) + \bigg(\frac{n_0 \tanh \xi }{\varepsilon_0 + P_0} \bigg)\!\! \left(I_{51}^{(1)} - I_{52}^{(1)}\right) - \frac{I^{(1)}_{41}\, \tanh \xi}{\left(m^2\, I_{10}^{(0)} - 2\, I_{31}^{(0)}\right)}\! \left(\!I_{31}^{(0)} - \frac{n_0 I_{41}^{(0)}}{\varepsilon_0 + P_0} \!\right) \!\! \Bigg],~~~~~~\label{eq:betan2}\\
	B_{n}^{(3)} &=& \frac{8 \mathfrak{s}^2}{3\, m^2} \cosh \xi \left[ - \tanh \xi\, I_{42}^{(1)} + \bigg(\frac{n_0 \tanh \xi}{\varepsilon_0 + P_0}\bigg) I_{52}^{(1)}\right],
	\label{eq:betan3}\\
	B_{n}^{(4)} &=& \frac{8 \mathfrak{s}^2}{3\, m^2} \cosh \xi \left[\frac{I^{(1)}_{41}\, \tanh \xi}{\left(m^2 I_{10}^{(0)}-2 I_{31}^{(0)}\right)}\left(I_{31}^{(0)}-\frac{n_0 I_{41}^{(0)}}{\varepsilon _0+P_0}\right)\right],
	\label{eq:betan4}\\
	B_{n}^{(5)} &=& \frac{8 \mathfrak{s}^2}{3\, m^2} \cosh \xi \Bigg[ - \tanh \xi\, I_{42}^{(1)} + \bigg(\frac{n_0 \tanh \xi}{\varepsilon_0 + P_0} \bigg) I_{52}^{(1)} - \frac{I_{41}^{(1)}\, \tanh \xi}{m^2\, I_{10}^{(0)} - \left(I_{30}^{(0)} + I_{31}^{(0)}\right)} \left(I_{31}^{(0)} - \frac{n_0 I_{41}^{(0)}}{\varepsilon_0 + P_0} \right)\Bigg], \label{eq:betan5}\\
	B_{n}^{(6)} &=& \frac{8 \mathfrak{s}^2}{3\, m^2} \cosh \xi \Bigg[ \frac{I_{41}^{(1)}\, \tanh \xi}{m^2\, I_{10}^{(0)} - \left(I_{30}^{(0)} + I_{31}^{(0)}\right)} \left(I_{31}^{(0)} - \frac{n_0 I_{41}^{(0)}}{\varepsilon_0 + P_0} \right)\Bigg], \label{eq:betan6}
\end{eqnarray}
\begin{eqnarray}
    B_{\Sigma}^{(1)} &=& - \frac{4 \mathfrak{s}^2}{3} \cosh \xi\, I_{21}^{(1)}, \label{eq:betaSig1}\\
    B_{\Sigma}^{(2)} &=& - \frac{8 \mathfrak{s}^2}{3\, m^2} \cosh \xi \left(I_{41}^{(1)} + \frac{I_{41}^{(1)}\, I_{31}^{(0)}}{m^2 I_{10}^{(0)}-2 I_{31}^{(0)}}\right), \label{eq:betaSig2}\\
    B_{\Sigma}^{(3)} &=& - \frac{8 \mathfrak{s}^2}{3\, m^2} \cosh \xi\, I_{42}^{(1)}, \label{eq:betaSig3}\\
    B_{\Sigma}^{(4)} &=& - \frac{8 \mathfrak{s}^2}{3\, m^2} \cosh \xi \left( \frac{I_{41}^{(1)}\, I_{31}^{(0)}}{m^2I_{10}^{(0)} - \left(I_{30}^{(0)} + I_{31}^{(0)}\right)}\right),\label{eq:betaSig4}\\
    B_{\Sigma}^{(5)} &=& \frac{8 \mathfrak{s}^2}{3\, m^2} \cosh \xi \left(\frac{I_{41}^{(1)}\, I_{31}^{(0)}}{m^2\, I_{10}^{(0)} - 2\, I_{31}^{(0)}}\right). \label{eq:betaSig5}
\end{eqnarray}


\section{Landau matching Conditions}
\label{sec:LanCon}

In this section we show that $\delta N^{\mu}$, $\delta T^{\mu\nu}$, and $\delta S^{\lambda,\mu\nu}$ given by Eqs.~(\ref{delN4}),  (\ref{delT3}), and (\ref{deltaS1}) satisfy the relations (\ref{eq:lm1}), (\ref{eq:lm2}), and (\ref{eq:lm3}).

\subsection{Proving \texorpdfstring{$u_{\mu}\delta N^{\mu}=0$}{}}

Projecting Eq.~\eqref{delN4} along $u_{\mu}$ we obtain
\beq
    u_{\mu}\delta N^{\mu} &=& - 4 I_{20}^{(1)} \dot{\xi} \tau_{\rm eq} \cosh \xi + 4 \left(I_{31}^{(1)} \beta \theta + I_{30}^{(1)} \dot{\beta}\right) \tau_{\rm eq} \sinh \xi. \label{D1}
\eeq
Using the recurrence relation (\ref{eq:ir1}) we can write down
\beq
    I_{20}^{(1)} = I_{10}^{(0)} = n_0, \qquad
    I_{31}^{(1)} = I_{21}^{(0)} = -P_0, \qquad
    I_{30}^{(1)} = I_{20}^{(0)} = \varepsilon_0. \label{D1.1}
\eeq
Substituting the above values for $I_{20}^{(1)}$, $I_{31}^{(1)}$, $I_{30}^{(1)}$ and the values of $\dot{\xi}$ and $\dot{\beta}$ from Eqs. (\ref{eq:xidot1}) and (\ref{eq:betadot1}) into Eq.~(\ref{D1}), we can show that the right-hand side of Eq.~\eqref{D1} vanishes.

\subsection{Proving \texorpdfstring{$u_{\mu}\delta T^{\mu\nu}=0$}{}}

Projecting Eq.~\eqref{delT3} along $u_{\mu}$ we obtain
\beq
    u_{\mu} \delta T^{\mu\nu} &=& - 4\tau_{\rm eq}\sinh \xi \left[I_{30}^{(1)} \dot{\xi} u^{\nu} + I_{31}^{(1)} \nabla^{\nu} \xi \right]
    + 4\tau_{\rm eq}\cosh \xi \left[I_{40}^{(1)} \dot{\beta} u^{\nu} + I_{41}^{(1)}\left(\beta \dot{u}^{\nu} + \nabla^{\nu} \beta +\beta \theta u^{\nu}\right)\right]. \label{D3a}
\eeq
Using Eq.~\eqref{eq:udot1}, the above equation can be written as 
\beq
    u_{\mu}\delta T^{\mu\nu} = -4\tau_{\rm eq}\!  \left[I_{30}^{(1)} \dot{\xi} \sinh \xi \!- I_{40}^{(1)} \dot{\beta} \cosh \xi \!- I_{41}^{(1)} \beta \theta \cosh \xi\right]\! u^\nu
    \!- 4 \tau_{\rm eq}\! \left[I_{31}^{(1)} \sinh \xi \!- I_{41}^{(1)} \cosh \xi\, \frac{n_0 \tanh \xi}{\varepsilon_0 + P_0}\right] \!\nabla^{\nu} \xi. \label{D3}
\eeq
Using the recurrence relations (\ref{eq:ir1}) and (\ref{eq:ir2}) we can write
\beq
    I_{30}^{(1)} = I_{20}^{(0)}, \qquad
    I_{40}^{(1)} = I_{30}^{(0)}=n_0, \qquad
    I_{41}^{(1)} = I_{31}^{(0)}= -\frac{1}{\beta} \left(I_{20}^{(0)} - I_{21}^{(0)}\right). \label{D5}
\eeq
Using the above relations along with the values of $\dot{\xi}$ and $\dot{\beta}$ from Eqs.~(\ref{eq:xidot1}) and (\ref{eq:betadot1}), we see that the first square bracket term on the right-hand side of Eq.~\eqref{D3} vanishes; see Eq.~(\ref{eq:enddot1}) for details. Using the relations (\ref{D1.1}) and (\ref{D5}), it can also be shown that the second square bracket term in Eq.~(\ref{D3}) is zero.

%
%
%
%
%
%
%

\subsection{Proving \texorpdfstring{$u_{\mu}\delta S^{\lambda,\mu\nu}=0$}{}}

Projecting Eq. (\ref{eq_dspin4}) along $u_{\lambda}$ we obtain
\beq
    u_{\lambda } \delta S^{\lambda, \mu \nu} &=& \frac{4 \mathfrak{s}^2}{3} \tau_{\text{eq}} \Bigg[\! -\sinh \xi \bigg\{\! I_{10}^{(0)}\, \dot{\xi}\, \omega^{\mu \nu} + \frac{2}{m^2}\Big(I_{41}^{(1)}\left(u^{[\mu} \omega^{\nu]}{}_{\alpha} \nabla^{\alpha} \xi + u^{\alpha} \omega^{[\nu}{}_{\alpha} \nabla^{\mu]} \xi \right) + \dot{\xi} \left(I_{40}^{(1)} u^{\alpha} u^{[\mu} \omega^{\nu ]}{}_{\alpha} + I_{41}^{(1)} \Delta^{\alpha[\mu} \omega^{\nu]}{}_{\alpha} \right) \! \Big)\!\bigg\}\nn\\
    &+& \cosh \xi \bigg\{\! \left(I_{31}^{(1)} \beta \theta \omega^{\mu \nu} + I_{30}^{(1)} \dot{\beta} \omega^{\mu \nu} \right) + \frac{2}{m^2} \dot{\beta} I_{50}^{(1)} u^{\alpha} u^{[\mu} \omega^{\nu]}{}_{\alpha} + \frac{2}{m^2} I_{52}^{(1)} \left(\beta \theta \Delta^{\alpha [\mu} \omega^{\nu]}{}_{\alpha} + \beta (\nabla^{[\mu} u^{\alpha} + \nabla^{\alpha} u^{[\mu}) \omega^{\nu]}{}_{\alpha} \right)\nn\\
    &+& \frac{2}{m^2} I_{51}^{(1)} \left(\beta \theta u^{\alpha} u^{[\mu} \omega^{\nu]}{}_{\alpha } + \left(\beta \dot{u}^{\alpha} + \nabla^{\alpha} \beta \right) u^{[\mu} \omega^{\nu]}{}_{\alpha } + \left(\beta \dot{u}^{[\mu} + \nabla^{[\mu} \beta \right) \omega^{\nu]}{}_{\alpha} u^{\alpha} + \dot{\beta} \Delta^{\alpha [\mu} \omega^{\nu]}{}_{\alpha} \right)\nn\\
    &-& \!\left(\! I_{20}^{(1)} - \frac{2}{m^2} I_{41}^{(1)} \!\right) \!\!\! \left(\!\! \dot{\omega}^{\mu \nu} + \frac{2}{ \left(\! m^2 I_{20}^{(1)} - 2 I_{41}^{(1)} \!\right)} \left(I_{40}^{(1)} - I_{41}^{(1)}\right) u^{\alpha} u^{[\mu} \dot{\omega}^{\nu]}{}_{\alpha} \!\!\right)\! - \frac{2}{m^2} I_{41}^{(1)} \!\left( u^{[\mu} \nabla^{\alpha} + u^{\alpha} \nabla^{[\mu}\right) \omega^{\nu]}{}_{\alpha}\!\bigg\}\!\Bigg].
    \label{D8}
\eeq
Using Eq.~(\ref{eq:omunudot1}), the above equation can further be written as
\beq
    u_{\lambda} \delta S^{\lambda, \mu \nu} &=& \frac{4 \mathfrak{s}^2}{3} \tau_{\text{eq}} \Bigg[ \! - \sinh \xi\, I_{10}^{(0)} \dot{\xi} \omega^{\mu \nu} - \frac{2\, \sinh \xi }{m^2} \Big(I_{41}^{(1)} \left(u^{[\mu}\omega^{\nu]}{}_{\alpha} \nabla^{\alpha} \xi + u^{\alpha} \omega^{[\nu}{}_{\alpha} \nabla^{\mu]} \xi\right) + \dot{\xi} \left(I_{40}^{(1)} u^{\alpha} u^{[\mu} \omega^{\nu]}{}_{\alpha} + I_{41}^{(1)} \Delta^{\alpha [\mu} \omega^{\nu]}{}_{\alpha}\right)\Big)\nn\\
    &+& \cosh \xi \bigg(I_{31}^{(1)} \beta \theta \omega^{\mu \nu} + I_{30}^{(1)} \dot{\beta} \omega^{\mu \nu} + \frac{2}{m^2} \dot{\beta} I_{50}^{(1)} u^{\alpha} u^{[\mu} \omega^{\nu]}{}_{\alpha} + \frac{2}{m^2}I_{52}^{(1)} \left(\beta \theta \Delta^{\alpha [\mu} \omega^{\nu]}{}_{\alpha} + \beta (\nabla^{[\mu} u^{\alpha} + \nabla^{\alpha} u^{[\mu}) \omega^{\nu]}{}_{\alpha}\right)\nn\\
    &+& \frac{2}{m^2} I_{51}^{(1)} \left(\beta \theta u^{\alpha} u^{[\mu}\omega^{\nu]}{}_{\alpha} + \left(\beta \dot{u}^{\alpha} + \nabla^{\alpha} \beta \right) u^{[\mu} \omega^{\nu]}{}_{\alpha} + \left(\beta \dot{u}^{[\mu} + \nabla^{[\mu} \beta \right) \omega^{\nu]}{}_{\alpha} u^{\alpha} + \dot{\beta} \Delta^{\alpha [\mu} \omega^{\nu]}{}_{\alpha}\right)\nn\\
    &+& \left(I_{10}^{(0)} \theta\, \omega^{\mu \nu} + I_{10}^{(0)}\, \dot{\xi}\, \omega^{\mu \nu}\, \tanh \xi - \dot{\beta}\, I_{20}^{(0)}\, \omega^{\mu \nu} \right)\nn\\
    &+& \frac{2}{m^2} \Bigg\{\tanh \xi \left(I_{30}^{(0)} - 3 I_{31}^{(0)}\right) \dot{\xi} u^{\alpha} u^{[\mu} \omega^{\nu]}{}_{\alpha} + \tanh \xi\, I_{31}^{(0)} \left(\nabla_{\lambda} \xi + \dot{\xi} u_{\lambda}\right) \left(u^{[\mu}\omega^{\nu]\lambda} - \omega^{\mu \nu} u^{\lambda} + u^{\alpha} g^{\lambda[\mu} \omega^{\nu]}{}_{\alpha} \right) \nn\\
    &-& \left(\dot{\beta}\left({I}_{40}^{(0)} - 3 {I}_{41}^{(0)}\right) u^{\alpha} u^{[\mu} \omega^{\nu]}{}_{\alpha} + \left(\nabla_{\lambda} \beta + \dot{\beta} u_{\lambda} \right) I_{41}^{(0)} u^{[\mu} \omega^{\nu] \lambda} - \dot{\beta}{I}_{41}^{(0)} \omega^{\mu \nu} + \left(\nabla_{\lambda} \beta + \dot{\beta} u_{\lambda} \right) I_{41}^{(0)} u^{\alpha} g^{\lambda [\mu} \omega^{\nu]}{}_{\alpha} \right)\nn\\
    &+& \left(I_{30}^{(0)} - 3 I_{31}^{(0)}\right) \theta\, u^{\alpha} u^{[\mu} \omega^{\nu]}{}_{\alpha} + \left(I_{30}^{(0)} - 2 I_{31}^{(0)}\right) \dot{u}^{\alpha} u^{[\mu}\omega^{\nu]}{}_{\alpha} + \left(I_{30}^{(0)} - 2 I_{31}^{(0)}\right) u^{\alpha} \dot{u}^{[\mu} \omega^{\nu]}{}_{\alpha}\nn\\
    &+& I_{31}^{(0)} \! \left(\! \omega^{[\nu\lambda} \nabla_{\lambda} u^{\mu]} + u^{[\mu} \nabla_{\lambda} \omega^{\nu]\lambda} - \theta\, \omega^{\mu \nu} + \omega^{[\nu}{}_{\alpha} \nabla^{\mu]} u^{\alpha} + u^{\alpha} \nabla^{[\mu} \omega^{\nu]}{}_{\alpha} \right)\!\!\Bigg\} - \frac{2}{m^2} I_{41}^{(1)} \!\left(u^{[\mu} \nabla^{\alpha} + u^{\alpha}\nabla^{[\mu} \right)\! \omega^{\nu]}{}_{\alpha}\!\!\bigg)\!\Bigg]. \label{D9}
\eeq
Now using Eq.~(\ref{eq:udot1}) we rewrite the above equation as
\beq
    u_{\lambda} \delta S^{\lambda,\mu \nu} &=& \frac{4 \mathfrak{s}^2}{3} \tau_{\text{eq}} \Bigg[\omega^{\mu\nu} \bigg(-I_{10}^{(0)} \dot{\xi}\, \sinh \xi + \frac{2}{m^2} I_{41}^{(1)} \dot{\xi}\, \sinh \xi + I_{30}^{(1)} \dot{\beta} \cosh \xi + I_{31}^{(1)} \beta\, \theta\, \cosh \xi - \frac{2}{m^2} I_{51}^{(1)} \cosh \xi\, \dot{\beta} \nn\\
    &-& \frac{2}{m^2} I_{52}^{(1)} \beta\, \theta\, \cosh \xi + I_{10}^{(0)} \dot{\xi} \sinh \xi - I_{20}^{(0)} \dot{\beta} \cosh \xi + I_{10}^{(0)} \theta\, \cosh \xi - \frac{2}{m^2} I_{31}^{(0)} \dot{\xi}\, \sinh \xi \nn\\
    &+& \frac{2}{m^2} I_{41}^{(0)} \dot{\beta}\, \cosh \xi - \frac{2}{m^2} I_{31}^{(0)} \theta\, \cosh \xi \bigg)\nn\\
    &+& \frac{2}{m^2} \left(\nabla^{\alpha} \xi \right) u^{[\mu} \omega^{\nu]}{}_{\alpha} \bigg( - I_{41}^{(1)} \sinh \xi + \frac{n_0 \tanh \xi }{\varepsilon_0 + P_0} I_{41}^{(0)} \cosh \xi + I_{31}^{(0)} \sinh \xi - \frac{n_0 \tanh \xi }{\varepsilon_0 + P_0} I_{41}^{(0)} \cosh \xi \bigg)\nn\\
    &+& \frac{2}{m^2} \left(\nabla^{[\mu} \xi \right) u^{\alpha} \omega^{\nu]}{}_{\alpha} \bigg( - I_{41}^{(1)} \sinh \xi + \frac{n_0 \tanh \xi }{\varepsilon_0 + P_0} I_{51}^{(1)} \cosh \xi + I_{31}^{(0)} \sinh \xi - \frac{n_0 \tanh \xi }{\varepsilon_0 + P_0} I_{41}^{(0)} \cosh \xi \bigg)\nn\\
    &+& \frac{2}{m^2} u^{\alpha} u^{[\mu} \omega^{\nu]}{}_{\alpha} \bigg( - I_{40}^{(1)} \dot{\xi} \sinh \xi + I_{41}^{(1)} \dot{\xi} \sinh \xi + I_{50}^{(1)} \dot{\beta}\, \cosh \xi + I_{51}^{(1)} \beta\, \theta\, \cosh \xi - I_{51}^{(1)} \dot{\beta} \cosh \xi - I_{52}^{(1)} \beta\,  \theta\,  \cosh \xi \nn\\
    &+& \left(I_{30}^{(0)} - 3 I_{31}^{(0)}\right) \dot{\xi}\, \sinh \xi + 2 I_{31}^{(0)} \dot{\xi} \sinh \xi - \left(I_{40}^{(0)} - 3 I_{41}^{(0)}\right) \dot{\beta} \cosh \xi - 2 I_{41}^{(0)} \dot{\beta}\, \cosh \xi + \left(I_{30}^{(0)} - 3 I_{31}^{(0)}\right) \theta  \cosh \xi \bigg) \nn\\
    &+& \frac{2}{m^2} \left(\nabla^{[\mu} u^{\alpha} + \nabla^{\alpha} u^{[\mu} \right) \omega^{\nu]}{}_{\alpha} \bigg(I_{52}^{(1)} \beta\, \cosh \xi +I_{31}^{(0)} \cosh \xi \bigg)\nn\\
    &+& \frac{2}{m^2}\left(u^{[\mu} \nabla^{\alpha} + u^{\alpha} \nabla^{[\mu} \right) \omega^{\nu]}{}_{\alpha} \bigg(I_{31}^{(0)} \cosh \xi - I_{41}^{(1)} \cosh \xi\bigg)\Bigg].
\eeq
From this equation it can be clearly seen that the coefficient of all the tensor objects on the right-hand side cancels out. Thus, we confirm that
\beq
u_{\lambda } \delta S^{\lambda ,\mu \nu }&=&0.
\eeq


\section{Calculation of \texorpdfstring{$\nu ^{\alpha}$}{}, \texorpdfstring{$\pi^{\alpha\beta}$}{} and \texorpdfstring{$\Pi$}{} } 
\label{Ag}
By contracting Eq.~(\ref{delN4}) with $\Delta _{\mu }^{\alpha}$, the following expression for particle diffusion current can be obtained 
\beq
    \nu^{\alpha} &=& \Delta_{\mu}^{\alpha} \delta\text{N}^{\mu}\nn\\
    &=& - 4\tau_{\rm eq} (\nabla^{\alpha} \xi)\cosh \xi I_{21}^{(1)} + 4 \tau_{\rm eq} I_{31}^{(1)} \left(\beta \dot{u}^{\alpha} + \nabla^{\alpha} \beta\right) \sinh \xi. \label{eq:diffc1}
\eeq
Using Eq.~(\ref{eq:udot1}), the above equation can be cast in the following simpler form
\beq
    \nu^{\alpha} &=& - 4\tau_{\rm eq} \left(\nabla^{\alpha }\xi \right) \Bigg[\cosh \xi I_{21}^{(1)} - \left(\frac{n_0 \tanh \xi }{\varepsilon_0 + P_0}\right) I_{31}^{(1)} \sinh \xi \Bigg]. \label{eq:diffc2}
\eeq
Contracting Eq.~(\ref{delT3}) with $\Delta _{\mu \nu }^{\alpha \beta }$ yields
\beq
    \pi^{\alpha \beta} = \Delta_{\mu \nu}^{\alpha \beta} \delta T^{\mu \nu}
    &=& \Delta_{\mu \nu}^{\alpha \beta} \Bigg[\! - 4 \tau_{\rm eq} \sinh \xi \left(\nabla_{\rho} \xi + u_{\rho} \dot{\xi} \right) \left(I_{30}^{(1)} u^{\mu} u^{\nu} u^{\rho} + I_{31}^{(1)} \left(\Delta^{\mu \nu} u^{\rho} + \Delta^{\rho \mu} u^{\nu} + \Delta^{\nu \rho} u^{\mu}\right)\right)\nn\\
    &+& 4 \tau_{\rm eq} \cosh \xi \left(\beta\nabla_{\rho} u_\lambda + \beta u_{\rho} \dot{u}_{\lambda} + u_\lambda \nabla_{\rho} \beta + u_\lambda u_{\rho} \dot{\beta}\right) \bigg\{I_{40}^{(1)} u^{\lambda} u^{\mu} u^{\nu} u^{\rho}\nn\\
    &+& I_{41}^{(1)} \left(\Delta^{\mu \lambda} u^{\nu} u^{\rho} + \Delta^{\nu \lambda} u^{\mu} u^{\rho} + \Delta^{\mu \nu} u^{\lambda} u^{\rho} + \Delta^{\lambda \rho} u^{\mu} u^{\nu} + \Delta^{\mu \rho} u^{\lambda} u^{\nu} + \Delta^{\nu \rho} u^{\lambda} u^{\mu}\right)\nn\\
    &+& I_{42}^{(1)} \left(\Delta^{\mu \lambda} \Delta^{\nu \rho} + \Delta^{\mu \rho} \Delta^{\nu \lambda} + \Delta^{\lambda \rho} \Delta^{\mu \nu} \right)\!\bigg\}\!\Bigg].
\eeq
Doing simple algebraic manipulations where  Eqs.~(\ref{eq:identities1})--(\ref{eq:identities2e}) are used, we find
\beq
    \pi^{\alpha \beta} = 8 \tau_{\rm eq} \cosh \xi \beta I_{42}^{(1)} \sigma^{\alpha \beta},
\eeq
where
$\sigma^{\alpha \beta} = \frac{1}{2} \left(\nabla^{\beta} u^{\alpha} + \nabla^{\alpha} u^{\beta} - \frac{2}{3} \Delta^{\alpha \beta} \nabla^{\lambda} u_{\lambda}\right)$ is the shear flow tensor. Thus, the bulk pressure $\Pi$ can be expressed by the formula
\beq
    \Pi = - \frac{1}{3} \Delta_{\mu \nu} \delta{T}^{\mu \nu}
    &=& - \frac{1}{3} \Delta_{\mu \nu} \Bigg[ - 4 \tau_{\rm eq} \sinh \xi \left(\nabla_{\rho} \xi + u_{\rho} \dot{\xi} \right) \left\{I_{30}^{(1)} u^{\mu} u^{\nu} u^{\rho} + I_{31}^{(1)} \left(\Delta^{\mu \nu} u^{\rho} + \Delta^{\rho \mu} u^{\nu} + \Delta^{\nu \rho} u^{\mu}\right)\right\}\nn\\
    &+& 4 \tau _{\rm eq} \cosh \xi \left(\beta\nabla_{\rho} u_\lambda + \beta u_{\rho} \dot{u}_{\lambda} + u_\lambda \nabla_{\rho} \beta + u_\lambda u_{\rho} \dot{\beta}\right) \bigg\{I_{40}^{(1)} u^{\lambda} u^{\mu} u^{\nu} u^{\rho}\nn\\
    &+& I_{41}^{(1)} \left(\Delta^{\mu \lambda} u^{\nu} u^{\rho} + \Delta^{\nu \lambda} u^{\mu} u^{\rho} + \Delta^{\mu \nu} u^{\lambda} u^{\rho} + \Delta^{\lambda \rho} u^{\mu} u^{\nu} + \Delta^{\mu \rho} u^{\lambda} u^{\nu} + \Delta^{\nu \rho} u^{\lambda} u^{\mu}\right)\nn\\
    &+& I_{42}^{(1)} \left(\Delta^{\mu \lambda} \Delta^{\nu \rho} + \Delta^{\mu \rho} \Delta^{\nu \lambda} + \Delta^{\lambda \rho} \Delta^{\mu \nu}\right)\bigg\}\Bigg].
\eeq
Using the relations defined in Eq.~(\ref{eq:identities1}) we obtain
\beq
    \Pi = 4\tau_{\rm eq} \left[I_{31}^{(1)} \dot{\xi} \sinh \xi - \cosh \xi \left(I_{41}^{(1)} \dot{\beta} + \frac{5}{3} I_{42}^{(1)} \beta\, \nabla^{\lambda } u_{\lambda}\right)\right]. \label{eq:bulkP1}
\eeq
Now using the recurrence relation (\ref{eq:ir1}) we can write 
\beq
    I_{41}^{(1)} &=& I_{31}^{(0)} = - \frac{1}{\beta} \left(I_{20}^{(0)} - I_{21}^{(0)}\right) = - \frac{1}{\beta} (\varepsilon _0 + P_0), \\
    I_{31}^{(1)} &=& I_{21}^{(0)} = - P_0 = - \frac{n_0}{\beta}.  
\eeq
Substituting  $I_{41}^{(1)}$ and $I_{31}^{(1)}$ from the above equations and the convective derivatives $\dot{\xi}$ and  $\dot{\beta}$ from Eqs.~(\ref{eq:xidot1}) and (\ref{eq:betadot1}) into Eq.~(\ref{eq:bulkP1}) the following result for the bulk pressure can be obtained
\beq
    \Pi &=& - 4\tau_{\text{eq}} \Bigg[\frac{n_0 \left(\cosh \xi \sinh^2 \xi \left(\varepsilon_0 \left(P_0 + \varepsilon_0 \right) - n_0 T \left(P_0 \left(z^2 + 3\right) + 3 \varepsilon_0 \right)\right)\right)}{\beta  \left(\varepsilon_0^2 \sinh^2 \xi - n_0 T \cosh^2 \xi \left(P_0 \left(z^2 + 3\right) + 3 \varepsilon_0 \right)\right)}\nn\\
    &-&\frac{\cosh \xi}{\beta} \left(\frac{n_0 \left(P_0 + \varepsilon_0\right) \left(P_0 \cosh^2 \xi + \varepsilon_0\right)}{n_0 T \cosh^2 \xi \left(P_0 \left(z^2 + 3\right) + 3 \varepsilon_0\right) - \varepsilon_0^2 \sinh^2 \xi}\right) + \frac{5\beta}{3} I^{(1)}_{42}\Bigg]\theta.
\eeq

\twocolumngrid

\newpage
\bibliography{spin-lit}{}

\providecommand{\href}[2]{#2}\begingroup\raggedright\begin{thebibliography}{10}

\bibitem{STAR:2017ckg}
{\bf STAR} Collaboration, L.~Adamczyk {\em et al.}, ``{Global $\Lambda$ hyperon
  polarization in nuclear collisions: evidence for the most vortical fluid},''
  \href{http://dx.doi.org/10.1038/nature23004}{{\em Nature} {\bf 548} (2017)
  62--65},
\href{http://arxiv.org/abs/1701.06657}{{\tt arXiv:1701.06657 [nucl-ex]}}.

\bibitem{Adam:2018ivw}
{\bf STAR} Collaboration, J.~Adam {\em et al.}, ``{Global polarization of
  $\Lambda$ hyperons in Au+Au collisions at $\sqrt{s_{_{NN}}}$ = 200 GeV},''
  \href{http://dx.doi.org/10.1103/PhysRevC.98.014910}{{\em Phys. Rev.} {\bf
  C98} (2018)  014910},
\href{http://arxiv.org/abs/1805.04400}{{\tt arXiv:1805.04400 [nucl-ex]}}.

\bibitem{Acharya:2019vpe}
{\bf ALICE} Collaboration, S.~Acharya {\em et al.}, ``{Measurement of
  spin-orbital angular momentum interactions in relativistic heavy-ion
  collisions},'' \href{http://dx.doi.org/10.1103/PhysRevLett.125.012301}{{\em
  Phys. Rev. Lett.} {\bf 125} (2020) no.~1, 012301},
  \href{http://arxiv.org/abs/1910.14408}{{\tt arXiv:1910.14408 [nucl-ex]}}.

\bibitem{Kornas:2019}
{\bf HADES} Collaboration, F.~Kornas {\em et al.}, {\em {Lambda Polarization in
  Au+Au collisions at $\sqrt{s_{NN}}=$ 2.4 GeV measured with HADES}}.
\newblock talk given at the Strange Quark Matter, Bari, Italy, June 11-15,
  2019.

\bibitem{Liang:2004ph}
Z.-T. Liang and X.-N. Wang, ``{Globally polarized quark-gluon plasma in
  non-central A+A collisions},''
  \href{http://dx.doi.org/10.1103/PhysRevLett.94.102301,
  10.1103/PhysRevLett.96.039901}{{\em Phys. Rev. Lett.} {\bf 94} (2005)
  102301}, \href{http://arxiv.org/abs/nucl-th/0410079}{{\tt
  arXiv:nucl-th/0410079 [nucl-th]}}.
[Erratum: Phys. Rev. Lett.96,039901(2006)].

\bibitem{Liang:2004xn}
Z.-T. Liang and X.-N. Wang, ``{Spin alignment of vector mesons in non-central
  A+A collisions},''
  \href{http://dx.doi.org/10.1016/j.physletb.2005.09.060}{{\em Phys. Lett.}
  {\bf B629} (2005)  20--26},
\href{http://arxiv.org/abs/nucl-th/0411101}{{\tt arXiv:nucl-th/0411101
  [nucl-th]}}.

\bibitem{Voloshin:2004ha}
S.~A. Voloshin, ``{Polarized secondary particles in unpolarized high energy
  hadron-hadron collisions?},''
\href{http://arxiv.org/abs/nucl-th/0410089}{{\tt arXiv:nucl-th/0410089
  [nucl-th]}}.

\bibitem{Betz:2007kg}
B.~Betz, M.~Gyulassy, and G.~Torrieri, ``{Polarization probes of vorticity in
  heavy ion collisions},''
  \href{http://dx.doi.org/10.1103/PhysRevC.76.044901}{{\em Phys. Rev.} {\bf
  C76} (2007)  044901},
\href{http://arxiv.org/abs/0708.0035}{{\tt arXiv:0708.0035 [nucl-th]}}.

\bibitem{Becattini:2007sr}
F.~Becattini, F.~Piccinini, and J.~Rizzo, ``{Angular momentum conservation in
  heavy ion collisions at very high energy},''
  \href{http://dx.doi.org/10.1103/PhysRevC.77.024906}{{\em Phys. Rev.} {\bf
  C77} (2008)  024906},
\href{http://arxiv.org/abs/0711.1253}{{\tt arXiv:0711.1253 [nucl-th]}}.

\bibitem{Becattini:2013vja}
F.~Becattini, L.~Csernai, and D.~J. Wang, ``{$\Lambda$ polarization in
  peripheral heavy ion collisions},''
  \href{http://dx.doi.org/10.1103/PhysRevC.93.069901,
  10.1103/PhysRevC.88.034905}{{\em Phys. Rev.} {\bf C88} (2013) no.~3, 034905},
  \href{http://arxiv.org/abs/1304.4427}{{\tt arXiv:1304.4427 [nucl-th]}}.
[Erratum: Phys. Rev.C93,no.6,069901(2016)].

\bibitem{Becattini:2013fla}
F.~Becattini, V.~Chandra, L.~Del~Zanna, and E.~Grossi, ``{Relativistic
  distribution function for particles with spin at local thermodynamical
  equilibrium},'' \href{http://dx.doi.org/10.1016/j.aop.2013.07.004}{{\em
  Annals Phys.} {\bf 338} (2013)  32--49},
\href{http://arxiv.org/abs/1303.3431}{{\tt arXiv:1303.3431 [nucl-th]}}.

\bibitem{Becattini:2007nd}
F.~Becattini and F.~Piccinini, ``{The Ideal relativistic spinning gas:
  Polarization and spectra},''
  \href{http://dx.doi.org/10.1016/j.aop.2008.01.001}{{\em Annals Phys.} {\bf
  323} (2008)  2452--2473},
\href{http://arxiv.org/abs/0710.5694}{{\tt arXiv:0710.5694 [nucl-th]}}.

\bibitem{Becattini:2016gvu}
F.~Becattini, I.~Karpenko, M.~Lisa, I.~Upsal, and S.~Voloshin, ``{Global
  hyperon polarization at local thermodynamic equilibrium with vorticity,
  magnetic field and feed-down},''
  \href{http://dx.doi.org/10.1103/PhysRevC.95.054902}{{\em Phys. Rev.} {\bf
  C95} (2017) no.~5, 054902},
\href{http://arxiv.org/abs/1610.02506}{{\tt arXiv:1610.02506 [nucl-th]}}.

\bibitem{Becattini:2015ska}
F.~Becattini, G.~Inghirami, V.~Rolando, A.~Beraudo, L.~Del~Zanna, A.~De~Pace,
  M.~Nardi, G.~Pagliara, and V.~Chandra, ``{A study of vorticity formation in
  high energy nuclear collisions},''
  \href{http://dx.doi.org/10.1140/epjc/s10052-015-3624-1,
  10.1140/epjc/s10052-018-5810-4}{{\em Eur. Phys. J.} {\bf C75} (2015) no.~9,
  406}, \href{http://arxiv.org/abs/1501.04468}{{\tt arXiv:1501.04468
  [nucl-th]}}.
[Erratum: Eur. Phys. J.C78,no.5,354(2018)].

\bibitem{Karpenko:2016jyx}
I.~Karpenko and F.~Becattini, ``{Study of $\Lambda $ polarization in
  relativistic nuclear collisions at $\sqrt{s_\mathrm {NN}}=7.7$ –200 GeV},''
  \href{http://dx.doi.org/10.1140/epjc/s10052-017-4765-1}{{\em Eur. Phys. J.}
  {\bf C77} (2017) no.~4, 213},
\href{http://arxiv.org/abs/1610.04717}{{\tt arXiv:1610.04717 [nucl-th]}}.

\bibitem{Xie:2017upb}
Y.~Xie, D.~Wang, and L.~P. Csernai, ``{Global Lambda polarization in high
  energy collisions},''
  \href{http://dx.doi.org/10.1103/PhysRevC.95.031901}{{\em Phys. Rev.} {\bf
  C95} (2017) no.~3, 031901},
\href{http://arxiv.org/abs/1703.03770}{{\tt arXiv:1703.03770 [nucl-th]}}.

\bibitem{Pang:2016igs}
L.-G. Pang, H.~Petersen, Q.~Wang, and X.-N. Wang, ``{Vortical Fluid and
  $\Lambda$ Spin Correlations in High-Energy Heavy-Ion Collisions},''
  \href{http://dx.doi.org/10.1103/PhysRevLett.117.192301}{{\em Phys. Rev.
  Lett.} {\bf 117} (2016) no.~19, 192301},
  \href{http://arxiv.org/abs/1605.04024}{{\tt arXiv:1605.04024 [hep-ph]}}.

\bibitem{Becattini:2017gcx}
F.~Becattini and I.~Karpenko, ``{Collective Longitudinal Polarization in
  Relativistic Heavy-Ion Collisions at Very High Energy},''
  \href{http://dx.doi.org/10.1103/PhysRevLett.120.012302}{{\em Phys. Rev.
  Lett.} {\bf 120} (2018) no.~1, 012302},
  \href{http://arxiv.org/abs/1707.07984}{{\tt arXiv:1707.07984 [nucl-th]}}.

\bibitem{Becattini:2020ngo}
F.~Becattini and M.~A. Lisa, ``{Polarization and Vorticity in the Quark Gluon
  Plasma},'' \href{http://arxiv.org/abs/2003.03640}{{\tt arXiv:2003.03640
  [nucl-ex]}}.

\bibitem{Niida:2018hfw}
{\bf STAR} Collaboration, T.~Niida,
  \href{http://dx.doi.org/10.1016/j.nuclphysa.2018.08.034}{``Global and local
  polarization of $\lambda$ hyperons in au+au collisions at 200 gev from
  star,''} in {\em Global and local polarization of $\Lambda$ hyperons in Au+Au
  collisions at 200 GeV from STAR}, vol.~982, pp.~511--514.
\newblock 2019.
\newblock \href{http://arxiv.org/abs/1808.10482}{{\tt arXiv:1808.10482
  [nucl-ex]}}.

\bibitem{Adam:2019srw}
{\bf STAR} Collaboration, J.~Adam {\em et al.}, ``Polarization of $\lambda$
  ($\bar{\Lambda}$) hyperons along the beam direction in au+au collisions at
  $\sqrt{s\_{\_{NN}}}$ = 200 gev,''
  \href{http://dx.doi.org/10.1103/PhysRevLett.123.132301}{{\em Phys.Rev.Lett.}
  {\bf 123} (2019) no.~13, 132301}, \href{http://arxiv.org/abs/1905.11917}{{\tt
  arXiv:1905.11917 [nucl-ex]}}.

\bibitem{Li:2017dan}
H.~Li, H.~Petersen, L.-G. Pang, Q.~Wang, X.-L. Xia, and X.-N. Wang, ``{Local
  and global $\Lambda$ polarization in a vortical fluid},''
  \href{http://dx.doi.org/10.1016/j.nuclphysa.2017.04.008}{{\em Nucl. Phys. A}
  {\bf 967} (2017)  772--775}, \href{http://arxiv.org/abs/1704.03569}{{\tt
  arXiv:1704.03569 [nucl-th]}}.

\bibitem{Li:2017slc}
H.~Li, L.-G. Pang, and X.-L. Wang, Qun wand~Xia, ``{Global $\Lambda$
  polarization in heavy-ion collisions from a transport model},''
  \href{http://dx.doi.org/10.1103/PhysRevC.96.054908}{{\em Phys. Rev.} {\bf
  C96} (2017) no.~5, 054908},
\href{http://arxiv.org/abs/1704.01507}{{\tt arXiv:1704.01507 [nucl-th]}}.

\bibitem{Sun:2017xhx}
Y.~Sun and C.~M. Ko, ``{$\Lambda$ hyperon polarization in relativistic heavy
  ion collisions from a chiral kinetic approach},''
  \href{http://dx.doi.org/10.1103/PhysRevC.96.024906}{{\em Phys. Rev.} {\bf
  C96} (2017) no.~2, 024906},
\href{http://arxiv.org/abs/1706.09467}{{\tt arXiv:1706.09467 [nucl-th]}}.

\bibitem{Sun:2018bjl}
Y.~Sun and C.~M. Ko, ``{Azimuthal angle dependence of the longitudinal spin
  polarization in relativistic heavy ion collisions},''
  \href{http://dx.doi.org/10.1103/PhysRevC.99.011903}{{\em Phys. Rev. C} {\bf
  99} (2019) no.~1, 011903}, \href{http://arxiv.org/abs/1810.10359}{{\tt
  arXiv:1810.10359 [nucl-th]}}.

\bibitem{Fang:2016vpj}
R.-h. Fang, L.-g. Pang, Q.~Wang, and X.-n. Wang, ``{Polarization of massive
  fermions in a vortical fluid},''
  \href{http://dx.doi.org/10.1103/PhysRevC.94.024904}{{\em Phys. Rev. C} {\bf
  94} (2016) no.~2, 024904}, \href{http://arxiv.org/abs/1604.04036}{{\tt
  arXiv:1604.04036 [nucl-th]}}.

\bibitem{Florkowski:2019qdp}
W.~Florkowski, A.~Kumar, R.~Ryblewski, and R.~Singh, ``{Spin polarization
  evolution in a boost invariant hydrodynamical background},''
  \href{http://dx.doi.org/10.1103/PhysRevC.99.044910}{{\em Phys. Rev. C} {\bf
  99} (2019) no.~4, 044910}, \href{http://arxiv.org/abs/1901.09655}{{\tt
  arXiv:1901.09655 [hep-ph]}}.

\bibitem{Florkowski:2019voj}
W.~Florkowski, A.~Kumar, R.~Ryblewski, and A.~Mazeliauskas, ``{Longitudinal
  spin polarization in a thermal model},''
  \href{http://dx.doi.org/10.1103/PhysRevC.100.054907}{{\em Phys. Rev. C} {\bf
  100} (2019) no.~5, 054907}, \href{http://arxiv.org/abs/1904.00002}{{\tt
  arXiv:1904.00002 [nucl-th]}}.

\bibitem{Gao:2020vbh}
J.-H. Gao, G.-L. Ma, S.~Pu, and Q.~Wang, ``{Recent developments in chiral and
  spin polarization effects in heavy-ion collisions},''
  \href{http://arxiv.org/abs/2005.10432}{{\tt arXiv:2005.10432 [hep-ph]}}.

\bibitem{Li:2020vwh}
F.~Li and S.~Y. Liu, ``{Anomalous Lorentz transformation and side jump of a
  massive fermion},'' \href{http://arxiv.org/abs/2004.08910}{{\tt
  arXiv:2004.08910 [nucl-th]}}.

\bibitem{Liu:2020bbd}
S.~Y. Liu, Y.~Sun, and C.~M. Ko, ``{Local spin polarizations in relativistic
  heavy ion collisions},'' in {\em {28th International Conference on
  Ultrarelativistic Nucleus-Nucleus Collisions}}.
\newblock 2, 2020.
\newblock \href{http://arxiv.org/abs/2002.11752}{{\tt arXiv:2002.11752
  [nucl-th]}}.

\bibitem{Liu:2019krs}
S.~Y. Liu, Y.~Sun, and C.~M. Ko, ``{Spin Polarizations in a Covariant
  Angular-Momentum-Conserved Chiral Transport Model},''
  \href{http://dx.doi.org/10.1103/PhysRevLett.125.062301}{{\em Phys. Rev.
  Lett.} {\bf 125} (2020) no.~6, 062301},
  \href{http://arxiv.org/abs/1910.06774}{{\tt arXiv:1910.06774 [nucl-th]}}.

\bibitem{Ayala:2020ndx}
A.~Ayala, D.~de~la Cruz, L.~Hernández, and J.~Salinas, ``{Relaxation time for
  the alignment between the spin of a finite-mass quark/antiquark and the
  thermal vorticity in relativistic heavy-ion collisions},''
  \href{http://arxiv.org/abs/2003.06545}{{\tt arXiv:2003.06545 [hep-ph]}}.

\bibitem{Ivanov:2020qqe}
Y.~Ivanov, ``{Global polarization in heavy-ion collisions based on axial
  vortical effect},'' \href{http://arxiv.org/abs/2006.14328}{{\tt
  arXiv:2006.14328 [nucl-th]}}.

\bibitem{Liu:2020ymh}
Y.-C. Liu and X.-G. Huang, ``{Anomalous chiral transports and spin polarization
  in heavy-ion collisions},''
  \href{http://dx.doi.org/10.1007/s41365-020-00764-z}{{\em Nucl. Sci. Tech.}
  {\bf 31} (2020) no.~6, 56}, \href{http://arxiv.org/abs/2003.12482}{{\tt
  arXiv:2003.12482 [nucl-th]}}.

\bibitem{Huang:2020xyr}
X.-G. Huang, ``{Vorticity and Spin Polarization --- A Theoretical
  Perspective},'' \href{http://arxiv.org/abs/2002.07549}{{\tt arXiv:2002.07549
  [nucl-th]}}.

\bibitem{Deng:2020ygd}
X.-G. Deng, X.-G. Huang, Y.-G. Ma, and S.~Zhang, ``{Vorticity in low-energy
  heavy-ion collisions},''
  \href{http://dx.doi.org/10.1103/PhysRevC.101.064908}{{\em Phys. Rev. C} {\bf
  101} (2020) no.~6, 064908}, \href{http://arxiv.org/abs/2001.01371}{{\tt
  arXiv:2001.01371 [nucl-th]}}.

\bibitem{Montenegro:2020paq}
D.~Montenegro and G.~Torrieri, ``{Linear response theory of relativistic
  hydrodynamics with spin},'' \href{http://arxiv.org/abs/2004.10195}{{\tt
  arXiv:2004.10195 [hep-th]}}.

\bibitem{Ivanov:2019ern}
Y.~B. Ivanov, V.~Toneev, and A.~Soldatov, ``{Estimates of hyperon polarization
  in heavy-ion collisions at collision energies $\sqrt{s_{NN}}=$ 4--40 GeV},''
  \href{http://dx.doi.org/10.1103/PhysRevC.100.014908}{{\em Phys. Rev. C} {\bf
  100} (2019) no.~1, 014908}, \href{http://arxiv.org/abs/1903.05455}{{\tt
  arXiv:1903.05455 [nucl-th]}}.

\bibitem{Becattini:2018duy}
F.~Becattini, W.~Florkowski, and E.~Speranza, ``{Spin tensor and its role in
  non-equilibrium thermodynamics},''
  \href{http://dx.doi.org/10.1016/j.physletb.2018.12.016}{{\em Phys. Lett.}
  {\bf B789} (2019)  419--425},
\href{http://arxiv.org/abs/1807.10994}{{\tt arXiv:1807.10994 [hep-th]}}.

\bibitem{Florkowski:2017ruc}
W.~Florkowski, B.~Friman, A.~Jaiswal, and E.~Speranza, ``{Relativistic fluid
  dynamics with spin},''
  \href{http://dx.doi.org/10.1103/PhysRevC.97.041901}{{\em Phys. Rev.} {\bf
  C97} (2018) no.~4, 041901},
\href{http://arxiv.org/abs/1705.00587}{{\tt arXiv:1705.00587 [nucl-th]}}.

\bibitem{Florkowski:2017dyn}
W.~Florkowski, B.~Friman, A.~Jaiswal, R.~Ryblewski, and E.~Speranza,
  ``{Spin-dependent distribution functions for relativistic hydrodynamics of
  spin-1/2 particles},''
  \href{http://dx.doi.org/10.1103/PhysRevD.97.116017}{{\em Phys. Rev.} {\bf
  D97} (2018) no.~11, 116017},
\href{http://arxiv.org/abs/1712.07676}{{\tt arXiv:1712.07676 [nucl-th]}}.

\bibitem{Florkowski:2018fap}
W.~Florkowski, R.~Ryblewski, and A.~Kumar, ``{Relativistic hydrodynamics for
  spin-polarized fluids},''
  \href{http://dx.doi.org/10.1016/j.ppnp.2019.07.001}{{\em Prog. Part. Nucl.
  Phys.} {\bf 108} (2019)  103709},
\href{http://arxiv.org/abs/1811.04409}{{\tt arXiv:1811.04409 [nucl-th]}}.

\bibitem{Bhadury:2020puc}
S.~Bhadury, W.~Florkowski, A.~Jaiswal, A.~Kumar, and R.~Ryblewski,
  ``{Relativistic dissipative spin dynamics in the relaxation time
  approximation},'' \href{http://arxiv.org/abs/2002.03937}{{\tt
  arXiv:2002.03937 [hep-ph]}}.

\bibitem{Weickgenannt:2020aaf}
N.~Weickgenannt, E.~Speranza, X.-l. Sheng, Q.~Wang, and D.~H. Rischke,
  ``{Generating spin polarization from vorticity through nonlocal
  collisions},'' \href{http://arxiv.org/abs/2005.01506}{{\tt arXiv:2005.01506
  [hep-ph]}}.

\bibitem{Speranza:2020ilk}
E.~Speranza and N.~Weickgenannt, ``{Spin tensor and pseudo-gauges: from nuclear
  collisions to gravitational physics},''
  \href{http://arxiv.org/abs/2007.00138}{{\tt arXiv:2007.00138 [nucl-th]}}.

\bibitem{Hattori:2019ahi}
K.~Hattori, Y.~Hidaka, and D.-L. Yang, ``{Axial Kinetic Theory and Spin
  Transport for Fermions with Arbitrary Mass},''
  \href{http://dx.doi.org/10.1103/PhysRevD.100.096011}{{\em Phys. Rev. D} {\bf
  100} (2019) no.~9, 096011}, \href{http://arxiv.org/abs/1903.01653}{{\tt
  arXiv:1903.01653 [hep-ph]}}.

\bibitem{Yang:2020hri}
D.-L. Yang, K.~Hattori, and Y.~Hidaka, ``{Quantum kinetic theory for spin
  transport: general formalism for collisional effects},''
\href{http://arxiv.org/abs/2002.02612}{{\tt arXiv:2002.02612 [hep-ph]}}.

\bibitem{Hattori:2019lfp}
K.~Hattori, M.~Hongo, X.-G. Huang, M.~Matsuo, and H.~Taya, ``{Fate of spin
  polarization in a relativistic fluid: An entropy-current analysis},''
  \href{http://dx.doi.org/10.1016/j.physletb.2019.05.040}{{\em Phys. Lett.}
  {\bf B795} (2019)  100--106},
\href{http://arxiv.org/abs/1901.06615}{{\tt arXiv:1901.06615 [hep-th]}}.

\bibitem{Shi:2020htn}
S.~Shi, C.~Gale, and S.~Jeon, ``{Relativistic Viscous Spin Hydrodynamics from
  Chiral Kinetic Theory},'' \href{http://arxiv.org/abs/2008.08618}{{\tt
  arXiv:2008.08618 [nucl-th]}}.

\bibitem{Gallegos:2020otk}
A.~Gallegos and U.~Gürsoy, ``{Holographic spin liquids and Lovelock
  Chern-Simons gravity},'' \href{http://arxiv.org/abs/2004.05148}{{\tt
  arXiv:2004.05148 [hep-th]}}.

\bibitem{Mathisson:1937zz}
M.~Mathisson, ``{Neue mechanik materieller systemes},''
{\em Acta Phys. Polon.} {\bf 6} (1937)  163--2900.

\bibitem{Itzykson:1980rh}
C.~Itzykson and J.~B. Zuber, {\em {Quantum Field Theory}}.
\newblock International Series In Pure and Applied Physics. McGraw-Hill, New
  York, 1980.
\newblock
\url{http://dx.doi.org/10.1063/1.2916419}.
\newblock

\bibitem{Weickgenannt:2019dks}
N.~Weickgenannt, X.-L. Sheng, E.~Speranza, Q.~Wang, and D.~H. Rischke,
  ``{Kinetic theory for massive spin-1/2 particles from the Wigner-function
  formalism},'' \href{http://dx.doi.org/10.1103/PhysRevD.100.056018}{{\em Phys.
  Rev.} {\bf D100} (2019) no.~5, 056018},
\href{http://arxiv.org/abs/1902.06513}{{\tt arXiv:1902.06513 [hep-ph]}}.

\bibitem{Florkowski:2018ahw}
W.~Florkowski, A.~Kumar, and R.~Ryblewski, ``{Thermodynamic versus kinetic
  approach to polarization-vorticity coupling},''
  \href{http://dx.doi.org/10.1103/PhysRevC.98.044906}{{\em Phys. Rev.} {\bf
  C98} (2018)  044906},
\href{http://arxiv.org/abs/1806.02616}{{\tt arXiv:1806.02616 [hep-ph]}}.

\bibitem{Hehl:1976vr}
F.~W. Hehl, ``{On the Energy Tensor of Spinning Massive Matter in Classical
  Field Theory and General Relativity},''
\href{http://dx.doi.org/10.1016/0034-4877(76)90016-1}{{\em Rept. Math. Phys.}
  {\bf 9} (1976)  55--82}.

\bibitem{Tinti:2020gyh}
L.~Tinti and W.~Florkowski, ``{Particle polarization, spin tensor and the
  Wigner distribution in relativistic systems},''
  \href{http://arxiv.org/abs/2007.04029}{{\tt arXiv:2007.04029 [nucl-th]}}.

\bibitem{DeGroot:1980dk}
S.~R. De~Groot, {\em {Relativistic Kinetic Theory. Principles and
  Applications}}.
\newblock
1980.
\newblock

\bibitem{Florkowski:2010cf}
W.~Florkowski and R.~Ryblewski, ``{Highly-anisotropic and strongly-dissipative
  hydrodynamics for early stages of relativistic heavy-ion collisions},''
  \href{http://dx.doi.org/10.1103/PhysRevC.83.034907}{{\em Phys. Rev.} {\bf
  C83} (2011)  034907},
\href{http://arxiv.org/abs/1007.0130}{{\tt arXiv:1007.0130 [nucl-th]}}.

\bibitem{Martinez:2010sc}
M.~Martinez and M.~Strickland, ``{Dissipative Dynamics of Highly Anisotropic
  Systems},'' \href{http://dx.doi.org/10.1016/j.nuclphysa.2010.08.011}{{\em
  Nucl. Phys.} {\bf A848} (2010)  183--197},
\href{http://arxiv.org/abs/1007.0889}{{\tt arXiv:1007.0889 [nucl-th]}}.

\end{thebibliography}\endgroup
\bibliographystyle{utphys}

\end{document}